\documentclass[AMA,Times1COL]{WileyNJDv5}
\usepackage{moreverb}

\usepackage{amsfonts,amssymb}
\DeclareSymbolFontAlphabet{\mathbb}{AMSb}
\usepackage{float}
\usepackage{fancyhdr}
\usepackage{fancybox}
\usepackage{ifthen}
\usepackage{url}
\usepackage{lscape,afterpage}
\usepackage{xspace}
\usepackage{epstopdf} 	
\usepackage{subfig}
\usepackage{setspace}
\usepackage{xr}
\usepackage[utf8]{inputenc}
\usepackage{graphicx}
\usepackage{amssymb, amsmath, amsthm}
\usepackage{makecell}
\usepackage{mathtools}
\usepackage{thmtools}
\usepackage{bbm}
\usepackage{wrapfig}
\usepackage{color,soul}
\usepackage{multirow}
\usepackage{hyperref}
\usepackage[tablesfirst]{endfloat}

\makeatletter
\newcommand*{\addFileDependency}[1]{
\typeout{(#1)}
\@addtofilelist{#1}
\IfFileExists{#1}{}{\typeout{No file #1.}}
}\makeatother

\newcommand*{\myexternaldocument}[1]{%
\externaldocument{#1}%
\addFileDependency{#1.tex}%
\addFileDependency{#1.aux}%
}

\myexternaldocument{appendix}

\newcommand{\Z}{{\mathbb{Z}}}

\newcommand{\E}{{\mathbb{E}}}

\allowdisplaybreaks
\usepackage{etoolbox}

\newcommand\BibTeX{{\rmfamily B\kern-.05em \textsc{i\kern-.025em b}\kern-.08em
T\kern-.1667em\lower.7ex\hbox{E}\kern-.125emX}}

\articletype{Research Article}%

\begin{document}

\title{Regression Not-to-the-Mean: An Oddity of Regression, Illustrated with the Risk of Overdose Deaths}

\author[1]{Kelly C. Kung*}

\author[2]{Natasha K. Martin}

\author[1]{Judith J. Lok}

\authormark{Kung, Martin, Lok}
\titlemark{Regression Not-to-the-Mean: An oddity of regression, illustrated with the risk of overdose deaths}

\address[1]{\orgdiv{Department of Mathematics and Statistics}, \orgname{Boston University}, \orgaddress{\state{Massachusetts}, \country{U.S.A.}}}

\address[2]{\orgdiv{Division of Infectious Diseases and Global Public Health}, \orgname{University of California San Diego}, \orgaddress{\state{California}, \country{U.S.A.}}}

\corres{*Kelly C. Kung, Judith J. Lok, CCDS 439,
665 Commonwealth Avenue,
Boston MA 02215. \email{kkung@bu.edu}}

\presentaddress{Judith J. Lok \\
CCDS 439 \\
665 Commonwealth Avenue \\
Boston MA 02215}

\abstract[Abstract]{
Recent works in econometrics have shown that there can be issues with applying a constant treatment effect model in longitudinal settings with staggered treatment and heterogeneous treatment effects.
We focus on the issue that the estimated constant treatment effect may be a weighted average, with some negative weights, of treatment effects that are heterogeneous across treatment durations.
When this issue arises, the estimated
constant treatment effect and estimated heterogeneous treatment effects may result in conflicting results.
Through the example of estimating the effect of drug-induced homicide (DIH) prosecutions reported by media on unintentional drug-overdose deaths in the United States, we illustrate how the negative weighting issue can lead to conflicting results in practice. 
Moreover, although research has shown that the negative weight issue may arise in linear regression models, we show this issue may also arise in logistic regression models.
Using a linear link, we estimated a constant treatment effect risk ratio of 0.977 (95\% CI: (0.866, 1.101)) and an average risk ratio of 0.728 (range: 0.507-0.979) over different treatment durations.
Using a logistic link, we estimated a constant treatment risk ratio effect of 1.064 (95\% CI: (0.972, 1.165)) and an average risk ratio of 0.739 (range: 0.538-1.008) over different treatment durations.
Under both models, the estimated constant treatment effect is either smaller in magnitude or has a different sign than almost all estimated heterogeneous treatment effects, suggesting a negative weighting issue is present.
Our results suggest additional care is needed when applying constant treatment effect models in longitudinal settings.
}


\maketitle

\renewcommand\thefootnote{}
\footnotetext{\textbf{Abbreviations:} DIH, drug-induced homicide; OLS, ordinary least squares; GAM, generalized additive model; CDC, Centers for Disease Control and Prevention.}

\renewcommand\thefootnote{\fnsymbol{footnote}}
\setcounter{footnote}{1}

\section{Introduction}
\label{sec:issues_w_constant_eff}

The difference-in-differences model is a well-known causal inference model that is used to estimate treatment effects from observational data. \cite{snow1856mode, bertrand2004much}
In practice, the difference-in-differences model is often used to estimate the effects of policies by comparing the results before and after the policies were enacted, assuming that in the absence of treatment, the expected change in outcomes over time is the same in treatment and control groups. \cite{powell2018medical, abouk2019association}
To estimate the treatment effect, one typically uses a linear regression model that accounts for state and time effects and contains a treatment variable.
An example of such a model is the two-way fixed effects (TWFE) regression model which accounts for both state and time fixed effects. 
The TWFE model is commonly used to estimate treatment effects in longitudinal data settings where there may be multiple groups and time periods.\cite{de2023two, doi:10.1086/719588}

Although the TWFE model is widely used to estimate treatment effects, recent literature in econometrics has shown that there are potential issues when using a constant treatment effect model in longitudinal settings.\cite{de2018fuzzy, de2020two, roth2022s, borusyak2024revisiting, sun2021estimating}
We focus on the longitudinal setting with a staggered adoption of treatment in which the treatment is applied to different groups at different points in time, as is often the case for policy adoption across different states.
Furthermore, we focus on the setting where treatment is absorbing so once a unit is treated, it will stay treated.
In particular, we examine the issue of negative weighting of heterogeneous treatment effects that depend on treatment duration that can arise in constant treatment effect models with staggered, absorbing treatment adoption.\cite{de2018fuzzy, de2020two, borusyak2024revisiting, sun2021estimating}

Under the Stable Unit Treatment Value Assumption, parallel trends assumption, and the assumption of no anticipation effects, 
the constant treatment effect estimator is conditionally unbiased for a weighted average of heterogeneous treatment effects. 
However, the weighted average is not ``proper'' since some weights may be negative.\cite{de2020two, borusyak2024revisiting, sun2021estimating}
The weights depend on the residuals of a linear regression model where we regress the treatment variables on the state and time fixed effects.
Weights are more likely to be negative for states that are treated for longer periods of time.
Hence, the estimated treatment effect from the constant treatment effect model may be a biased estimate, with a negative weighting of long-run causal effects.

Although such difference-in-differences models have been used in many different applications, there is not much awareness of this potential negative weighting issue, especially outside of the econometrics literature.
We show, through an example of estimating the effect of drug-induced homicide (DIH) prosecutions reported by the media on drug-overdose deaths, that this negative weighting issue may occur in practice.
DIH prosecutions were introduced to combat the rising rates of drug-overdose deaths by aiming to punish individuals who deliver drugs that result in drug-overdose deaths.\cite{drugInducedHom1986}
While the DIH prosecutions were originally intended to target drug dealers, in practice, friends and family members of drug-overdose death victims are also prosecuted.
As a result, there has been a growing interest in understanding the effect of DIH prosecutions on drug-overdose deaths.\cite{carroll2021drug, lee2022longitudinal, carroll2022discussion}

We estimate the effect of DIH prosecutions reported by the media on drug-overdose deaths using models with 1) a linear link function and 2) a logistic link function under the assumptions of a) constant treatment effect and b) heterogeneous treatment effects depending on the treatment duration.
In this DIH prosecutions application, the estimated treatment effects under the constant treatment effect assumption are generally either smaller in magnitude or have a different sign than the estimated heterogeneous treatment effects, suggesting that the negative weighting of heterogeneous treatment effects issue is present.
Additionally, we compute the weights using the residuals from a linear regression model of the treatment variable on the state and time fixed effects, showing that the weights tend to be negative for states that have had at least one DIH prosecution reported by the media earlier in the analysis period. 
This result follows the theory from the econometrics literature.
The DIH application shows the importance of model specification in practice and how using a constant treatment effect model can, in certain settings, lead to erroneous results.

Section \ref{sec:theoretical_background} summarizes the theoretical findings and explains the occurrence of the negative weights. 
Section \ref{dih_intro} introduces drug-induced homicide prosecutions and the data used to estimate their effects on drug-overdose deaths.
Sections \ref{sec:linear_link_model} and \ref{sec:logistic_link_models} present the models and results using a model with a linear link and a model with a logistic link, respectively.
Section \ref{sec:discussion} discusses these results. 

\section{Negative Weighting of Treatment Effects}\label{sec:theoretical_background}


To illustrate the negative weighting issue, consider longitudinal data with groups (here, we focus on state-level data) $s$ and time periods $t$. 
While the results generalize to settings with multiple units $i$ per group $s$, our example of drug-induced homicide prosecutions focuses on state-level longitudinal data where there is one unit per group, namely the state. 
Accordingly, we use group-level notation $s$ rather than unit-group-level notation.
Consider data generated by a general Ordinary Least Squares (OLS) model:
\begin{align}\label{eq:data_generating_model}
    Y_{st} = \alpha_s + \gamma_t + A_{st}\beta_{st} + \epsilon_{st},
\end{align}
where $Y_{st}$ indicates the outcome for state $s$ at time $t$, $\alpha_s$ and $\gamma_t$ indicate the state and time fixed effects, respectively, $A_{st}$ is a treatment variable where once a state is treated, it remains treated (i.e. $A_{st} = \textbf{1}\{K_{st} \geq 0\}$ where $K_{st}$ is the number of time periods since state $s$ was first treated), and $\beta_{st}$ is the heterogeneous treatment effect. 
Furthermore, the fitting assumes that $\epsilon_{st}$ is such that $\E\left(\epsilon_{st} \mid A_{st}\right) = 0$ and $var\left(\epsilon_{st} \mid A_{st}\right) < \infty$.
Note that the treatment variables $A_{st}$ may be correlated within the state $s$, but we assume that they are independent between different states. 
Suppose we fit the data with a difference-in-differences OLS model with a constant treatment effect:
\begin{align}\label{eq:ols_outcome_model}
    Y_{st} = {\alpha}_s + {\gamma}_t + A_{st} \beta^{constant}  + {\upsilon}_{st},
\end{align}
where $Y_{st}, \alpha_s, \gamma_t, A_{st}$ are defined as before, $\beta^{constant}$ is the constant treatment effect, and we assume that ${\upsilon}_{st}$ is such that $\E\left({\upsilon}_{st} \mid A_{st}\right) = 0$ and $var\left({\upsilon}_{st} \mid A_{st} \right) < \infty$.

One can show that the constant treatment effect estimator $\hat{\beta}^{constant}$ is conditionally unbiased for a weighted average of the heterogeneous effects $\beta_{st}$.
Following the setup in previous work, consider the following assumptions: \textit{Stable Unit Treatment Value Assumption (SUTVA)}, \textit{parallel trends}, and \textit{no anticipation effects}.\cite{de2020two, borusyak2024revisiting} 
\begin{assumption}[Stable Unit Value Treatment Value Assumption]\label{sutva}
The potential outcome of state $s$ at time period $t$ only depend on the treatment of state $s$ and time period $t$.
\end{assumption}
\noindent By the consistency property of SUTVA, the observed outcome for state $s$ and time period $t$ is given by the potential outcome under treatment $A_{st}$, i.e. $Y_{st} = Y_{st}(A_{st})$.
\begin{assumption}[Parallel Trends Assumption]\label{parallel_trends}
There exists non-stochastic $\alpha_s$ and $\gamma_t$ such that $\E(Y_{st}(0) | \textbf{A}) = \alpha_s + \gamma_t$ for all states $s$ and time periods $t$ where $\textbf{A}$ is the set of all treatments $A_{st}$.
\end{assumption}
\noindent The parallel trends assumption states that, in the absence of treatment, the expected outcome is given by the state and time effects. 
Furthermore, the average change in outcomes over time periods $t$ and $t'$ where $t \neq t'$ is the same across states $s$.
In particular, this difference is given by $\gamma_t - \gamma_{t'}$.
\begin{assumption}[No Anticipation Effects]\label{no_anticipation_effects}
$Y_{st} = Y_{st}(0)$ for all states $s$ and time periods $t < T_s^*$, where $T_s^*$ is the time that state $s$ is first treated.
\end{assumption}
\noindent That is, the no anticipation effects assumption states that the untreated outcome does not depend on whether or when the state $s$ will be treated in the future. 
Furthermore, the outcome under no treatment is given by $Y_{st}(0)$ for state $s$ and time periods $t < T_s^*$.

Under these assumptions, we can formally state the main theorem that is applied in our paper.
\begin{theorem}\label{thm_1}
Assume that SUTVA (Assumption \ref{sutva}), parallel trends assumption (Assumption \ref{parallel_trends}), and no anticipation effects assumption (Assumption \ref{no_anticipation_effects}) hold.
The OLS estimator $\hat{\beta}^{constant}$ in Equation~\eqref{eq:ols_outcome_model} is conditionally unbiased, given the set of treatments \textbf{A} for states $s$ and time periods $t$, for the weighted average of heterogeneous treatment effects $\beta_{st}$ given in Equation~\eqref{eq:data_generating_model}.\cite{de2020two, borusyak2024revisiting}  
That is, 
\begin{align}\label{tx_effect_as_wt_avg}
    \E\left(\hat{\beta}^{constant} \bigg| \textbf{A}\right) = \sum_{st: A_{st} > 0} w_{st} {\beta}_{st},
\end{align}
where 
\begin{align}\label{eq:weights_eq}
w_{st} = \frac{\tilde{A}_{st}A_{st}}{\sum_{st}\tilde{A}_{st}^2},
\end{align}
where $\tilde{A}_{st}$ are the residuals from an OLS fit of the treatment variable $A_{st}$ on the state and time fixed effects and $\sum_{st: A_{st} > 0} w_{st} = 1$.
\end{theorem}

The proof of Theorem \ref{thm_1} can be found in Appendix \ref{appendix:proof_of_thm_1}. 
The proof uses the Frisch-Waugh-Lovell Theorem to show that the estimate of the constant treatment effect can be re-written in terms of residuals of the treatment effect on the states and time fixed effects.\cite{de2020two, borusyak2024revisiting, frisch1933partial, proof_for_sec_2}
Then, using the consistency property of SUTVA, the no anticipation effects assumption, and Equation~\eqref{eq:data_generating_model}, we show that the expectation of $Y_{st}$ can be rewritten as the expectation of the potential outcome under no treatment plus the potential outcome under the treatment $A_{st}$ multiplied by the treatment effects $\beta_{st}$.
Using the parallel trends assumption and the OLS property that the residuals of an OLS model of the treatment variable on the state and time fixed effects are orthogonal to the space spanned by the state and time indicators, we arrive at Equation \eqref{tx_effect_as_wt_avg}.
To show that the weights sum to one, we substitute the treatment variable $A_{st}$ by the summation of the residual $\tilde{A}_{st}$ and the projection of $A_{st}$ on the state and time fixed effects and apply the orthogonality property of OLS residuals to the projection.

The weights $w_{st}$ may however not be ``proper'' in the sense that some $w_{st}$ may be negative in settings where treatment is staggered and treatment effects are heterogeneous. \cite{de2020two, borusyak2024revisiting, sun2021estimating}
To understand why negative weights may occur, one has to understand how treatment effects are estimated in difference-in-differences models.
Typically in difference-in-differences, one compares outcomes of the treated group after treatment occurred with the outcomes of the control group where states are untreated.
However, when the treatment is staggered, the control group effectively used can consist of already-treated states.
OLS uses these treated states to estimate the time-fixed effects; a so-called ``forbidden comparison''. \cite{de2020two, borusyak2024revisiting, sun2021estimating}
When the true treatment effects are homogeneous, the time effects are accurately estimated.
However, when the true treatment effects are heterogeneous, these forbidden comparisons lead to biased estimates of the treatment effect. \cite{de2020two, borusyak2024revisiting, sun2021estimating}
In particular, in settings with staggered treatment, there may be an overweighting of short-term effects and a negative weighting (or at least underweighting) of long-term effects. \cite{de2020two, borusyak2024revisiting, sun2021estimating}

To understand this, note from Equation \eqref{eq:weights_eq} that whether the weights are negative depends on $\tilde{A}_{st}$, i.e. the residual for the regression where one regresses $A_{st}$ on the fixed state and time effects.
The estimated fixed state effects and time effects for $A_{st}$, denoted by $\hat{\eta}$, tend to be higher for states that were treated earlier in the analysis because these states have more treated time periods, and for later time periods since these time periods have more treated states. \cite{de2020two}
Hence, since $\tilde{A}_{st} = A_{st} - X_{st}\hat{\eta}$, where $X_{st}$ are the state and time indicators, treated states that were treated earlier are more likely to have negative weights $w_{st}$ in Equation~\eqref{eq:weights_eq}, especially at later time periods.
This results in potential negative weighting of long-term effects.
As a result of the negative weights, it is possible that the estimated constant treatment effect is positive even if all the $\beta_{st}$ are negative.

We show, through the example of estimating DIH prosecutions reported by the media on unintentional drug-overdose deaths, that this negative weighting phenomenon can happen in practice. 
Furthermore, although the theory of the negative weighting has been developed for OLS models, we show that negative weighting can also occur in logistic regression models.

\section{Drug-Induced Homicide Prosecutions}\label{dih_intro}
\subsection{Background}
In 1986, the U.S. Congress passed the Anti-Drug Abuse Act, which includes a law in which anyone who knowingly or intentionally distributes controlled substances resulting in death is subject to a sentencing enhancement of a 20-year mandatory minimum in prison. \cite{drugInducedHom1986}
A number of states also passed similar drug-induced homicide (DIH) laws to punish those who distributed drugs resulting in death. 
However, until 2000, these laws were almost never invoked in drug law enforcement. \cite{healthinjusticeaction}
States adopted DIH laws, with the aim that such policies might deter drug trafficking and prevent overdoses as a result. \cite{njLaw}
Per the wording of DIH laws, though, anyone who distributes controlled substances, including friends and family members of individuals who died from drug overdose, can be prosecuted. \cite{people_v_boand} 
Thus, the message intended for high-level drug dealers may be reaching friends and family members of individuals who died from drug overdose and low-level drug dealers instead, as these are the individuals who are usually prosecuted. \cite{alliance2017overdose, walker_2017}

As a result, due to legal liability concerns, DIH prosecutions may be discouraging people from seeking help during a drug overdose and may in fact be aggravating the overdose risk.\cite{carroll2021drug}
There has been recent research arguing that DIH prosecutions do not prevent drug-overdose deaths \cite{carroll2021drug} and, in response, research that argues that DIH prosecutions actually prevent drug-overdose deaths.\cite{lee2022longitudinal, carroll2022discussion}
Heterogeneity in state-level adoption of DIH prosecutions in response to the overdose crisis provides an opportunity for quantitative modeling of the effect of DIH prosecutions on fatal overdose patterns.
Here, we estimate the effect of DIH prosecutions reported by the media on drug-overdose deaths.

\subsection{Data}\label{sec:dih_data}
\subsubsection{Outcome: Unintentional drug-overdose deaths}
To assess the effect of DIH prosecutions reported by the media on drug-overdose deaths, we obtained monthly outcome data on unintentional drug-overdose deaths for people who were at least 18 years old for all 50 U.S. states from 1999 to 2019 from the Centers for Disease Control and Prevention (CDC), using International Classification of Diseases, Tenth Revision codes X40-X44. 
However, the CDC suppresses the number of drug-overdose deaths if it is below ten, leading to missing monthly data. 
Although the monthly data were sometimes missing, we were able to obtain most of the yearly data since almost all of the numbers of unintentional drug-overdose deaths surpassed ten when aggregated yearly.
Using the yearly number of drug-overdose deaths, combined with linear interpolation, we imputed the missing monthly number of unintentional drug-overdose deaths in 2000-2019 such that the total number of imputed deaths for a state $s$ in year $u$ is equal to the number of unaccounted deaths for that state and for that year (see Appendix \ref{interpolation} for details). 
We then aggregated the number of unintentional drug-overdose deaths into six-month intervals (from January to June and July to December) and analyzed the data as such. 
Since we observed the yearly number of drug-overdose deaths in general, the imputation process of the missing monthly unintentional drug-overdose deaths essentially just allocated the observed yearly number of overdose deaths to the two six-month intervals.  
We denote the risk of unintentional drug-overdose deaths in a state $s$ and six-month time interval $t$ as $Y_{st}$.
Instead of estimating the effects of the various interventions on a risk difference scale, we estimate the effects on a risk ratio scale since the intervention effect could be proportionally higher in states with a higher overdose death risk.\cite{powell2018medical, abouk2019association}

\subsubsection{DIH Prosecutions and Other Policy Measures}\label{sec:define_intervention_var}
In the absence of centralized criminal justice data tracking DIH prosecutions, we used data detailing mass media coverage of DIH prosecutions in 2000-2019 collected by the Health in Justice Action Lab. \cite{health_in_justice_lab}
We filtered the data to exclude DIH prosecutions reported by the media where the individuals who died from drug overdose were known to be younger than 18 years old.
We focus on DIH prosecutions reported by the media because media reports are a good medium to spread information about DIH prosecutions since they are readily available and widely accessible.
Furthermore, media reports capture the intended effect of DIH prosecutions---to deter people from using and distributing drugs by instilling fear of being prosecuted for homicide. 

\begin{figure}[!htb]
\centerline{\includegraphics{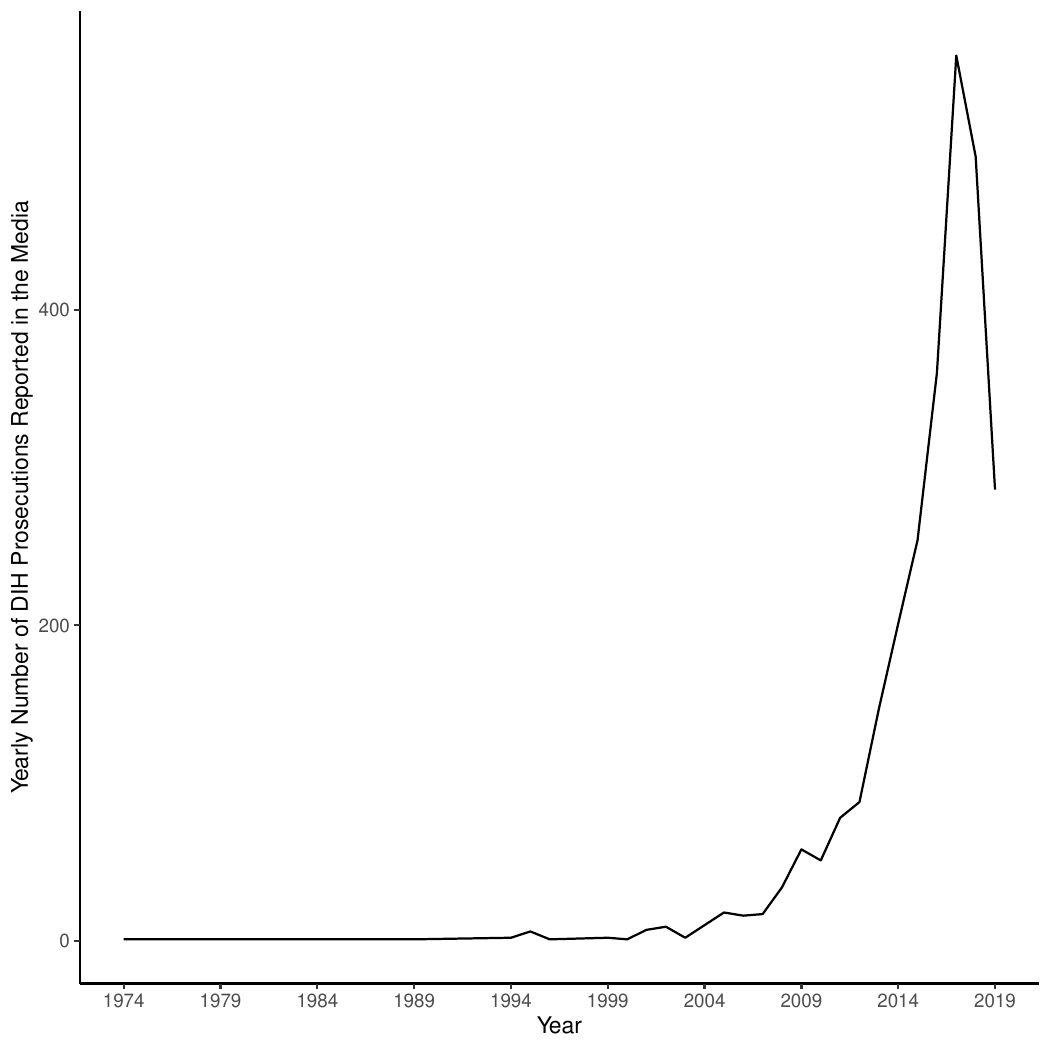}}
    \caption{Total number of yearly drug-induced homicide (DIH) prosecutions reported by the media in all 50 U.S. states from 1974 to 2019 where individuals who died were 18 and up.}
    \label{fig:dih_prosecutions_per_yr}
\end{figure}

For each state, the \textit{intervention date} for DIH prosecutions reported by the media was the first charge date that we found for DIH prosecutions reported by the media in the state in 2000-2019.
We denote the intervention date for state $s$ as $T_s^*$.
The intervention date for the intervention of interest, DIH prosecutions reported by the media, for each state can be found in Appendix \ref{appen:intervention_dates} Table \ref{appendix_tab:dih_int_dates}. 
We focus on the years 2000-2019 because 1) the CDC data on unintentional drug-overdose deaths date back only to 1999 and 2) prior to 2000, states rarely prosecuted people for drug-induced homicide (Figure \ref{fig:dih_prosecutions_per_yr}).

To address the growing drug overdose risk during this time period, states also adopted other policies. 
To account for potential confounding due to other policy measures, we obtained data on the following additional policies that have been seen in prior research to potentially impact drug-overdose deaths. \cite{abouk2019association, doi:10.1086/719588, mcclellan2018opioid}
From the Prescription Drug Abuse Policy System, we collected data on the presence of naloxone access laws---one law where pharmacists can dispense naloxone without a prescription and one law where pharmacists cannot dispense naloxone without a prescription, the presence of legalized medical marijuana laws and recreational marijuana laws, the presence of 911 Good Samaritan laws, and the presence of Prescription Drug Monitoring Programs in each state.\cite{pdaps_nal, pdaps_mml, pdaps_rml, pdaps_gsl, pdaps_pdmp}  
Data on whether a state expanded Medicaid as part of the Affordable Care Act were collected from the Henry J. Kaiser Family Foundation.\cite{kff_db} 
We made several adjustments to the policy intervention dates upon further research and also included policy intervention dates for some states that were not captured in the above datasets.
Intervention dates for the relevant policy measures for each state is presented in Appendix \ref{appen:intervention_dates} Table \ref{appendix_tab:other_polices_int_date}.

\subsection{Intervention Definition}\label{sec:intervention_def}
Our primary focus is to estimate the effect of having at least one DIH prosecution media report versus never having any DIH prosecutions reported by the media on unintentional drug-overdose deaths.
Here, we describe the intervention variable in the context of DIH prosecutions reported by the media.
We define the intervention variables for the other relevant policy measures and the DIH prosecutions reported by the media in the same way. 
Denote the six-month time interval $t$ as $\mathcal{I}_{t} = \left(D_{t,1}, D_{t,L}\right)$, where $D_{t,1}$ and $D_{t,L}$ indicate the first and last dates of time interval $t$, respectively.
For each time interval $t$, the intervention variable is defined as the proportion of days in time interval $t$ in which the intervention was in effect:
\begin{align}\label{eq:tx_primary}
    A_{st} = \begin{cases} 0 & \text{if $D_{t,L} < T_s^*$} \\
    1 & \text{if $D_{t,1} > T_s^*$}\\
    \frac{\text{\# of days exposed to intervention in $\mathcal{I}_{t}$}}{\text{number of days in $\mathcal{I}_{t}$}} & \text{if $T_s^* \in \mathcal{I}_t$}. \end{cases}
\end{align}
That is, if by the end of time interval $t$, there had not been any DIH prosecutions reported by the media in state $s$, then $A_{st} = 0$.
If by the beginning of time interval $t$, there had been at least one DIH prosecution reported by the media in state $s$, then $A_{st} = 1$.
Otherwise, if the intervention date for state $s$ occurred in $\mathcal{I}_t$, $A_{st}$ is equal to the proportion of days in the time interval in which the state was exposed to the intervention.
Furthermore, we let $K_{st} \in \Z$ be the number of six-month time intervals that the intervention has been in effect.
If the intervention occurred in time interval $t$, then  $K_{st} = 0$.
If time interval $t$ is before $T_s^*$, then $K_{st} < 0$, and if time interval $t$ is after $T_s^*$, then $K_{st} > 0$.

\section{Negative Weighting of Drug-Induced Homicide Prosecution Effects in Models with Linear Link Function}\label{sec:linear_link_model}
To analyze the effect of DIH prosecutions reported by the media on the unintentional drug-overdose death risk when assuming a constant treatment effect, we use a difference-in-differences-like generalized additive model (GAM).
Under SUTVA, the parallel trends assumption, and no anticipation effects assumption, we can estimate potential outcomes for the treatment group had the treatment not occurred, for post-treatment time periods, using the control group.
The treatment effect in the treated is the difference between the observed outcome for the treatment group and the potential outcome had the treatment group not received treatment.
In a classic difference-in-differences setting, there are only two time periods (before and after treatment) and two groups of states (treated states and control states).
In our analysis, we apply the differences-in-differences model to a setting with forty time periods and fifty states which can be treated at different time periods.
Hence, we also have to account for state and time effects, which we estimate using a semi-parametric generalized additive model.

The generalized additive model (GAM) assumes that a function $g$ of the mean outcome is given by the sum of functions of $M < \infty$ predictors $X_1, \dots, X_m$:\cite{hastie1990generalized}
\begin{align}
    g(\E(Y)) = \sum_{m}^M f_m(X_m),
\end{align}
where $g$ is a link function and the $f_m$ can be parametric or non-parametric functions such as regression splines.
For example, the OLS model is a specific example of GAMs, in which $g$ is the identity link function and $f_m(X_m) = \beta_m X_m$. 
By using a GAM, we can estimate treatment effects using both non-parameteric and parametric functions.
We use a GAM to estimate smoothed time effects in the different U.S. regions using cubic regression splines.
The list of states and their U.S. regions were obtained from the U.S. Census (see Table \ref{appendix_tab:us_regions} in Appendix \ref{appen:states_and_us_regions}).\cite{us_census_bureau_2018}

To estimate the treatment effects, we use models with a linear link function (Section \ref{sec:linear_link_model}) and a logistic link function (Section \ref{sec:logistic_link_models}).

\subsection{GAM Model with Linear Link Function}\label{sec:gam_model_linear_link}
We first estimate the effect of DIH prosecutions reported by the media on drug-overdose deaths using models with a linear link function.
We estimate the effect of DIH prosecutions reported by the media on drug-overdose deaths under two different assumptions: 1) the treatment effect is constant and 2) the treatment effect depends on the treatment duration.
Note that while there may be other sources of treatment heterogeneity, we focus specifically on treatment duration heterogeneity in this paper.

Under the assumption of constant treatment effect, we fit the following GAM to predict the log overdose death risk:
\begin{align}\label{eq:constant_tx_model_eq_lin_link_func}
    \E\left(\log Y_{st} \mid s,t, X_{st}, A_{st}\right) = \alpha_s + \gamma_{r(s)}(t) +  A_{st}\beta +  X_{st}\delta,
\end{align}
where $\log Y_{st}$ is the log risk of unintentional drug-overdose deaths in state $s$ at time interval $t$, $\alpha_s$ indicates the state fixed effect of state $s$, $\gamma_{r(s)}(t)$ indicates the smoothed time effects which can differ by the U.S. regions of the states (denoted by $r(s)$) and are estimated using cubic regression splines, $A_{st}$ indicates the intervention of having any DIH prosecutions reported by the media versus none (given by Equation~\eqref{eq:tx_primary}), and $X_{st}$ indicates the other various policy measures (defined similar to Equation~\eqref{eq:tx_primary}).

Under the assumption that the effect of DIH prosecutions reported by the media on drug-overdose deaths may depend on the treatment duration, we estimate the treatment effects using an event study model.
Event study analyses are typically used in settings where treatment is staggered. \cite{doi:10.1086/719588, borusyak2024revisiting, sun2021estimating}
Event study models differ from the constant treatment effect model in that instead of a single variable that indicates whether a state is treated, event study models contain lead and lag indicators which estimate the effects leading up to and following treatment initiation, respectively.
Under a general event study model for an outcome $Y_{st}^*$, one typically assumes that
\begin{align}\label{eq:general_event_study_model} 
    \E({Y}^*_{st}) = {\alpha}^*_i + {\gamma}^*_t + \sum_{k=B}^{C} {\beta}^*_k \mathbb{I}\{{K}^*_{st} = k\} +  {X}^*_{st}{\delta}^*,
\end{align}
where ${Y}^*_{st}$ is the outcome for state $i$ at time $t$, ${\alpha}^*_i$ and ${\gamma}^*_t$ are state and time fixed effects, ${X}^*_{st}$ is a vector of other potential confounding variables, and ${K}^*_{st}$ is the time relative to the treatment time.
Values of $k < 0$ indicate periods before the treatment, and values of $k \geq 0$ indicate periods at or after treatment.
Under a \textit{fully specified} model, all leads and lags are included except for one (commonly $k = -1$). \cite{doi:10.1086/719588, borusyak2024revisiting}
Another common specification of the event study model chooses $B = 0, C > 0$.
Under this specification of the event study model, one assumes that there are no pre-treatment anticipation effects. 

We focus on the latter model specification where $B = 0$, i.e. we assume there are no pre-treatment anticipation effects.
To verify the assumption that there are no pre-treatment anticipation effects, we first check if for $k < 0$, coefficients $\beta^*_k$ in the fully specified model
are zero.
Typically, the values of $\beta^*_k$ are checked both visually and statistically by hypothesis tests. 
Moreover, in practice, the test of no pre-treatment anticipation effects is often used to check the validity of the parallel trends assumption. \cite{abouk2019association, doi:10.1086/719588, borusyak2024revisiting}
If there are no pre-treatment effects, i.e. $\beta^*_k = 0$ for $k < 0$, then we can estimate the post-treatment effects using the following event study model:
\begin{align}\label{eq:ols_model_dih}
    \E(\log Y_{st}|s, t, X_{st}, K_{st}) &= \alpha_s + \gamma_{r(s)}(t) + \sum_{k = 0}^{39} \mathbb{I}\{K_{st} = k\} \beta_k + X_{st}\delta,
\end{align}
where the variables $Y_{st}, \alpha_s, \gamma_{r(s)}(t), X_{st}$ are as defined as in Equation~\eqref{eq:constant_tx_model_eq_lin_link_func} and $K_{st}$ indicates the number of six-month time intervals relative to the intervention time, up to a maximum of 39 six-month intervals (or 19.5 years).

Under a GAM with a linear link function, one typically assumes independence between observations.
However, this assumption may not hold in our setting of drug-overdose deaths since there may be dependencies between and within states between the different time intervals.
Following Bertrand, Duflo, and Mullainathan, we assume that there are no dependencies between states.\cite{bertrand2004much}
To account for dependencies within states, we use a sandwich estimator for the state-clustered standard errors of the parameters of the GAM with a linear link function (see Appendix \ref{appendix:sandwich_est_event_study} for details).\cite{van1998asymptotic}

We use the event study model given by Equation~\eqref{eq:ols_model_dih} as a benchmark comparison to the constant treatment effect model given by Equation~\eqref{eq:constant_tx_model_eq_lin_link_func} to illustrate the bias that arises when treatment duration heterogeneity is present.
The event study model serves as a natural benchmark because it is commonly used in drug-overdose literature and relaxes the treatment effect homogeneity assumption of the constant treatment effect model by allowing treatment effects to vary across treatment durations.\cite{abouk2019association, pacula2015assessing}
However, recent literature has shown that the event model of Equation~\eqref{eq:ols_model_dih} itself has limitations when other treatment effect heterogeneity is present, such as treatment effect heterogeneity across states.\cite{borusyak2024revisiting, sun2021estimating, de2026difference}
We show an example of the presence of such limitations in Appendix \ref{appendix:robust_event_study_estimator} where we compare the treatment effect estimates from the traditional event study model in Equation~\eqref{eq:ols_model_dih} -- refitted with a fixed time effect rather than a smoothed time effect to accommodate the robust estimator -- against the heterogeneity-robust estimator of de Chaisemartin and D'Haultf{\oe}uille.\cite{de2026difference} 
Differences between the two treatment effect estimates suggest that the traditional event study estimator in Equation~\eqref{eq:ols_model_dih} may be misspecified in our setting, with the treatment effects depending not only on the treatment duration but also on the state. 

Both the fit of the traditional event study model of Equation~\eqref{eq:ols_model_dih} and the fit of the more robust estimator in Appendix \ref{appendix:robust_event_study_estimator} show substantially conflicting results with the constant treatment effect model of Equation~\eqref{eq:constant_tx_model_eq_lin_link_func}, with the constant treatment effect estimates falling outside the range of the 5th and the 95th percentile of the estimated heterogeneous treatment effects (details in Section \ref{sec:linear_link_model_results} and Appendix \ref{appendix:robust_event_study_estimator}).
Despite the limitations of the current event study model, the comparison we make between the event study model in Equation~\eqref{eq:ols_model_dih} against the constant treatment effect model in Equation~\eqref{eq:constant_tx_model_eq_lin_link_func} remains informative, as it highlights the biases that may occur when using a constant treatment effect model that imposes treatment effect homogeneity.

We also conduct two sensitivity analyses where we 1) exclude the last five years of the analysis data since there may be biased long-run causal effects due to the negative weighting issue and 2) include the number of states with at least one DIH prosecution reported by the media by the start of time interval $t$ as a predictor to measure additional time effects (see Appendix \ref{appendix:sensitivity_analysis} for details).

\subsection{Results for Models with Linear Link Function}\label{sec:linear_link_model_results}
\begin{figure}[!htb]
\centerline{\includegraphics{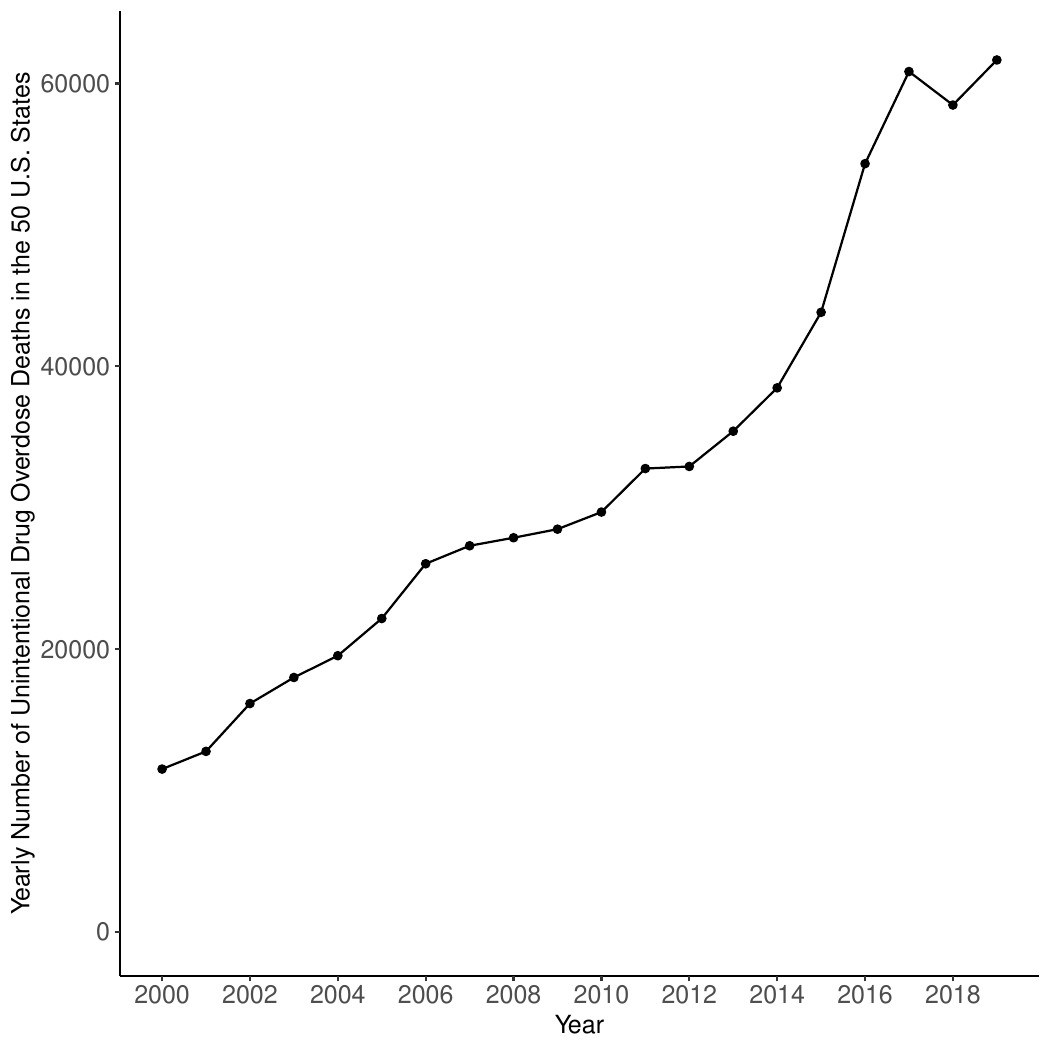}}
    \caption{Total number of yearly unintentional drug-overdose deaths in all 50 U.S. states from 2000 to 2019 where individuals who died were 18 and up.}
    \label{fig:data_outcome_tx}
\end{figure}

\begin{figure}[!htb]
\centerline{\includegraphics{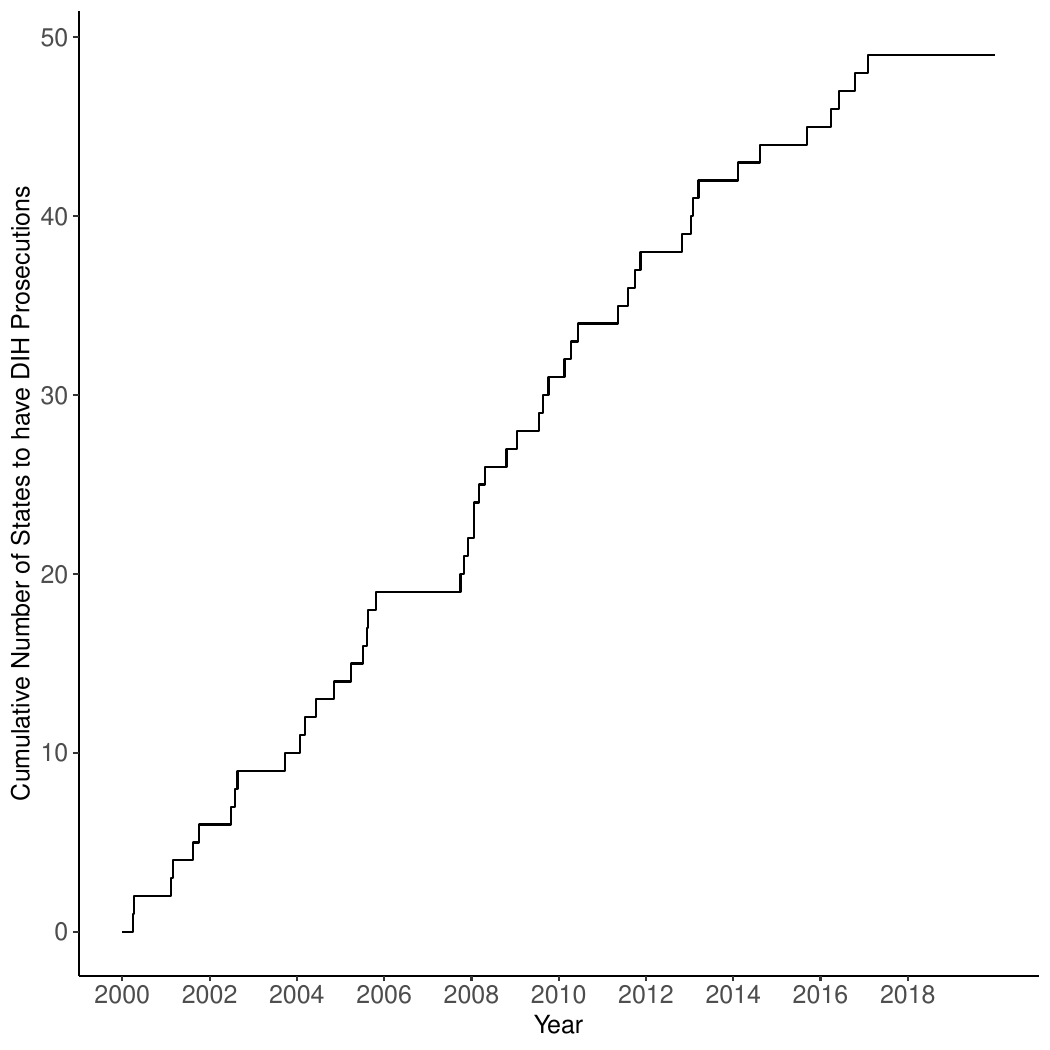}}
    \caption{Cumulative number of states with at least one drug-induced homicide (DIH) prosecution reported by the media from 2000 to 2019. Only Hawaii did not have any drug-induced homicide prosecutions by the end of 2019.}
    \label{fig:dih_prosecutions_states}
\end{figure}

Figure \ref{fig:data_outcome_tx} presents the yearly total number of unintentional drug-overdose deaths in the 50 U.S. states in 2000-2019 for individuals who were at least 18 years old. 
The total number of unintentional drug-overdose deaths in the 50 U.S. states for those at least 18 years of age ranged from 11,514 deaths in 2000 to 61,665 deaths in 2019. 
In total, there were approximately 658,216 unintentional drug-overdose deaths for those at least 18 years of age in the U.S. from 2000 to 2019.
Figure \ref{fig:dih_prosecutions_states} shows the cumulative number of states that had at least one DIH prosecution media reported by the media from 2000 to 2019. 
Hawaii was the only state without any DIH prosecutions reported by the media by the end of 2019.

\begin{figure}[!htb]
\centerline{\includegraphics{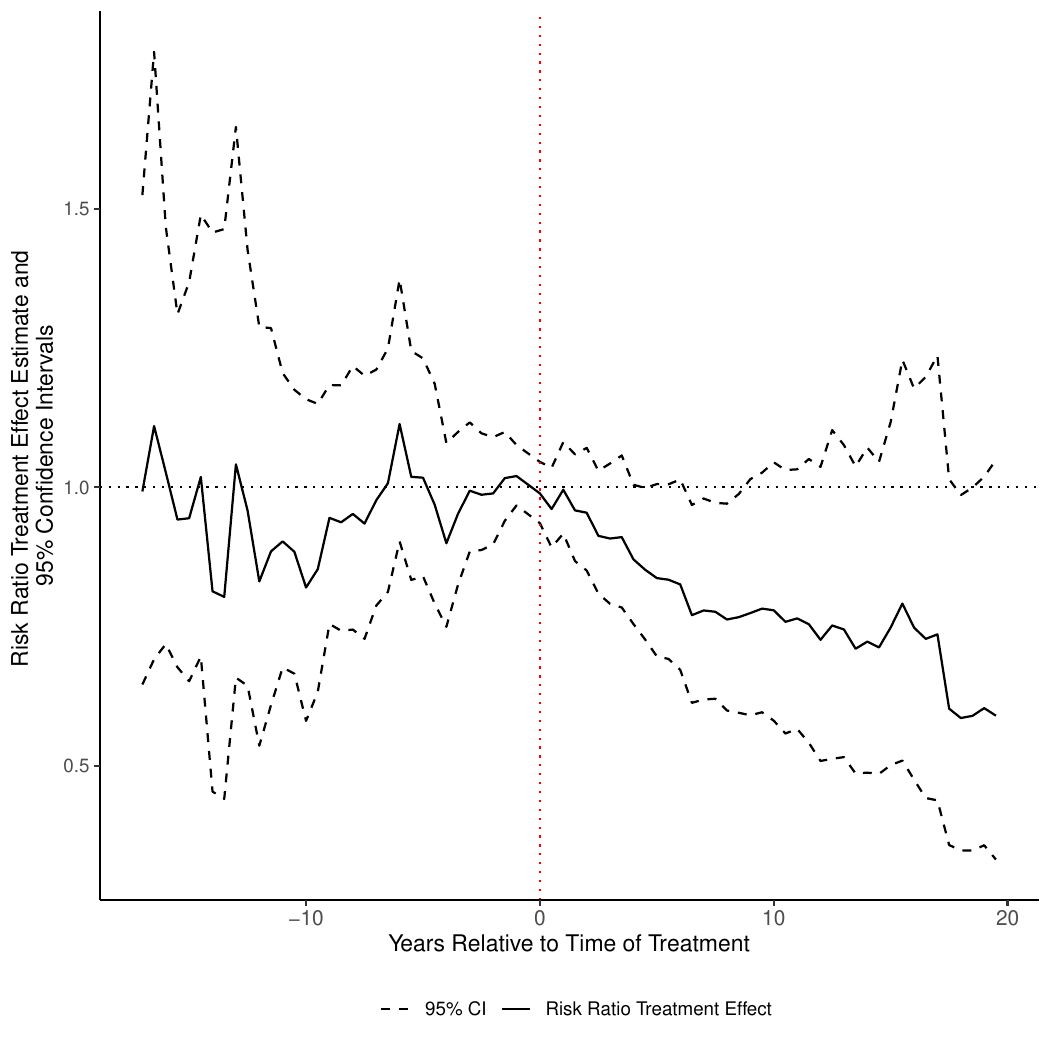}}
    \caption{Estimated treatment effects with 95\% confidence intervals for each time interval before and after the first drug-induced homicide (DIH) prosecution reported by the media in the state, where the reference time interval is the six-month interval right before the treatment. Generalized additive model (GAM) with linear link function, treatment effect depending on the treatment duration.}
    \label{fig:pre_tx_trend}
\end{figure}

We first check for any pre-treatment trends.
Figure \ref{fig:pre_tx_trend} shows the coefficients and 95\% confidence intervals for the periods leading up to and following the first time a state had at least one DIH prosecution reported by the media. 
The six-month time interval right before the time of the first DIH prosecution reported by the media in the state is the reference category.
From the coefficients and the 95\% confidence intervals, we conclude that there are likely no pre-treatment anticipation trends---no clear trends and all coefficients prior to the treatment were not statistically significantly different from zero.
Hence, we focus on estimating post-treatment effects as given by Equation~\eqref{eq:ols_model_dih}.

Table \ref{tab:linear_link_models_RR} shows the estimated risk ratios and 95\% confidence intervals for the effects of the different relevant policy measures and the DIH prosecutions reported by the media 1) when we assume a constant treatment effect and 2) when we assume the treatment effect depends on the treatment duration.
Under both models, there are no statistically significant results for the various policy measures.
Under both models, the naloxone access law that allows pharmacists to dispense naloxone without a prescription, the recreational marijuana law, and the prescription drug monitoring program are associated with a protective effect. 
On the other hand, the medical marijuana law, the 911 Good Samaritan law, and the medicaid expansion are associated with a harmful effect.
The naloxone access law that does not allow pharmacists to dispense naloxone without a prescription is associated with a protective effect under the constant treatment effect model, but it is associated with a harmful effect under the model where the treatment effect depends on the treatment duration. 

\begin{table}[!htb]
	\caption{Estimated risk ratios and 95\% confidence intervals for relevant policy measures and drug-induced homicide (DIH) prosecutions reported by the media for generalized additive models (GAM) with linear link function 1) when assuming a constant treatment effect of DIH prosecution media reports (left column) and 2) when assuming that the treatment effect of DIH prosecution media reports depends on the treatment duration (right column).} 
 \centering
\begin{tabular}{|lcc|}
			\hline
			 & \thead{Linear GAM with \\
			 Constant Treatment Effect \\ (95\% Confidence Interval)} & 
			 \thead{Linear GAM with Treatment Effect \\ 
    Depending
    on the Treatment Duration\\ (95\% Confidence Interval)} \\
			\hline
			Naloxone Access Law: pharmacists can dispense without prescription & 0.916 (0.806, 1.041) & 0.936 (0.832, 1.052)\\
            Naloxone Access Law: pharmacists cannot dispense without prescription & 0.995 (0.876, 1.131) & 1.002 (0.890, 1.128)\\
            Medical Marijuana Law & 1.203 (0.970, 1.492) & 1.224 (0.973, 1.541)\\
            Recreational Marijuana Law & 0.893 (0.755, 1.057) & 0.910 (0.763, 1.085)\\
            911 Good Samaritan Law & 1.060 (0.941, 1.194) & 1.068 (0.948, 1.204)\\
            Prescription Drug Monitoring Program & 0.859 (0.712, 1.035) & 0.841 (0.699, 1.011)\\
            Medicaid expansion & 1.096 (0.945, 1.272) & 1.087 (0.933, 1.267)\\
            DIH prosecutions reported by media & 0.977 (0.866, 1.101) & See Table \ref{tab:event_study_models_coef}, Left\\
			\hline
		\end{tabular}	
  \label{tab:linear_link_models_RR}
\end{table}

\begin{figure}[!htb]
\centerline{\includegraphics{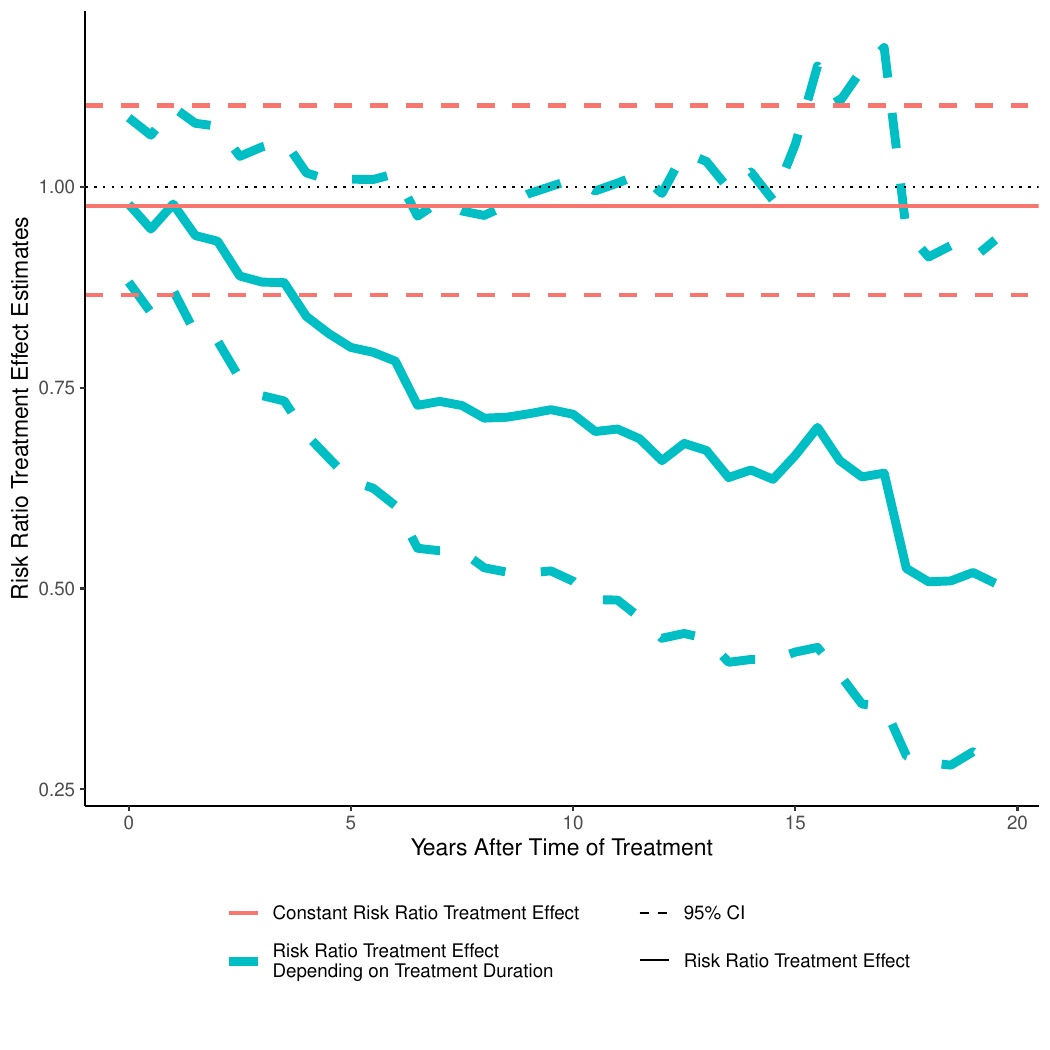}}
    \caption{Estimated risk ratios with 95\% confidence intervals for each time interval after the first drug-induced homicide (DIH) prosecution reported by the media in the state. Linear generalized additive models (GAMs) with 1) constant treatment effect (red, thin lines) and 2) treatment effect depending on the treatment duration (blue, thick lines).}
    \label{fig:linear_link_event_study_post_tx}
\end{figure}

Under the constant treatment effect model, we estimate a slightly protective non-significant risk ratio for DIH prosecutions reported by the media of 0.977, 95\% CI: (0.866, 1.101).
Table \ref{tab:event_study_models_coef} (left) presents the values of the estimated risk ratios under the assumption that the treatment effect depends on the treatment duration, with 95\% confidence intervals.
The estimated risk ratios range from 0.507 to 0.979, with an average risk ratio for DIH prosecutions reported by the media of 0.728 over the different treatment durations, suggesting a general protective effect (although most are not statistically significant).
Figure \ref{fig:linear_link_event_study_post_tx} shows the estimated risk ratios for each year since the first DIH prosecution reported by the media in the state for the models assuming a constant treatment effect and assuming that the treatment effect depends on the treatment duration.
The estimated constant treatment effect (risk ratio: 0.977) falls outside of the range of the 5th and 95th percentiles of the estimated treatment effects that depend on the treatment duration (0.510 and 0.949, respectively). 
Hence, most of the estimated treatment effects that depend on the treatment duration are more extreme than the estimated constant treatment effect. 
The sensitivity analyses yield similar results.
This suggests that the negative weighting (or at least underweighting) issue is present.  

\begin{table}[!htb]
    \caption{Estimated risk ratios and 95\% confidence intervals for each six-month time interval after the first drug-induced homicide (DIH) prosecution was reported by the media in the state for generalized additive models (GAMs) with linear (left column) and logistic (right column) link function when assuming treatment effect depends on the treatment duration.} 
    \centering
    \begin{tabular}{|ccc|}
			\hline
			 \thead{Number of six-month time intervals after first \\DIH prosecution reported by the media} & \thead{Risk Ratio: Linear GAM \\ (95\% Confidence Interval)} & 
			 \thead{Risk Ratio: Logistic GAM \\ (95\% Confidence Interval)} \\
			\hline
			0 & 0.979 (0.882, 1.086) & 0.977 (0.911, 1.048)\\
                1  & 0.948 (0.844, 1.065) & 1.008 (0.927, 1.095)\\
                2 & 0.979 (0.871, 1.099) & 1.002 (0.927, 1.083)\\
                3 & 0.940 (0.818, 1.079) & 0.955 (0.869, 1.050)\\
                4 & 0.933 (0.808, 1.076) & 0.947 (0.863, 1.038)\\
                5 & 0.889 (0.762, 1.038) & 0.902 (0.816, 0.997)\\
                6 & 0.882 (0.740, 1.050) & 0.886 (0.783, 1.001)\\
                7 & 0.881 (0.734, 1.058) & 0.898 (0.794, 1.016)\\
                8 & 0.839 (0.692, 1.018) & 0.861 (0.749, 0.989)\\
                9 & 0.818 (0.663, 1.009) & 0.841 (0.722, 0.978)\\
                10 & 0.800 (0.634, 1.010) & 0.828 (0.701, 0.977)\\
                11 & 0.794 (0.625, 1.009) & 0.820 (0.688, 0.976)\\
                12 & 0.783 (0.603, 1.017) & 0.814 (0.680, 0.974)\\
                13 & 0.728 (0.550, 0.964) & 0.762 (0.636, 0.913)\\
                14 & 0.733 (0.547, 0.982) & 0.760 (0.627, 0.921)\\
                15 & 0.728 (0.546, 0.970) & 0.768 (0.634, 0.929)\\
                16 & 0.712 (0.526, 0.965) & 0.750 (0.611, 0.919)\\
                17 & 0.713 (0.521, 0.977) & 0.746 (0.605, 0.921)\\
                18 & 0.718 (0.519, 0.992) & 0.729 (0.584, 0.910)\\
                19 & 0.723 (0.522, 1.001) & 0.740 (0.591, 0.926)\\
                20 & 0.717 (0.509, 1.010) & 0.705 (0.555, 0.896)\\
                21 & 0.696 (0.486, 0.995) & 0.688 (0.537, 0.881)\\
                22 & 0.699 (0.486, 1.005) & 0.688 (0.538, 0.881)\\
                23 & 0.686 (0.464, 1.016) & 0.681 (0.521, 0.889)\\
                24 & 0.659 (0.438, 0.992) & 0.666 (0.502, 0.886)\\
                25 & 0.681 (0.444, 1.044) & 0.655 (0.485, 0.886)\\
                26 & 0.672 (0.438, 1.032) & 0.642 (0.477, 0.863)\\
                27 & 0.638 (0.408, 0.998) & 0.629 (0.464, 0.853)\\
                28 & 0.648 (0.412, 1.019) & 0.633 (0.462, 0.867)\\
                29 & 0.636 (0.411, 0.984) & 0.618 (0.456, 0.838)\\
                30 & 0.666 (0.421, 1.053) & 0.626 (0.459, 0.855)\\
                31 & 0.701 (0.427, 1.151) & 0.677 (0.484, 0.948)\\
                32 & 0.659 (0.393, 1.107) & 0.653 (0.464, 0.920)\\
                33 & 0.639 (0.356, 1.146) & 0.623 (0.430, 0.903)\\
                34 & 0.644 (0.353, 1.174) & 0.625 (0.417, 0.937)\\
                35 & 0.525 (0.292, 0.945) & 0.550 (0.373, 0.811)\\
                36 & 0.508 (0.283, 0.913) & 0.538 (0.365, 0.792)\\
                37 & 0.510 (0.280, 0.927) & 0.572 (0.386, 0.849)\\
                38 & 0.520 (0.297, 0.911) & 0.537 (0.355, 0.813)\\
                39 & 0.507 (0.275, 0.934) & 0.575 (0.360, 0.919)\\
			\hline
    \end{tabular}	
    \label{tab:event_study_models_coef}
\end{table}

\begin{figure}
    \centerline{\includegraphics{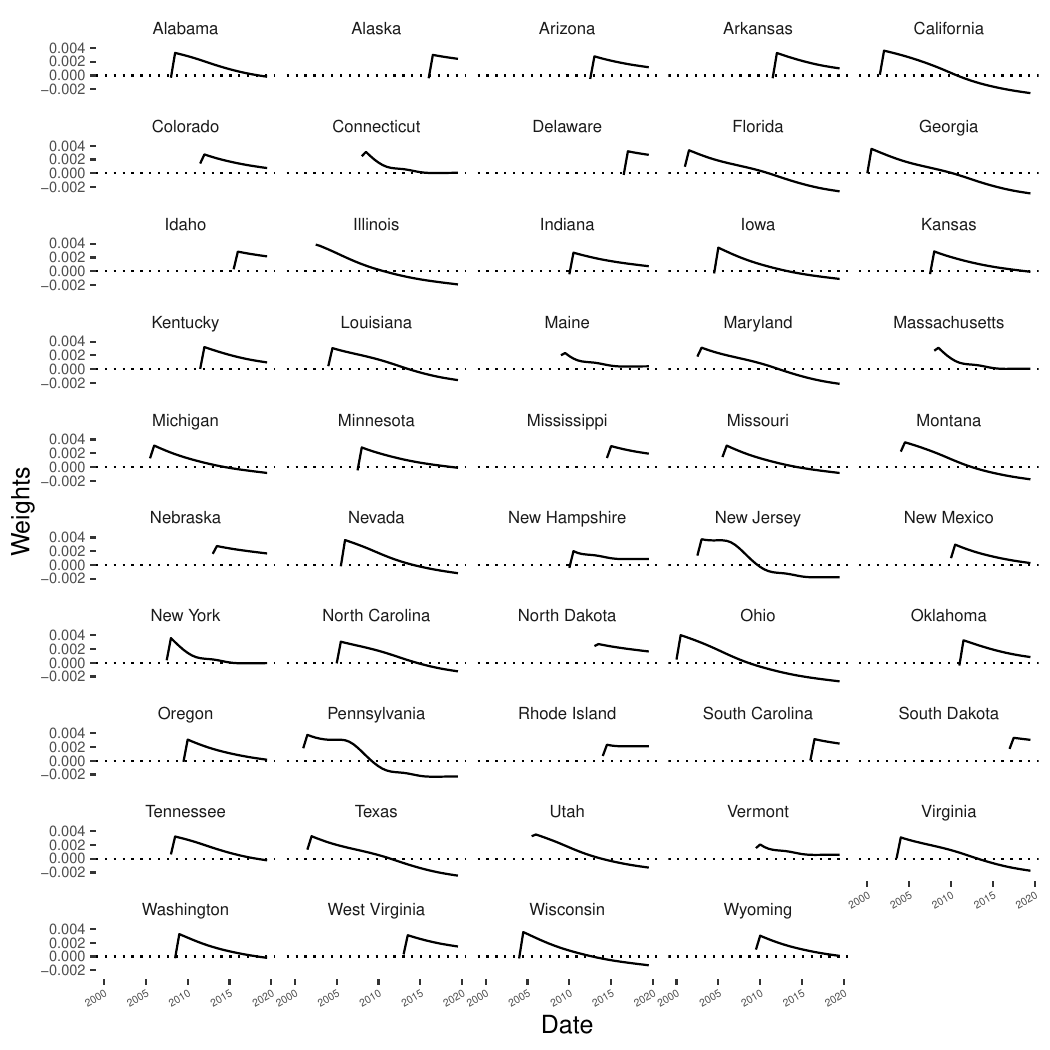}}
    \caption{The estimated weights for the weighted average of heterogeneous treatment effects for each state with at least one drug-induced homicide (DIH) prosecution reported by the media for time intervals after the intervention date, calculated using Equation~\eqref{eq:weights_eq}: $w_{st} = \tilde{A}_{st}/\sum_{st}\tilde{A}_{st}^2$.}
    \label{fig:weights_for_ols}
\end{figure}

To verify whether the negative weighting issue is present, we estimated the weights for the weighted average of heterogeneous treatment effects for each state with at least one DIH prosecution reported by the media for time intervals after the treatment time (see Figure \ref{fig:weights_for_ols}). 
We computed these weights using Equation~\eqref{eq:weights_eq}: $w_{st} = \frac{\tilde{A}_{st}A_{st}}{\sum_{st}\tilde{A}_{st}^2}$, where
the residuals $\tilde{A}_{st}$ are computed from a GAM with a linear link where we regress the treatment on the fixed state effects and smoothed time effects that may differ by the U.S. regions.
In general, weights tend to increase at first, shortly after the state had at least one DIH prosecution reported by the media. 
This may be due to the definition of the intervention variable, where it is less than one during the first time period where it was exposed to the intervention.
However, the weights tend to decrease as the time of exposure to the intervention increases.
As mentioned in Section \ref{sec:theoretical_background} and in the econometrics literature, negative weights tend to occur for states that were treated earlier in the analysis period for time intervals that occurred later in the analysis period (as seen in Ohio, Georgia, Pennsylvania, Florida, to name a few).
Hence, the conflicting results from the models under the two different treatment effect assumptions are likely due to a negative weighting of larger, later, treatment effects from the model where the treatment effects depends on the treatment duration.

\section{Negative Weighting of Drug-Induced Homicide Prosecution Effects in Models with Logistic Link Function}\label{sec:logistic_link_models}
Since the probability of a drug-overdose death for a random person is small (with a maximum probability of approximately 0.0003 in any state in a six-month time interval), the odds ratio is essentially equivalent to the risk ratio. 
Hence, we can also estimate the effect of DIH prosecutions reported by the media on the risk ratio scale using a model with a logistic link function.

\subsection{GAM Model with Logistic Link Function}
Similar to Section \ref{sec:linear_link_model}, we estimate the effect of DIH prosecutions reported by the media on unintentional drug-overdose deaths under the assumption of constant treatment effect and under the assumption that the treatment effect depends on the treatment duration.
Variables for the outcome, intervention of interest, and other policy measures are defined the same as in Section \ref{sec:linear_link_model}.
As before, we estimate the treatment effect using a GAM where we estimate smoothed time effects over the U.S. regions using a cubic regression splines model.
Therefore, assuming that the treatment effect is constant, we fit the following logistic GAM to predict the overdose death risk:
\begin{align}\label{eq:constant_tx_model_eq}
    logit(\E\left(Y_{st} \mid s,t, X_{st}, A_{st}\right)) = \alpha_s + \gamma_{r(s)}(t) +  A_{st}\beta + X_{st}\delta.
\end{align}
When we assume that the treatment effect depends on the treatment duration, we fit the following logistic GAM event study model:
\begin{align}\label{eq:logistic_event_study}
    logit(\E\left(Y_{st} \mid s,t, X_{st}, A_{st}\right)) = \alpha_s + \gamma_{r(s)}(t) +  \sum_{k=0}^{39}\mathbb{I}\{K_{st} = k\}\beta_k + X_{st}\delta. 
\end{align}
We also conduct the same sensitivity analyses as in Section \ref{sec:linear_link_model}, but where the models are now fit with a logistic link function.

As with the GAMs with a linear link function, one typically assumes independence between observations for the logistic GAMs.
Since there may be dependencies within states (as in Section \ref{sec:linear_link_model}, we assume independence between states), we use a sandwich estimator for the variance of the parameters (see Appendix \ref{sec:appendix_sandwich_est_constant_tx} for more details). 

\subsection{Results for Models with Logistic Link Function}\label{dih_constant_tx_eff_results}
Table \ref{tab:logistic_models_results} shows the estimated risk ratios with 95\% confidence intervals for the DIH prosecution reported by the media and the various policy measures.
Under the assumption of constant treatment effect, no coefficients are statistically significant. 
Similar to the model with a linear link function, the naloxone access law that allows pharmacists to dispense naloxone without a prescription, the recreational marijuana law, and the prescription drug monitoring program are associated with a protective effect. 
Moreover, the naloxone access law where the pharmacist cannot dispense without a prescription, the medical marijuana law, the 911 Good Samaritan law, and the medicaid expansion are associated with a harmful effect.
Under the GAM with a logistic link function, assuming a constant treatment effect, we estimate a risk ratio of 1.064 (95\% CI: (0.972, 1.165), suggesting that DIH prosecutions reported by the media are associated with a harmful effect.

\begin{table}[!htb]
    \caption{Estimated risk ratios and 95\% confidence intervals for relevant policy measures and drug-induced homicide (DIH) prosecutions reported by the media for generalized additive models (GAMs) with logistic link function when 1) assuming a constant treatment effect of DIH prosecution media reports and 2) when assuming that the treatment effect of DIH prosecution media reports depends on the treatment duration.} 
    \centering
    \begin{tabular}{|lcc|}
			\hline
			 & \thead{Logistic GAM with \\
			 Constant Treatment Effect \\ (95\% Confidence Interval)} & 
			 \thead{Logistic GAM with Treatment Effect \\ 
    Depending on the Treatment Duration\\ (95\% Confidence Interval)} \\
			\hline
			Naloxone Access Law: pharmacists can dispense without prescription & 0.974 (0.888, 1.070) & 0.980 (0.904, 1.062)\\
            Naloxone Access Law: pharmacists cannot dispense without prescription & 1.008 (0.914, 1.111) & 1.017 (0.927, 1.116)\\
            Medical Marijuana Law & 1.058 (0.945, 1.183) & 1.047 (0.930, 1.178)\\
            Recreational Marijuana Law & 0.963 (0.847, 1.094) & 0.964 (0.841, 1.106)\\
            911 Good Samaritan Law & 1.035 (0.952, 1.125) & 1.054 (0.972, 1.144)\\
            Prescription Drug Monitoring Program & 0.981 (0.857, 1.123) & 0.958 (0.851, 1.077)\\
            Medicaid expansion & 1.103 (0.979, 1.244) & 1.102 (0.987, 1.231)\\
            DIH prosecutions reported by media & 1.064 (0.972, 1.165) & See Table \ref{tab:event_study_models_coef}, Right\\
			\hline
    \end{tabular}	
    \label{tab:logistic_models_results}
\end{table}

\begin{figure}[!htb]
    \centerline{\includegraphics{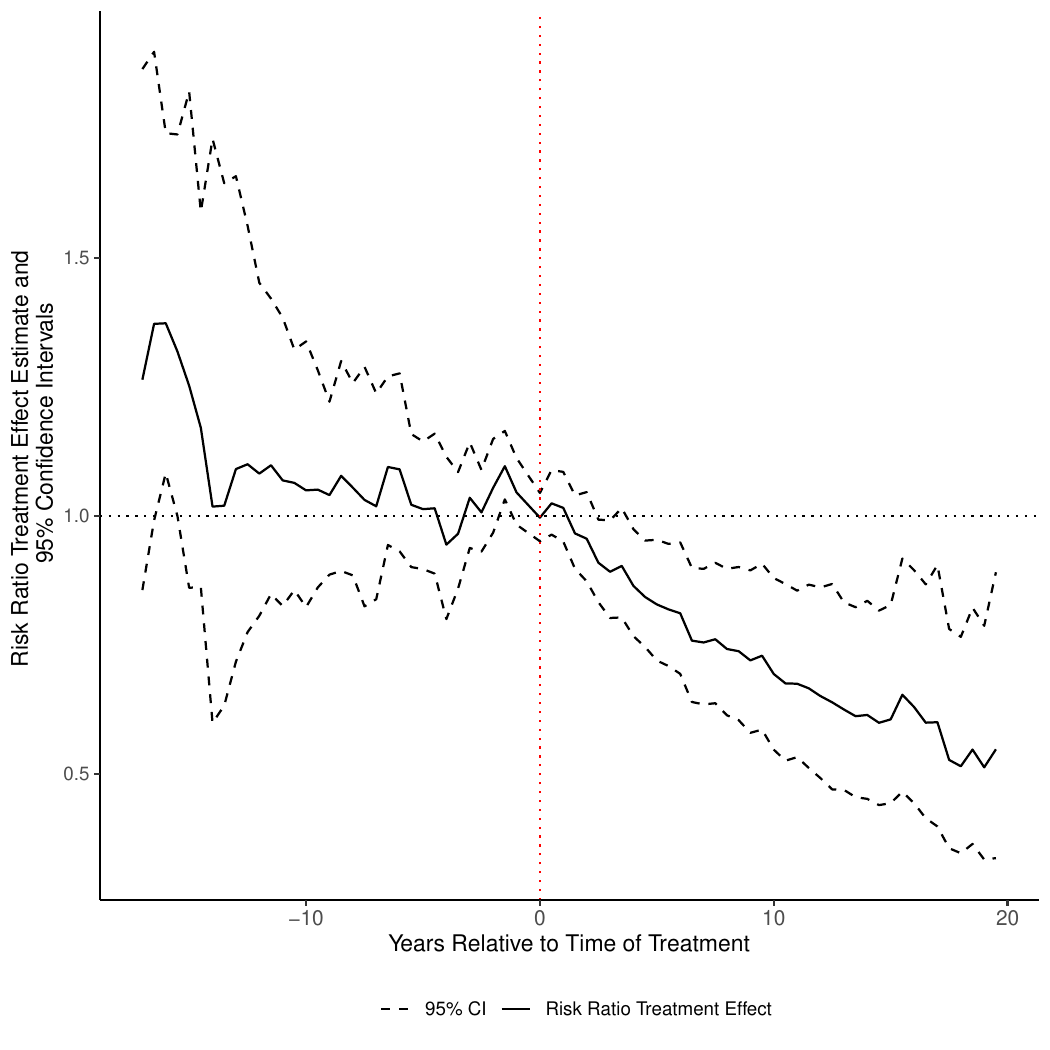}}
    \caption{Estimated risk ratios with 95\% confidence intervals for each time interval before and after the first drug-induced homicide (DIH) prosecution reported by the media in the state, where the reference time interval is the six-month interval right before the treatment. Generalized additive model (GAM) with logistic link function, treatment effect depending on the treatment duration.}
    \label{fig:pre_tx_trend_logistic}
\end{figure}

Under the model where we assume that the treatment effect may depend on the treatment duration, we first verify that there were no pre-treatment effects. 
Figure \ref{fig:pre_tx_trend_logistic} shows the estimated treatment effects with 95\% confidence intervals for each six-month time interval before and after the first time a state had at least one DIH prosecution reported by the media.
There are three coefficients before the intervention time that are positive and statistically significant.
However, it is unlikely that the time intervals 15.5 and 16 years before the intervention time have an effect on the unintentional drug-overdose deaths.
The coefficient for the time interval that is 1.5 years before the intervention time (indicated by coefficient $\hat{\beta}_{-3}$) also indicates an increase in unintentional drug-overdose deaths before the intervention. 
Since the post-treatment coefficients seen in Figure \ref{fig:pre_tx_trend_logistic} suggest a protective effect as the treatment duration increases, the positive coefficient for $\hat{\beta}_{-3}$ just before the intervention may suggest a conservative estimate of the effect of DIH prosecutions reported by the media.
We proceed with estimating the post-treatment effects while assuming that all pre-treatment effects are zero.

\begin{figure}[!htb]
    \centerline{\includegraphics{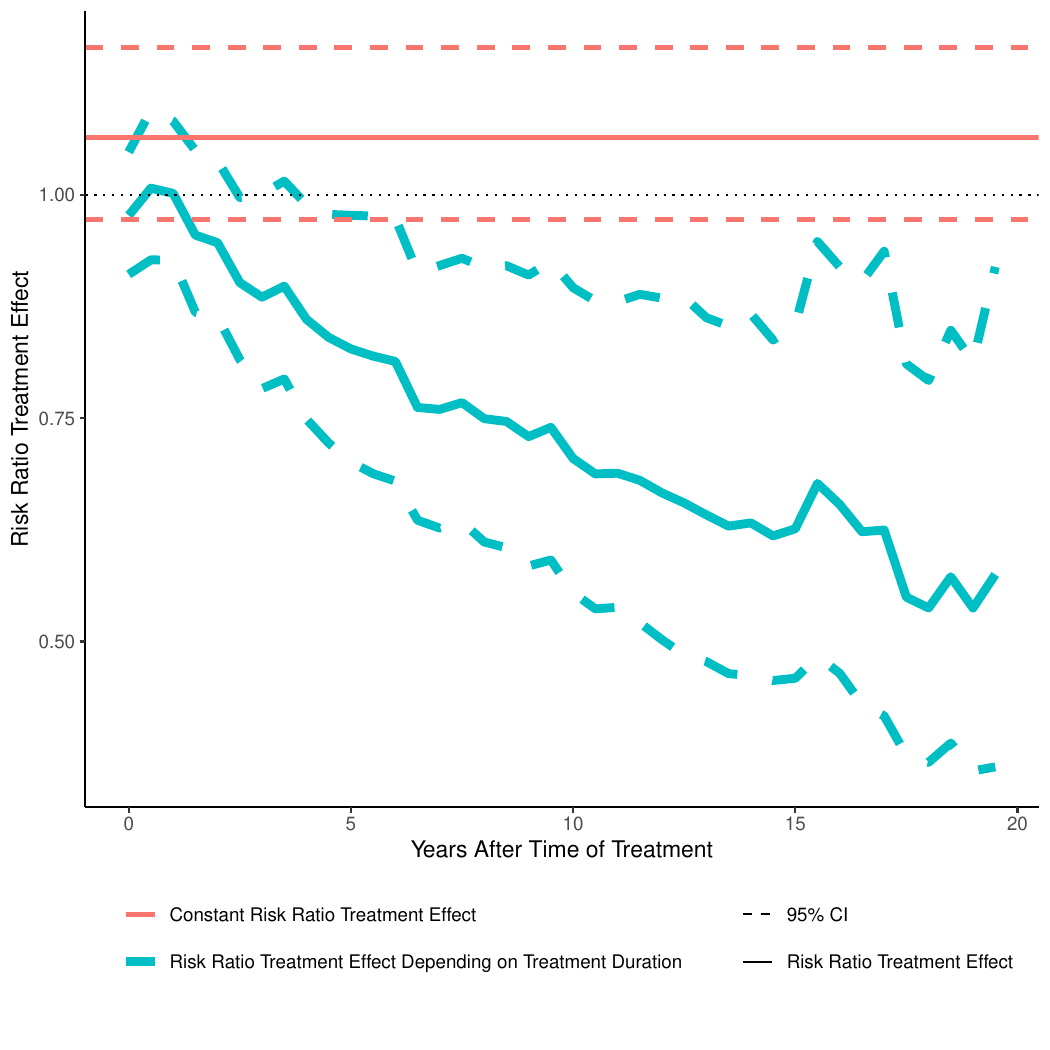}}
    \caption{Estimated risk ratios with 95\% confidence intervals for each time interval after the first drug-induced homicide (DIH) prosecution reported by the media in the state. Logistic generalized additive models (GAMs) models with 1) constant treatment effect (red, thin lines) and 2) treatment effect depending on the treatment duration (blue, thick lines).}
    \label{fig:logistic_link_event_study_post_tx}
\end{figure}

Risk ratio estimates with 95\% confidence intervals for the different relevant policy measures can be found in Table \ref{tab:logistic_models_results}.
Results are similar to the coefficients under the model assuming a constant treatment effect.
Risk ratio estimates with 95\% confidence intervals for each six-month interval after the first DIH prosecution reported by the media assuming a constant treatment effect and assuming that the treatment effect depends on the treatment duration are shown in Figure \ref{fig:logistic_link_event_study_post_tx} (values of risk ratio estimates and 95\% confidence intervals can be found in Table \ref{tab:event_study_models_coef}, right).
Under the assumption that the treatment effect depends on the treatment duration, we now estimate a generally statistically significant and protective effect where, as the treatment duration increases, the magnitude of the protective effect also increases.
The risk ratio estimates range from 0.538 to 1.008, with a mean of 0.739 over the different treatment durations. 
The estimated constant treatment effect (risk ratio: 1.064) has a sign opposite to almost all of the estimated treatment effects that depend on the treatment duration. 
The sensitivity analyses yield similar results.
The conflicting results between the constant treatment effect (harmful effect) and the estimated treatment effects that depend on the treatment duration (protective effect) suggest that the negative weighting issue is present here.

\section{Discussion}\label{sec:discussion}
We found that the different model specifications lead to different conclusions for the effect of having any DIH prosecutions reported by the media versus having none on unintentional drug-overdose deaths, depending on whether the model assumes a constant treatment effect or if the treatment effect depends on the treatment duration.
Under both the linear and logistic link functions, the estimated constant treatment effects are either smaller in magnitude or have a different sign than most of the estimated treatment effects that depend on the treatment duration. 
Under a linear link function, we estimated a smaller protective effect under a constant treatment effect assumption than almost all of the estimated treatment effects under the model where the treatment effect depends on the treatment duration. 
Under a logistic link function, we estimated a harmful effect under a constant treatment effect model, but mostly protective effects under the assumption that the treatment effect may depend on the treatment duration.

The conflicting results support the recent econometrics literature which have found potential issues when applying difference-in-differences methods with a constant treatment effect to settings where states are treated at different times and where treatment effects are heterogeneous. \cite{de2018fuzzy, de2020two, roth2022s, borusyak2024revisiting}
Our results suggest that in our application, the estimated constant treatment effect is an ``improper'' weighted average of the heterogeneous treatment effects where some of the weights are negative.
A possible explanation for the negative weights in our case is that there are many states that had at least one DIH prosecution reported by the media early on in the analysis period. 
Negative weights tend to occur for later time periods and for states that were treated early in the analysis period as seen in Figure \ref{fig:weights_for_ols}.\citep{de2020two, borusyak2024revisiting}
Figure \ref{fig:dih_prosecutions_per_yr} shows that by 2010, approximately half of the states had at least one DIH prosecution reported by the media, so approximately half of the states had been treated for at least nine years by the end of 2019.

Our results shed light on potential issues in applying difference-in-differences to settings with staggered adoption of treatment when treatment effects are heterogeneous.
To our knowledge, previous work on analyzing effects of interventions on drug-overdose deaths conducted both difference-in-differences and event study analyses, but none have encountered conflicting results such as the ones we found. \citep{powell2018medical, doi:10.1086/719588, pacula2015assessing}
It was not until upon further investigation that we found recent work (dating within the past few years) in the econometrics literature that highlighted these potential issues with models assuming constant treatment effects such as difference-in-differences models.\cite{de2018fuzzy, de2020two, roth2022s, borusyak2024revisiting, de2026difference}
The difference-in-differences model has been a very common method to estimate treatment effects.
However, our work and the recent econometrics literature suggest that there are settings where traditional difference-in-differences methods are not appropriate.
We hope that our work serves as an example of what could happen when one does not carefully consider all aspects of the constant treatment effect difference-in-differences model before applying the model to estimate treatment effects.

Although we found some statistically significant results under the logistic regression model under the assumption that the treatment effect depends on the treatment duration, the other models did not yield statistically significant results. 
Furthermore, we found a harmful effect for DIH prosecutions reported by the media from one of the analyses, albeit the result was not statistically significant.
Hence, we cannot definitively conclude that DIH prosecutions reported by the media have a protective or harmful effect.

There are also other potential effects of DIH prosecutions worth considering.
First, DIH prosecutions and their coverage in the mass media may actually discourage people from seeking emergency medical help. 
In two previous, separate studies, participants stated that police involvement was the main reason people did not make a 911 call or had a delay in doing so. \citep{baca2007heroin, pollini2006response}
In addition, awareness of 911 Good Samaritan laws is broadly lacking, and criminal justice measures such as DIH prosecutions may hurt efforts to increase public understanding and the lifesaving spirit of 911 Good Samaritan Laws. \citep{carroll2021drug, schneider2020knowledge}
Furthermore, there are additional concerns about the impact of criminal justice contact on the health risk of people who use drugs. 
A recent systematic literature review of global research found robust associations between police contact and HIV infection and risk behaviors. \citep{baker2019policing}
Baker et al. also highlighted the structural impact of law enforcement as a barrier to protective measures, including impact on Opioid Agonist Therapy (OAT) nonattendance, Site Engagement Program (SEP) avoidance, and healthcare avoidance.\cite{baker2019policing}
Lastly, there are also ethical concerns about law enforcement activities producing and reinforcing disparities by race and vulnerability. 
The perceived risks of legal repercussions for seeking help during overdose events are likely affected by existing racial disparities in contact with the criminal legal system. \citep{beletsky2011roles}A recent study explored the relationship between structural vulnerability and police abuse and harassment among people who inject drugs, and they found that a certain subgroup: men who were experiencing homelessness, were from rural areas, had traded sex, and dropped out of high school were most likely to experience police abuse and harassment. \citep{friedman2020intersectional}

Our analyses had several limitations. 
First, there may have been misclassification in the exposure to DIH prosecutions reported by the media. 
Specifically, it is possible that there was an underreporting of the number of states being exposed to DIH prosecutions reported by the media at any given time, since the exposure variable depended on the media reports collected by the Health in Justice Action Lab. 
If a media report was missing from the data, then the state would be considered as not being exposed to DIH prosecutions reported by the media.
Previous research showed that in scenarios of underreporting (here, it is likely that we have differential misclassification), the estimate of the treatment effect is biased towards the null. \citep{ferrao2014adjusting, kesmodel2018information, moradzadeh2018impact} 
Additionally, we only had information about the charge date of the DIH prosecution and not about the date of the media report, which may lead to misspecification of the intervention date.
Second, we only considered treatment as whether a state was ever exposed to DIH prosecutions reported by the media and not the number of DIH prosecutions reported by the media in each time interval.
Incorporating the prosecution intensity would provide a more nuanced characterization of exposure and is an interesting direction for future research.
Third, we did not account for interference effects from neighboring states, but the unintentional drug-overdose deaths in a state might depend on whether a neighboring state is prosecuting people for drug-induced homicide.
We also did not control for some other policies/programs that could have an effect on drug-overdose deaths, such as the number of OAT programs, the number of clinicians who prescribe buprenorphine, access to syringe service programs, access to OAT in prison/jails, etc. \citep{santo2021association, adams2020making} 
Finally, beyond the limitations of our specific analysis, future methodological work could investigate whether the negative weighting issue identified in this paper extends to other model specifications beyond the linear and logistic link functions considered here, such as weighted linear models.

The results from our analyses show the potential danger of using difference-in-differences models in settings where the treatment is staggered and treatment effects are heterogeneous over treatment duration.
We hope that our work sheds light on the potential negative weighting issue and calls for greater care when using constant treatment effect models to estimate treatment effects.

\newpage

\section*{Acknowledgements}
\ack{The authors gratefully thank the NSF [grant number DMS-1854934] for funding this study. 
The content is solely the responsibility of the authors and does not necessarily represent the official views of the NSF.
The authors also acknowledge the work of interns and staff of the Health In Justice Action Lab in compiling the drug induced prosecutions media dataset, especially Siri Nelson (Northeastern University).
Lastly, the authors are grateful for Leo Beletsky, JD (Northeastern University) for his support and sharing his expertise on the topic of drug overdoses and drug-induced homicide prosecutions.  
}

\section*{Conflicts of Interest}
Natasha K. Martin has received unrestricted research grants to the university from Gilead and Abbvie unrelated to this work. Kelly C. Kung and Judith J. Lok have no conflicts of interest to declare.

\begin{sloppypar}
{\doublespacing
\bibliography{Documents/main_bibliography}

\begin{thebibliography}{10}
\providecommand \doibase [0]{http://dx.doi.org/}%

\bibitem{Hirt1974}
Hirt CW, Amsden AA, Cook JL. An arbitrary {L}agrangian-{E}ulerian computing
  method for all flow speeds. {\it J {C}omput {P}hys.}
  1974\string;14(3)\string:227--253.

\bibitem{Liska2010}
Liska R, Shashkov M, Vachal P, et al. Optimization-based synchronized
  flux-corrected conservative interpolation (remapping) of mass and momentum
  for arbitrary {L}agrangian-{E}ulerian methods. {\it J {C}omput {P}hys.}
  2010\string;229(5)\string:1467--1497.

\bibitem{Taylor1937}
Taylor GI, Green AE. Mechanism of the production of small eddies from large
  ones. {\it P {R}oy {S}oc {L}ond {A} {M}at.}
  1937\string;158(895)\string:499--521.
\newblock \url{https://doi.org/10.1098/rspa.1937.0036},
  \url{http://rspa.royalsocietypublishing.org/content/158/895/499}.

\bibitem{Knupp1999}
Knupp PM. Winslow smoothing on two-dimensional unstructured meshes. {\it Eng
  {C}omput.} 1999\string;15\string:263--268.

\bibitem{Kamm2000}
Kamm J. Evaluation of the {S}edov-von {N}eumann-{T}aylor blast wave solution.
  Tech. Rep. Technical {R}eport LA-UR-00-6055, Los {A}lamos {N}ational
  {L}aboratory; The address:   2000.

\bibitem{Kucharik2003}
Kucharik M, Shashkov M, Wendroff B. An efficient linearity-and-bound-preserving
  remapping method. {\it J {C}omput {P}hys.}
  2003\string;188(2)\string:462--471.

\bibitem{Blanchard2015}
Blanchard G, Loubere R. High-Order {C}onservative {R}emapping with a posteriori
  {MOOD} stabilization on polygonal meshes. Details on how published;  2015.
\newblock Accessed January 13, 2016.
  \url{https://hal.archives-ouvertes.fr/hal-01207156}, the {HAL} {O}pen
  {A}rchive, hal-01207156.

\bibitem{Burton2013}
Burton DE, Kenamond MA, Morgan NR, Carney TC, Shashkov MJ, Author AB. An
  intersection based {ALE} scheme {(xALE)} for cell centered hydrodynamics
  {(CCH)}. In: Talk at {M}ultimat 2013, {I}nternational {C}onference on
  {N}umerical {M}ethods for {M}ulti-{M}aterial {F}luid {F}lows. The
  Organization.  September 2--6, 2013; San {F}rancisco.
\newblock LA-UR-13-26756.2.

\bibitem{Berndt2011}
Berndt M, Breil J, Galera S, Kucharik M, Maire PH, Shashkov M. Two-step hybrid
  conservative remapping for multimaterial arbitrary {L}agrangian-{E}ulerian
  methods. {\it J {C}omput {P}hys.} 2011\string;230(17)\string:6664--6687.

\bibitem{Kucharik2012}
Kucharik M, Shashkov M. One-step hybrid remapping algorithm for multi-material
  arbitrary {L}agrangian-{E}ulerian methods. {\it J {C}omput {P}hys.}
  2012\string;231(7)\string:2851--2864.

\bibitem{Breil2015}
Breil J, Alcin H, Maire PH. A swept intersection-based remapping method for
  axisymmetric {ReALE} computation. {\it Int {J} {N}umer {M}eth {F}l.}
  2015\string;77(11)\string:694--706.
\newblock Fld.3996.

\bibitem{Barth1997}
Barth TJ. Numerical methods for gasdynamic systems on unstructured meshes. In:
  Kroner D, Rohde C, Ohlberger M. \kern-2pt, eds. {\it An {I}ntroduction to
  {R}ecent {D}evelopments in {T}heory and {N}umerics for {C}onservation {L}aws,
  {P}roceedings of the {I}nternational {S}chool on {T}heory and {N}umerics for
  {C}onservation {L}aws}, 2~ed., Lecture {N}otes in {C}omputational {S}cience
  and {E}ngineering. Springer,  1997.

\bibitem{Lauritzen2011}
Lauritzen P, Erath C, Mittal R. On simplifying `incremental remap'-based
  transport schemes. {\it J {C}omput {P}hys.}
  2011\string;230(22)\string:7957--7963.

\bibitem{Klima2017}
Klima M, Kucharik M, Shashkov M. Local error analysis and comparison of the
  swept- and intersection-based remapping methods. {\it Commun {C}omput
  {P}hys.} 2017\string;21(2)\string:526--558.

\bibitem{Dukowicz2000}
Dukowicz JK, Baumgardner JR. Incremental remapping as a transport/advection
  algorithm. {\it J {C}omput {P}hys.} 2000\string;160(1)\string:318--335.

\bibitem{Kucharik2011}
Kucharik M, Shashkov M. Flux-based approach for conservative remap of
  multi-material quantities in {2D} arbitrary {L}agrangian-{E}ulerian
  simulations. In:  Fo\v{r}t J, F{\"{u}}rst J, Halama J, Herbin R, Hubert F.
  \kern-2pt, eds. {\it Finite {V}olumes for {C}omplex {A}pplications {VI}
  {P}roblems \& {P}erspectives}, 1~ed., Springer {P}roceedings in
  {M}athematics. Springer,  2011\string:623--631.

\bibitem{Kucharik2014}
Kucharik M, Shashkov M. Conservative multi-material remap for staggered
  multi-material arbitrary {L}agrangian-{E}ulerian methods. {\it J {C}omput
  {P}hys.} 2014\string;258\string:268--304.

\bibitem{Loubere2005}
Loubere R, Shashkov M. A subcell remapping method on staggered polygonal grids
  for arbitrary-{L}agrangian-{E}ulerian methods. {\it J {C}omput {P}hys.}
  2005\string;209(1)\string:105--138.

\bibitem{Caramana1998}
Caramana EJ, Shashkov MJ. Elimination of artificial grid distortion and
  hourglass-type motions by means of {L}agrangian subzonal masses and
  pressures. {\it J {C}omput {P}hys.} 1998\string;142(2)\string:521--561.

\bibitem{Hoch2009}
Hoch P. An arbitrary {L}agrangian-{E}ulerian strategy to solve compressible
  fluid flows. Tech. Rep. Technical {R}eport, CEA; The address:   2009.
\newblock Accessed January 13, 2016. HAL: hal-00366858.
  https://hal.archives-ouvertes.fr/docs/00/36/68/58/PDF/ale2d.pdf.

\bibitem{Shashkov1996}
Shashkov M. {\it Conservative {F}inite-{D}ifference {M}ethods on {G}eneral
  {G}rids}.
\newblock CRC {P}ress, 1996.

\bibitem{Benson1992}
Benson DJ. Computational methods in {L}agrangian and {E}ulerian hydrocodes.
  {\it Comput {M}ethod {A}ppl {M}.} 1992\string;99(2--3)\string:235--394.

\bibitem{Margolin2003}
Margolin LG, Shashkov M. Second-order sign-preserving conservative
  interpolation (remapping) on general grids. {\it J {C}omput {P}hys.}
  2003\string;184(1)\string:266--298.

\bibitem{Kenamond2013}
Kenamond MA, Burton DE. Exact intersection remapping of multi-material
  domain-decomposed polygonal meshes. In: Talk at {M}ultimat 2013,
  {I}nternational {C}onference on {N}umerical {M}ethods for {M}ulti-{M}aterial
  {F}luid {F}lows. The Organization.  September 2--6, 2013; San {F}rancisco.
\newblock LA-UR-13-26794.

\bibitem{Dukowicz1984}
Dukowicz J. Conservative rezoning (remapping) for general quadrilateral meshes.
  {\it J {C}omput {P}hys.} 1984\string;54(3)\string:411--424.

\bibitem{Margolin2002}
Margolin LG, Shashkov M. Second-order sign-preserving remapping on general
  grids. Tech. Rep. Technical Report LA-UR-02-525, Los {A}lamos {N}ational
  {L}aboratory; The address:   2002.

\bibitem{Mavriplis2003}
Mavriplis DJ. Revisiting the least-squares procedure for gradient
  reconstruction on unstructured meshes. In: AIAA 2003-3986. 16th {AIAA}
  {C}omputational {F}luid {D}ynamics {C}onference. The organization.  June
  23--26, 2003; Orlando, {F}lorida.

\bibitem{Scovazzi2008}
Scovazzi G, Love E, Shashkov M. Multi-scale {L}agrangian shock hydrodynamics on
  {Q1/P0} finite elements: {T}heoretical framework and two-dimensional
  computations. {\it Comput {M}ethod {A}ppl {M}.}
  2008\string;197(9--12)\string:1056--1079.

\end{thebibliography}
}
\end{sloppypar}

\end{document}


\appendix

\section{Proofs}
We show the proof to Theorem \ref{thm_1} in this section below, and it is based on the proofs in previous work\cite{proof_for_sec_2, borusyak2024revisiting}.

\subsection{Explanation of Frisch-Waugh-Lovell Theorem\cite{frisch1933partial}}\label{fwl_thm_explanation}
Let $Y$, $X_1$, and $X_2$ be matrices of dimensions $n \times 1$, $n \times p_1$, and $n \times p_2$, respectively.   
Consider a general Ordinary Least Squares (OLS) model:
\begin{align}\label{eq:fwl_original_eq}
    Y = X_1 \beta_1 + X_2 \beta_2 + \epsilon,
\end{align}
where $\beta_1$ is a vector of dimension $p_1$, $\beta_2$ is a vector of dimension $p_2$, and $\epsilon$ is a zero-mean error term.
Multiply both sides of the equation by $(I-X_1(X_1^T X_1)^{-1}X_1^T)$:
\begin{align*}
    Y &= X_1 \beta_1 + X_2 \beta_2 + {\epsilon} \\
    (I - X_1(X_1^T X_1)^{-1}X_1^T)Y &= (I - X_1(X_1^T X_1)^{-1}X_1^T)X_1 \beta_1 + (I - X_1(X_1^T X_1)^{-1}X_1^T)X_2 \beta_2 + (I - X_1(X_1^T X_1)^{-1}X_1^T){\epsilon} \\
    \tilde{Y} &= X_1 \beta_1 - X_1 \beta_1 + \tilde{X}_2 \beta_2 + \tilde{\epsilon} \\
    \tilde{Y} &= \tilde{X}_2 \beta_2 + \tilde{\epsilon},
\end{align*}
where $\tilde{Y}$ and $\tilde{X}_2$ are the residuals from OLS models regressing $Y$ and $X_{2}$ on $X_1$, respectively, and $\tilde{\epsilon}$ is a zero-mean error term.
Consider the new OLS model:
\begin{align}\label{eq:fwl_resid_eq}
    \tilde{Y} &= \tilde{X}_2 \beta_2 + \tilde{\epsilon}.
\end{align}
The Frisch-Waugh-Lovell theorem states that the estimate of $\beta_2$ based on the regression from Equation~\eqref{eq:fwl_original_eq} is the same as the estimate of $\beta_2$ based on the regression from Equation~\eqref{eq:fwl_resid_eq}.\cite{frisch1933partial} 
For the rest of the proof, we assume that $\hat{\beta}_2$ is a scalar.
From OLS, the estimator of $\beta_2$ based on Equation~\eqref{eq:fwl_resid_eq} is given as follows:
\begin{align}\label{eq:fwl_thm_beta_estimate}
    \hat{\beta}_2 = \frac{cov(\tilde{Y}, \tilde{X})}{var(\tilde{X})}.
\end{align}

\subsection{Proof of Theorem \ref{thm_1}}\label{appendix:proof_of_thm_1}
Assume that the Stable Unit Value Treatment Assumption (SUTVA) (Assumption \ref{sutva}) holds and there are no anticipation effects (Assumption \ref{no_anticipation_effects}). 
From the OLS model given by Equation~\eqref{eq:data_generating_model}, we have the following:
\begin{align}
\E(Y_{st}| \textbf{A}) &= \E \left(Y_{st}(A_{st})| \textbf{A}\right) \nonumber \\
    &=\E \left(Y_{st}(0) + \left(Y_{st}(A_{st}) - Y_{st}(0)\right) | \textbf{A}\right) \nonumber \\
    &=\E(Y_{st}(0) | \textbf{A}) + \E(Y_{st}(A_{st}) - Y_{st}(0) | \textbf{A}) \nonumber \\
    &= \E(Y_{st}(0)| \textbf{A}) + \frac{A_{st}}{A_{st}}\E(Y_{st}(A_{st}) - Y_{st}(0)| \textbf{A}) \nonumber \\
    &= \E(Y_{st}(0)| \textbf{A}) + A_{st}\beta_{st}\label{eq:substit_yit},
\end{align}
where $\textbf{A}$ is the set of treatments for all units.
The first and second equalities are the result of the consistency property of SUTVA where $Y_{st} = Y_{st}(A_{st})$ and adding and subtracting the same value $Y_{st}(0)$.
We use the linearity of expectations to derive the third equality.
The fourth equality is obtained by multiplying $\frac{A_{st}}{A_{st}}$ and adopting the convention that $\frac{0}{0} = 0$ if $A_{st} = 0$. 
From Equation~\eqref{eq:data_generating_model}, we assume that the treatment effects are given by $\beta_{st}$.
The final equality follows from defining the treatment effects as the conditional difference divided by $A_{st}$.

Using the Frisch-Waugh-Lovell theorem, we apply the results from Equation~\eqref{eq:fwl_thm_beta_estimate} to our setting to estimate $\beta^{constant}$.
Let $\tilde{A}_{st}$ denote the residuals of an OLS of $A_{st}$ on the unit and time fixed effects, and let $\tilde{Y}_{st}$ denote the residuals of an OLS regressing $Y_{st}$ on the unit and time fixed effects.
It follows that:
\begin{align}
    \hat{\beta}^{constant} &= \frac{cov(\tilde{Y}_{st}, \tilde{A}_{st})}{var(\tilde{A}_{st})} \nonumber \\
        &= \frac{\sum_{st}(\left(Y_{st} - \hat{\alpha}_s - \hat{\gamma}_t\right)\tilde{A}_{st})}{\sum_{st}\tilde{A}_{st}^2} \nonumber \\
        &= \frac{\sum_{st}Y_{st}\tilde{A}_{st} - \sum_{st}\left(\hat{\alpha}_s + \hat{\gamma}_t\right)\tilde{A}_{st}}{\sum_{st}\tilde{A}_{st}^2}\nonumber \\
        &= \frac{\sum_{st}Y_{st}\tilde{A}_{st}}{\sum_{st}\tilde{A}_{st}^2}\label{eq:substit_beta_hat},
\end{align}
where $\hat{\alpha}_s$ and $\hat{\gamma}_t$ are the estimated  unit and time fixed effects, respectively.
The first equality follows from Equation~\eqref{eq:fwl_thm_beta_estimate}. 
The second equality follows from the zero-mean property of OLS residuals and by substituting the residuals $\tilde{Y}_{st}$ with their definition of $Y_{st}$ minus the estimated fixed state and time effects.
The last equality follows because the OLS residuals $\tilde{A}_{st}$ are orthogonal to the space spanned by the unit and time indicators. 

Combining Equations~\eqref{eq:substit_yit} and \eqref{eq:substit_beta_hat}, it follows that:
\begin{align*}
    \E\left(\hat{\beta}^{constant} \bigg| \textbf{A}\right) &= \E\left(\frac{\sum_{st}Y_{st}\tilde{A}_{st}}{\sum_{st}\tilde{A}_{st}^2} \bigg| \textbf{A}\right)\\
        &= \frac{\sum_{st}\tilde{A}_{st}\E\left(Y_{st}|\textbf{A}\right)}{\sum_{st}\tilde{A}_{st}^2}\\
        &= \frac{\sum_{st}\tilde{A}_{st}\left(\E(Y_{st}(0) | \textbf{A}) + A_{st}\beta_{st}\right)}{\sum_{st}\tilde{A}_{st}^2} \\
        &= \frac{\sum_{st}\tilde{A}_{st}\left(\alpha_s + \gamma_t + A_{st}\beta_{st}\right)}{\sum_{st}\tilde{A}_{st}^2} \\
        &= \frac{\sum_{st}\tilde{A}_{st}\alpha_s + \sum_{st}\tilde{A}_{st}\gamma_t + \sum_{st}\tilde{A}_{st}A_{st}\beta_{st}}{\sum_{st}\tilde{A}_{st}^2} \\
        &= \frac{\sum_{st}\tilde{A}_{st}A_{st}\beta_{st}}{\sum_{st}\tilde{A}_{st}^2} \\
        &= \sum_{st: A_{st} > 0}\frac{\tilde{A}_{st}A_{st}}{\sum_{st}\tilde{A}_{st}^2}\beta_{st} + \sum_{s't': A_{s't'} = 0}\frac{\tilde{A}_{s't'}A_{s't'}}{\sum_{s't'}\tilde{A}_{s't'}^2}\beta_{s't'}  \\
        &= \sum_{st: A_{st} > 0}\frac{\tilde{A}_{st} A_{st}}{\sum_{st}\tilde{A}_{st}^2}\beta_{st}.
\end{align*}
The first equality is obtained by substituting Equation~\eqref{eq:substit_beta_hat} for $\hat{\beta}^{constant}$.
The second equality follows from the linearity of expectations and conditioning on $\textbf{A}$, where the residuals $\tilde{A}_{st}$ are deterministic given $\textbf{A}$.
The third equality substitutes the conditional expectation of $Y_{st}$ with the derivation from Equation~\eqref{eq:substit_yit} and $\beta_{st}$ given by Equation~\eqref{eq:data_generating_model} for the treatment effect.
Under the parallel trends assumption, the conditional expectation $\E(Y_{st}(0) \mid \mathbf{A})$ can be written as the sum of the state and time fixed effects, as specified in Equation~\eqref{eq:data_generating_model}. 
We substitute this into the expression to obtain the fourth equality.
The fifth equality is obtained assuming that the model for $Y_{st}(0)$ is correctly specified such that $\alpha_s$ and $\gamma_t$ are in the span of the unit and time indicators.
Since the residuals $\tilde{A}_{st}$ are orthogonal to the space spanned by the unit and time indicators, the resulting sums are zero.
The sixth equality splits the summand over states $s$ and time periods $t$ into instances where $A_{st} > 0$ and $A_{st} = 0$.
When $A_{st} = 0$ the entire term disappears, which gives the final equation. 
Let $w_{st} = \frac{\tilde{A}_{st}A_{st}}{\sum_{st}\tilde{A}_{st}^2}$.

Moreover, to prove that for $A_{st}> 0$, the sum of the weights $w_{st}$ of treated state-time combinations is equal to 1:
\begin{align*}
    \sum_{st: A_{st} > 0} w_{st} 
    = \sum_{st: A_{st} > 0} \frac{\tilde{A}_{st} A_{st}}{\sum_{st} \tilde{A}_{st}^2} 
    = \sum_{st: A_{st} > 0} \frac{\tilde{A}_{st} A_{st}}{\sum_{st} \tilde{A}_{st}^2} + \sum_{s't': A_{s't'} = 0} \frac{\tilde{A}_{s't'} A_{s't'}}{\sum_{s't'} \tilde{A}_{s't'}^2}
    = \frac{ \sum_{st} \tilde{A}_{st} A_{st}}{\sum_{st} \tilde{A}_{st}^2} 
    = \frac{ \sum_{st} \tilde{A}_{st} \left(\tilde{A}_{st} + Proj_{\alpha_s, \gamma_t}(A_{st})\right)}{\sum_{st} \tilde{A}_{st}^2} 
    = \frac{ \sum_{st} \tilde{A}_{st}^2 }{\sum_{st} \tilde{A}_{st}^2} = 1,
\end{align*}
where the second equality holds because $\sum_{st: A_{s't'} = 0} \frac{\tilde{A}_{s't'} A_{s't'}}{\sum_{s't'} \tilde{A}_{s't'}^2} = 0$.
The third equality holds by combining the summands in the second equality.
The fourth equality holds by $A_{st} = \tilde{A}_{st} + Proj_{\alpha_s, \gamma_t}(A_{st})$ (where $Proj_{\alpha_s, \gamma_t}(A_{st})$ is the projection of $A_{st}$ on the state and time fixed effects).
The fifth equality holds because the residuals $\tilde{A}_{st}$ are orthogonal to their projection. \textbf{QED.}
 
\section{Imputing Missing Drug Overdoses Outcome Data}\label{interpolation}
If the yearly total number of drug-overdose deaths is missing in a state, we first impute the missing yearly totals by dividing the total number of unaccounted drug-overdose deaths in that state from 2000 to 2019 equally by the number of years for which the yearly number of drug-overdose deaths were missing.
For state $s$ with missing monthly number of drug-overdose deaths, we linearly interpolate the number of unintentional drug-overdose deaths for month $m$ in year $u$ with missing outcome (indicated by $C_{s,u,m} = 1$) using yearly overdose death data.
Denote the linearly interpolated number of unintentional drug-overdose deaths in state $s$ for month $m$ in year $u$ by $y'_{s,u,m}$.
To ensure that the imputed monthly drug-overdose death data are consistent with the yearly CDC data, we weight the interpolated values using the yearly number of drug-overdose deaths.
Using the total number of drug-overdose deaths that are not accounted for by the monthly data, we scale the linearly interpolated outcomes, $y'_{s,u,m}$, by a factor so that the total number of imputed drug-overdose deaths in a year is equal to the total number of drug-overdose deaths that actually needs to be imputed for that year for that state.
Therefore, we compute the number of yearly overdose deaths that are not accounted for by the non-missing monthly data using the observed number of yearly drug-overdose deaths ($y_{s,u}$) and the observed, unsuppressed, number of monthly drug-overdose deaths ($y_{s,u,m})$:
\begin{align*}
    \#\text{ of unaccounted OD deaths in state $s$, year $u$} = y_{s,u} - \sum_{\tilde{m} = 1}^{12} y_{s,u,\tilde{m}}\left(1 - C_{s,u,\tilde{m}}\right).
\end{align*} 
Finally, the imputed number of unintentional drug-overdose deaths for state $s$ for year $u$ at month $m$ is:
\begin{align}\label{weight_lin}
    \tilde{y}_{s,u,m} = \frac{y_{s,u} - \sum_{\tilde{m} = 1}^{12} y_{s,u,\tilde{m}}\left(1 - C_{s,u,\tilde{m}}\right)}{\sum_{\tilde{m}=1}^{12} y'_{s,u,\tilde{m}}C_{s,u,\tilde{m}}} y'_{s,u,m}.
\end{align}
If there are no suppressed values in state $s$ in year $u$, we do not need to calculate this imputed value.
If we cannot linearly interpolate the value for a specific state-year-month combination (e.g. when there are a sequence of suppressed monthly drug-overdose deaths), we impute the number of overdose deaths by dividing the total number of unaccounted overdose deaths in state $s$ in year $u$ evenly amongst the months for which the outcome data are suppressed:
\begin{align}\label{equal_weight}
    \tilde{y}_{s,u,m} = \frac{y_{s,u} - \sum_{\tilde{m}=1}^{12} y_{s,u,\tilde{m}}\left(1 - C_{s,u,\tilde{m}}\right)}{\sum_{\tilde{m}=1}^{12} C_{s,u,\tilde{m}}}.
\end{align}

\section{Intervention Dates}\label{appen:intervention_dates}

\subsection{Drug-Induced Homicide Prosecutions Reported by the Media}
Table \ref{appendix_tab:dih_int_dates} contains the intervention dates of the first drug-induced homicide (DIH) prosecution reported by the media in each state between 2000-2019 where the individuals who died from drug overdose were aged 18 and up. 
These intervention dates were collected from the Health in Justice Action Lab.\cite{health_in_justice_lab}

\begin{table}[!htb]
	\caption{The intervention date (year-month) of the first drug-induced homicide (DIH) prosecutions reported by the media in each state and the number of DIH prosecutions reported by the media in each state.} 
	\begin{minipage}[b]{\linewidth}
         \centering
         \renewcommand{\arraystretch}{.8}
		
				\begin{tabular}{|c|c|c|}
			\hline
			 {State}          & {Intervention Date (year-month)} & {Total Number of DIH prosecutions in 2000-2019} \\ \hline
Alabama        & 2008-04          & 11  \\ 
Alaska         & 2016-06          & 7   \\ 
Arizona        & 2012-10          & 7   \\ 
Arkansas       & 2011-11          & 4   \\ 
California     & 2001-10          & 65  \\ 
Colorado       & 2011-08          & 29  \\ 
Connecticut    & 2008-01          & 43  \\ 
Delaware       & 2016-10          & 3   \\ 
Florida        & 2001-02          & 133 \\ 
Georgia        & 2000-04          & 27  \\ 
Hawaii         & NA               & 0  \\ 
Idaho          & 2015-09          & 8   \\ 
Illinois       & 2002-07          & 330 \\ 
Indiana        & 2010-04          & 55  \\ 
Iowa           & 2004-11          & 31  \\ 
Kansas         & 2007-12          & 6   \\ 
Kentucky       & 2011-09          & 41  \\ 
Louisiana      & 2004-03          & 63  \\ 
Maine          & 2009-01          & 17  \\ 
Maryland       & 2002-08          & 58  \\ 
Massachusetts  & 2008-01          & 33  \\ 
Michigan       & 2005-08          & 104 \\ 
Minnesota      & 2007-11          & 135 \\ 
Mississippi    & 2014-08          & 1   \\ 
Missouri       & 2005-08          & 40  \\ 
Montana        & 2004-01          & 10  \\ 
Nebraska       & 2013-01          & 1   \\ 
Nevada         & 2005-10          & 11  \\ 
New Hampshire  & 2010-06          & 38  \\ 
New Jersey     & 2002-08          & 134 \\ 
New Mexico     & 2010-02          & 4   \\ 
New York       & 2007-10          & 96 \\ 
North Carolina & 2005-04          & 121 \\ 
North Dakota   & 2013-01          & 51  \\ 
Ohio           & 2000-03          & 379 \\ 
Oklahoma       & 2011-05          & 39  \\ 
Oregon         & 2009-10          & 16  \\ 
Pennsylvania   & 2001-02          & 708 \\ 
Rhode Island   & 2014-02          & 2   \\ 
South Carolina & 2016-03          & 11  \\ 
South Dakota   & 2017-02          & 12  \\ 
Tennessee      & 2008-03          & 94  \\ 
Texas          & 2001-08          & 43  \\ 
Utah           & 2005-07          & 17  \\ 
Vermont        & 2009-07          & 12  \\ 
Virginia       & 2003-09          & 59  \\ 
Washington     & 2008-10          & 63  \\ 
West Virginia  & 2013-03          & 33  \\ 
Wisconsin      & 2004-06          & 364 \\ 
Wyoming        & 2009-08          & 19  \\ \hline
		\end{tabular}
	\end{minipage}
	\label{appendix_tab:dih_int_dates}
\end{table}

\subsection{Other Policies}
Table \ref{appendix_tab:other_polices_int_date} contains the intervention dates for the other policy measures we controlled for in our analysis. 
We obtained data on the intervention dates from the Prescription Drug Abuse Policy System (PDAPs) \cite{pdaps_nal, pdaps_mml, pdaps_rml, pdaps_gsl, pdaps_pdmp} and the Henry J. Kaiser Family Foundation. \cite{kff_db} 
We made several adjustments to the intervention dates upon further research, as detailed in the following sections.

\subsubsection{Medical Marijuana}
PDAPs collects data on states in which a medical marijuana law was in effect from January 1, 2014 to February 1, 2017. 
We updated the data to include states that had a medical marijuana law in effect after February 2017 and adjusted some intervention dates upon further research:

\begin{itemize}
    \item \textbf{Louisiana:} PDAPs reports a missing value for Louisiana. We found that Louisiana legalized medical marijuana in 1991. However, until June 2015, there was no framework set up to determine how the medical marijuana was to be dispensed, cultivated, etc. 
    In June 29, 2015, Louisiana's governor, Bobby Jindal, signed a bill to set up a framework for dispensing medical marijuana---we used June 29, 2015 as the intervention date for the medical marijuana law in Louisiana. \cite{Litten_2015}
    \item \textbf{Missouri:} PDAPs reports a missing value for Missouri. Missouri legalized medical marijuana on November 6, 2018. \cite{Angell_2018}
    \item \textbf{Oklahoma:} PDAPs reports a missing value for Oklahoma. Oklahoma legalized medical marijuana on June 26, 2018. \cite{Ballotpedia}
    \item \textbf{Utah:} PDAPs reports a missing value for Utah. Utah legalized medical marijuana on November 6, 2018. \cite{MPP_2023a}
    \item \textbf{West Virginia:} PDAPs reports a missing value for West Virginia. West Virginia legalized medical marijuana on April 19, 2017. \cite{MPP_2023b}
    \item \textbf{Minnesota:} PDAPs reports an intervention date of May 30, 2004 for Minnesota. Upon further research, Minnesota legalized medical marijuana on May 30, 2014, \cite{MPP_2023c} so we believe there was a typo in the dataset.
\end{itemize}

\subsubsection{Recreational Marijuana}
PDAPs collects data on states in which a recreational marijuana law was in effect from October 2014 to February 1, 2017. 
We updated the data to include states that had a recreational marijuana law in effect after February 2017 and adjusted some intervention dates upon further research:

\begin{itemize}
    \item \textbf{Colorado:} PDAPs reports an intervention date of July 1, 2014 for Colorado. Colorado was the one of the first states to legalize recreational marijuana on November 6, 2012. \cite{Gurman_2012}
    \item \textbf{Washington:} PDAPs reports an intervention date of June 12, 2014 for Washington. Washington also legalized recreational marijuana on November 6, 2012. \cite{Gurman_2012} 
    \item \textbf{California:} PDAPs reports an intervention date of January 1, 2017 for California. California legalized recreational marijuana on November 8, 2016. \cite{Ballotpedia_cali}
    \item \textbf{Michigan:} PDAPs reports a missing value for Michigan. Michigan legalized recreational marijuana on December 6, 2018. \cite{Gray_2018} 
    \item \textbf{Vermont:} PDAPs reports a missing value for Vermont. Vermont legalized recreational marijuana on July 1, 2018. \cite{Zezima_2018} 
\end{itemize}

\subsubsection{Medicaid Expansion}
We obtained the intervention dates for medicaid expansion in each state from the Henry J. Kaiser Family Foundation. 
We adjusted some intervention dates upon further research.
Specifically, there were several states that had an early start on medicaid expansion, \cite{KFF_2012} and these dates were not reflected in the dataset (these states had a reported intervention date of January 1, 2014 in the dataset):

\begin{itemize}
    \item \textbf{California:} California expanded medicaid on November 1, 2010.
    \item \textbf{Connecticut:} Connecticut expanded medicaid on April 1, 2010.
    \item \textbf{Minnesota:} Minnesota expanded medicaid on March 1, 2010.
    \item \textbf{New Jersey:} New Jersey expanded medicaid on April 14, 2011. 
    \item \textbf{Washington:} Washington expanded medicaid on January 3, 2011. 
\end{itemize}

\begin{table}[!htb]
	\caption{The intervention dates (year-month) for the other policy measures up to December 2019: naloxone access law where pharmacists a) can dispense naloxone without prescription (NAL: can dispense) and b) cannot dispense naloxone without prescription (NAL: cannot dispense) c) medical marijuana law (MML) d) recreational marijuana law (RML) e) Prescription Drug Monitoring Program (PDMP) f) 911 Good Samaritan Law (911 GSL) g) Medicaid expansion (Medicaid). Values of NA indicates ``not applicable'', i.e. the policy measure was not in effect during the period studied. } 
	\begin{minipage}[b]{\linewidth}
    \centering
    \renewcommand{\arraystretch}{.7}
    \resizebox*{!}{.8\textheight}{
		\begin{tabular}{|c|ccccccc|}
			\hline
{State}          & \thead{{a) NAL: can} \\{dispense}} & \thead{{b) NAL: cannot} \\ {dispense}} & {c) MML}     & {d) RML}     & {e) PDMP}   & \thead{{f) 911} \\{GSL}}     & {g) Medicaid} \\ \hline
Alabama             & 2015-06              & NA                  & NA      & NA      & 2005-05 & 2015-06 & NA       \\ 
Alaska                 & 2016-03              & NA                  & 1999-03 & 2015-02 & 2008-09 & 2008-09 & 2015-09  \\ 
Arizona                & 2016-08              & NA                  & 2010-12 & NA      & 2007-09 & 2018-04 & 2014-01  \\ 
Arkansas                & 2015-07              & NA                  & 2016-11 & NA      & 2011-03 & 2015-07 & 2014-01  \\ 
California             & 2014-01              & 2008-01             & 1996-11 & 2016-11 & 1939-01 & 2013-01 & 2010-11  \\ 
Colorado              & 2015-04              & 2013-05             & 2000-12 & 2012-11 & 2005-06 & 2012-05 & 2014-01  \\ 
Connecticut             & 2015-06              & 2003-10             & 2012-05 & NA      & 2006-06 & 2011-10 & 2010-04  \\ 
Delaware                & 2014-08              & NA                  & 2011-07 & NA      & 2010-07 & 2013-08 & 2014-01  \\ 
Florida                 & 2016-07              & 2015-06             & 2017-01 & NA      & 2009-06 & 2012-10 & NA       \\ 
Georgia                 & 2014-04              & NA                  & NA      & NA      & 2011-05 & 2014-04 & NA       \\ 
Hawaii                     & 2016-06              & NA                  & 2000-06 & NA      & 1943-01 & 2015-07 & 2014-01  \\ 
Idaho                   & 2015-07              & NA                  & NA      & NA      & 1967-01 & 2018-07 & 2020-01       \\ 
Illinois             & 2010-01              & NA                  & 2014-01 & NA      & 1961-01 & 2012-06 & 2014-01  \\ 
Indiana                & 2015-04              & NA                  & NA      & NA      & 1997-01 & 2014-03 & 2015-02  \\ 
Iowa                  & 2016-05              & NA                  & NA      & NA      & 2006-05 & 2018-07 & 2014-01  \\ 
Kansas                  & 2017-07              & NA                  & NA      & NA      & 2008-07 & NA      & NA       \\ 
Kentucky                 & 2013-06              & NA                  & NA      & NA      & 1998-07 & 2015-03 & 2014-01  \\ 
Louisiana                & 2015-08              & NA                  & 2015-06      & NA      & 2006-07 & 2014-08 & 2016-07  \\ 
Maine                   & 2015-10              & 2014-04             & 1999-12 & 2017-01 & 2003-06 & 2019-09      & 2019-01  \\ 
Maryland                 & 2015-10              & 2013-10             & 2013-10 & NA      & 2011-05 & 2009-10 & 2014-01  \\ 
Massachusetts          & 2014-07              & 2012-08             & 2013-01 & 2016-12 & 1992-01 & 2012-08 & 2014-01  \\ 
Michigan               & 2017-03              & 2014-10             & 2008-12 & 2018-12      & 1988-01 & 2017-01 & 2014-04  \\ 
Minnesota              & 2014-05              & NA                  & 2014-05 & NA      & 2007-07 & 2014-07 & 2010-03  \\ 
Mississippi            & 2015-07              & NA                  & NA      & NA      & 2005-01 & 2015-07 & NA       \\ 
Missouri               & 2016-08              & NA                  & 2018-11      & NA      & NA      & 2017-08 & 2021-08       \\ 
Montana                & 2017-05              & NA                  & 2004-11 & NA      & 2011-07 & 2017-05 & 2016-01  \\ 
Nebraska               & NA                   & 2015-05             & NA      & NA      & 2011-04 & 2017-08 & 2020-10       \\ 
Nevada                 & 2015-10              & NA                  & 2001-10 & 2017-01 & 1995-06 & 2015-10 & 2014-01  \\ 
New Hampshire          & 2015-06              & NA                  & 2013-07 & NA      & 2012-06 & 2015-09 & 2014-08  \\ 
New Jersey            & 2013-07              & NA                  & 2010-10 & NA      & 2008-01 & 2013-05 & 2011-04  \\ 
New Mexico             & 2014-03              & 2001-04             & 2007-07 & NA      & 2004-07 & 2007-06 & 2014-01  \\ 
New York               & 2014-06              & 2006-04             & 2014-07 & NA      & 1972-01 & 2011-09 & 2014-01  \\ 
North Carolina           & 2013-04              & NA                  & NA      & NA      & 2005-08 & 2013-04 & NA       \\ 
North Dakota           & 2015-08              & NA                  & 2016-12 & NA      & 2005-12 & 2015-08 & 2014-01  \\ 
Ohio                   & 2015-07              & 2014-03             & 2016-09 & NA      & 2005-05 & 2016-09 & 2014-01  \\ 
Oklahoma                & 2014-11              & 2013-11             & 2018-06      & NA      & 1990-05 & 2018-11      & 2021-07       \\ 
Oregon                & 2013-06              & NA                  & 1998-12 & 2015-07 & 2009-07 & 2016-01 & 2014-01  \\ 
Pennsylvania           & 2014-12              & NA                  & 2016-05 & NA      & 1972-01 & 2014-12 & 2015-01  \\ 
Rhode Island         & 2014-03              & 2012-06             & 2006-01 & NA      & 1978-01 & 2012-06 & 2014-01  \\ 
South Carolina          & 2016-06              & 2015-06             & NA      & NA      & 2006-06 & 2017-06 & NA       \\ 
South Dakota           & 2016-07              & NA                  & NA      & NA      & 2010-03 & 2017-07 & NA       \\ 
Tennessee               & 2014-07              & NA                  & NA      & NA      & 2003-01 & 2015-07 & NA       \\ 
Texas                  & 2015-09              & NA                  & NA      & NA      & 1981-09 & NA      & NA       \\ 
Utah                   & 2016-05              & 2014-05             & 2018-11      & NA      & 1995-01 & 2014-03 & 2020-01       \\ 
Vermont                 & 2013-07              & NA                  & 2004-07 & 2018-07      & 2006-05 & 2013-06 & 2014-01  \\ 
Virginia              & 2015-04              & 2013-07             & NA      & NA      & 2002-04 & 2015-07 & 2019-01  \\ 
Washington             & 2015-07              & 2010-06             & 1998-12 & 2012-11 & 2007-07 & 2010-06 & 2011-01  \\ 
West Virginia           & 2016-06              & 2015-05             & 2017-04      & NA      & 1995-07 & 2015-06 & 2014-01  \\ 
Wisconsin              & 2015-12              & 2014-04                  & NA      & NA      & 2010-05 & 2014-04 & NA       \\ 
Wyoming                & 2017-07              & NA                  & NA      & NA      & 2003-03 & NA      & NA       \\ \hline
		\end{tabular}}
	\end{minipage}
	\label{appendix_tab:other_polices_int_date}
\end{table}

\clearpage
\section{States and their U.S. Regions}\label{appen:states_and_us_regions}

Table \ref{appendix_tab:us_regions} contains the states under different U.S regions. 
These regions are used to estimate different smoothed time effects across the different U.S. regions.

\begin{table}[!htb]
	\caption{The states considered in the analysis with their U.S. regions, which were obtained from the U.S. Census. \cite{us_census_bureau_2018}} 
	\begin{minipage}[b]{\linewidth}
        \centering
		\begin{tabular}{|c|c|c|c|}
			\hline
{Northeast} & {Midwest} & {South} & {West} \\ \hline
Connecticut        & Illinois         & Alabama        & Alaska        \\

Maine              & Iowa             & Arkansas       & Arizona       \\

Massachusetts      & Indiana          & Delaware       & California    \\

New Hampshire      & Kansas           & Florida        & Colorado      \\

New Jersey         & Michigan         & Georgia        & Hawaii        \\

New York           & Minnesota        & Kentucky       & Idaho         \\

Pennsylvania       & Missouri         & Louisiana      & Montana       \\

Rhode Island       & Nebraska         & Maryland       & New Mexico    \\

Vermont            & North Dakota     & Mississippi    & Nevada        \\

                   & Ohio             & North Carolina & Oregon        \\
                   
                   & South Dakota     & Oklahoma       & Utah          \\
                   
                   & Wisconsin        & South Carolina & Washington    \\
                   
                   &                  & Tennessee      & Wyoming       \\
                   
                   &                  & Texas          &               \\
                   
                   &                  & Virginia       &               \\
                   
                   &                  & West Virginia  &               \\ \hline

		\end{tabular}
	\end{minipage}
	\label{appendix_tab:us_regions}
\end{table}

\section{Sandwich Estimator for the Variance}

\subsection{Models with Linear Link Function}\label{appendix:sandwich_est_event_study}
\sloppy
Denote the vector of state fixed effects, time effects, other relevant policy measures, and the DIH prosecutions reported by the media for state $s$ at time interval $t$ by
\begin{align}\label{eq:z_for_constant_tx}
\Vec{Z}_{st}^T = \begin{pmatrix}
\mathbb{I}\{\text{state} = s\} &
\mathbb{I}\{\text{time} = t\} &
A_{st} &
X_{st} 
\end{pmatrix} 
\end{align} or 
\begin{align}\label{eq:z_st_for_event_study}
\Vec{Z}_{st}^T = \begin{pmatrix}
\mathbb{I}\{\text{state} = s\} &
\mathbb{I}\{\text{time} = t\} & 
K_{st} &
X_{st} 
\end{pmatrix},
\end{align}
depending on whether we are assuming a constant treatment effect or we are assuming that the treatment effect depends on the treatment duration.

First, the Mean Value Theorem states that for a function $f$ that is continuous on $[a,b]$ and differentiable on $(a,b)$ for $a,b \in \mathbb{R}$ and $a<b$:
\begin{align*}
    f(b) - f(a) = f'(c)(b-a)
\end{align*}
for some value $c \in (a,b)$.
From the unbiased estimating equations, we let $f(\theta) = \sum_{st} \Vec{Z}_{st} \left(\log Y_{st} - 
\log p_{st,\theta} \right)$, where $Y_{st}$ is the risk of unintentional drug-overdose death in state $s$ at time interval $t$ and $\log p_{st,\theta^*} = \Vec{Z}_{st}^T \theta^*$ is the underlying probability of unintentional drug-overdose deaths given by true parameter $\theta^*$.
Then,
\begin{align}\label{eq:mean_value_thm_eq}
    f'(\tilde{\theta})\left( \hat{\theta} - \theta^*\right) = \sum_{st} \Vec{Z}_{st} \left(\log Y_{st} - \Vec{Z}_{st}^T \hat{\theta}\right) - \sum_{st} \Vec{Z}_{st} \left(\log Y_{st} - \Vec{Z}_{st}^T {\theta}^* \right),
\end{align}
where $\tilde{\theta}$ is between $\hat{\theta}$ and $\theta^*$.
Since $\hat{\theta} \overset{p} \to \theta^*$ and $\tilde{\theta}$ is between $\hat{\theta}$ and $\theta^*$, then $\tilde{\theta} \overset{p} \to \theta^*$.
Note that the derivative $f'$ is given by:
\begin{align*}
    f'(\theta) &= \frac{\partial}{\partial \theta} \sum_{st} \Vec{Z}_{st} \left(\log Y_{st} - \Vec{Z}_{st}^T \theta \right) \\
    &= - \sum_{st} \frac{\partial}{\partial \theta} \Vec{Z}_{st}  \Vec{Z}_{st}^T \theta \\
    &= - \sum_{st} \Vec{Z}_{st} \Vec{Z}_{st}^T.
\end{align*}
Substituting in $- \sum_{st} \Vec{Z}_{st}  \Vec{Z}_{st}^T$ for the derivative, we can rearrange Equation~\eqref{eq:mean_value_thm_eq}:
\begin{align}\label{eq:mean_value_thm_for_model_constant_tx}
    \sum_{st} \Vec{Z}_{st} \left(\log Y_{st} - \Vec{Z}_{st}^T \hat{\theta}\right)  &=  \sum_{st} \Vec{Z}_{st} \left(\log Y_{st} - \Vec{Z}_{st}^T \theta^* \right) -\sum_{st} \Vec{Z}_{st}  \Vec{Z}_{st}^T  \left(\hat{\theta} - \theta^*\right).
\end{align}
Note that the left hand side of Equation~\eqref{eq:mean_value_thm_for_model_constant_tx} is 0 since $\hat{\theta}$ solves the unbiased estimating equation.
Then, 
\begin{align}\label{eq:theta_distribution_constant_tx}
    \left(\hat{\theta} - \theta^*\right)  = \underbrace{\left(\sum_{st} \Vec{Z}_{st}  \Vec{Z}_{st}^T \right)^{-1}}_{=C^{-1}} \sum_{st} \Vec{Z}_{st} \left(\log Y_{st} - \Vec{Z}_{st}^T \theta^* \right).
\end{align}
Note that $C$ is a positive semi-definite matrix since $\Vec{Z}_{st}\Vec{Z}_{st}^T$ is positive semi-definite.
Since $C^{-1}$ is easy to estimate, we focus on the limiting distribution of $\sum_{st} \Vec{Z}_{st} \left(\log Y_{st} - \Vec{Z}_{st}^T \theta^* \right)$.
We make the following assumption:
\begin{assumption} \label{assumption:expectation_assumption}
Given the state fixed effects, time effects, other relevant policy measures, and the DIH prosecutions reported by the media for state $s$ at time interval $t$, denoted as ${\Vec{Z}}_{st}$, 
\begin{align}
    \mathbb{E}\left(\log Y_{st} \mid {\Vec{Z}}_{st} = {\Vec{z}}_{st} \right) =  \log {p_{st,\theta^*}},
\end{align}
where $p_{st,\theta^*}$, given by the true parameter $\theta^*$, is a correctly specified model for $p_{st}$.
\end{assumption}

First, we find the expected value of $\sum_{st} \Vec{Z}_{st} \left(\log Y_{st} - \Vec{Z}_{st}^T \theta^* \right)$.
By linearity of expectations, we focus on the individual $st$ terms:
\begin{align*}
    \E\left(\Vec{Z}_{st} \left(\log Y_{st} - \Vec{Z}_{st}^T \theta^* \right)\right)
    &= \E\left(\Vec{Z}_{st} \E\left[ \log Y_{st} - \log p_{st,\theta^*}\mid {\Vec{Z}}_{st} = \Vec{z}_{st}
    \right]\right) 
    = 0,
\end{align*}
where the first equality holds by the Law of Total Expectations and substituting in $\log p_{st,\theta^*}$ for $\Vec{Z}_{st}^T \theta^*$, and the last line holds by Assumption \ref{assumption:expectation_assumption}.
Since the expected value is zero, we focus on the second moment to find the variance.
\begin{align}\label{appen:second_moment}
    &\E\left(\left( \sum_{st} \Vec{Z}_{st} \left(\log Y_{st} - \Vec{Z}_{st}^T \theta^* \right)\right) \left( \sum_{st} \Vec{Z}_{st} \left(\log Y_{st} - \Vec{Z}_{st}^T \theta^*\right)\right)^T\right)  \nonumber \\
    &= \sum_{st} \sum_{s't'} \E\bigg( \Vec{Z}_{st} \left(\log Y_{st} - \Vec{Z}_{st}^T \theta^* \right) \left(\log Y_{s't'} - \Vec{Z}_{s't'}^T \theta^*\right)  \Vec{Z}_{s't'}^T \bigg).
\end{align}
The Sandwich estimator of the variance is then given by:
\begin{align}
    \hat{\Sigma} &= \frac{N}{N-d}\left(\sum_{st} \Vec{Z}_{st}  \Vec{Z}_{st}^T \right)^{-1} \left(\sum_{st} \sum_{s't'} \bigg( \Vec{Z}_{st} \left(\log Y_{st} - \Vec{Z}_{st}^T \hat{\theta}\right) \left(\log Y_{s't'} - \Vec{Z}_{s't'}^T \hat{\theta} \right) \Vec{Z}_{s't'}^T \bigg) \right) \left(\left(\sum_{st} \Vec{Z}_{st}  \Vec{Z}_{st}^T \right)^{-1}\right)^T,
\end{align}
where we estimate $\theta^*$ using $\hat{\theta}$.
In addition, we include a factor of $\frac{N}{N-d}$ for bias correction,\citep{li2015small} where $N$ is equal to the number of state-time combinations and $d$ is equal to the number of parameters. 
In the case that the following assumption holds true, the sandwich estimator of the variance is simplified:
\begin{assumption}\label{assumption:conditional_ind_btwn_states}
Following Bertrand, Duflo, and Mullainathan, \cite{bertrand2004much} we assume conditional independence between states. 
That is, consider states $s$ and $s'$, where $s \neq s'$, and $t' \leq t$.
Given $Y_{st}$ and ${\Vec{Z}}_{st}$ for state $s$ and time interval $t$ and $\Vec{Z}_{s't'}$ and $Y_{s't'}$ for state $s'$ at time interval $t'$,
\begin{align}
     Y_{st} \indep \left(\Vec{Z}_{s't'}, Y_{s't'}\right) \mid {\Vec{Z}}_{st}.
\end{align}
\end{assumption}

Under Assumption \ref{assumption:conditional_ind_btwn_states}, terms with $s \neq s'$ in Equation~\eqref{appen:second_moment} are zero: for $s \neq s'$, and without the loss of generality (WLOG) $t' \leq t$,
\begin{align*}
    &\E\left( \Vec{Z}_{st} \left(\log Y_{st} - \Vec{Z}_{st}^T \theta^* \right)\left(\log Y_{s't'} - \Vec{Z}_{s't'}^T \theta^*\right) \Vec{Z}_{s't'}^T \right) \\
    &={\E\left( \E\left[\Vec{Z}_{st} \left(\log Y_{st} - \Vec{Z}_{st}^T \theta^*\right) \left(\log Y_{s't'} - \Vec{Z}_{s't'}^T \theta^*\right) \Vec{Z}_{s't'}^T \mid {\Vec{Z}}_{st}, 
    \Vec{Z}_{s't'}, Y_{s't'} \right]\right)} \\
    &=  \E\left( \Vec{Z}_{st} \E\left[\log Y_{st} - \Vec{Z}_{st}^T \theta^* \mid {\Vec{Z}}_{st},
    \Vec{Z}_{s't'}, Y_{s't'} \right] \left(\log Y_{s't'} - \Vec{Z}_{s't'}^T \theta^*\right) \Vec{Z}_{s't'}^T \right) \\
    &=  \E\left( \Vec{Z}_{st} \E\left[\log Y_{st} - \Vec{Z}_{st}^T \theta^* \mid {\Vec{Z}}_{st}
    \right] \left(\log Y_{s't'} - \Vec{Z}_{s't'}^T \theta^*\right) \Vec{Z}_{s't'}^T \right) \\
    &= 0,
\end{align*}
where the first equality holds by Law of Total Expectations, the third equality holds by Assumption \ref{assumption:conditional_ind_btwn_states}, and the last equality holds by Assumption \ref{assumption:expectation_assumption}.
Hence, when $s \neq s'$ and WLOG $t' \leq t$, the expectation is equal to 0.
Therefore, the expectation is only non-zero for $s = s'$, i.e.
\begin{align*}
    &\E\left(\left( \sum_{st} \Vec{Z}_{st} \left(\log Y_{st} - \Vec{Z}_{st}^T \theta^* \right)\right) \left( \sum_{st} \Vec{Z}_{st} \left(\log Y_{st} - \Vec{Z}_{st}^T \theta^*\right) \right)\right)^T  \nonumber \\
    &= \sum_{s} \E\left( \left( \sum_{t} \Vec{Z}_{st} \left(\log Y_{st} - \Vec{Z}_{st}^T \theta^*\right)\right) \left( \sum_{t'} \Vec{Z}_{st'} \left(\log Y_{st'} -\Vec{Z}_{st'}^T \theta^*\right) \right)^T\right).
\end{align*}
Hence, the Sandwich estimator for the variance of the parameters for the generalized additive models (GAMs) with a linear link function under Assumption \ref{assumption:conditional_ind_btwn_states} is given by:
\begin{align}
    \hat{\Sigma} = \frac{N}{N-d} \left(\sum_{st} \Vec{Z}_{st}  \Vec{Z}_{st}^T \right)^{-1}
    \left(\sum_s \left(\sum_{t}  \Vec{Z}_{st} \left(\log Y_{st} - \Vec{Z}_{st}^T \hat{\theta}\right)\right) 
    \left(\sum_{t'}  \Vec{Z}_{st'} \left(\log Y_{st'} - \Vec{Z}_{st'}^T \hat{\theta}\right)\right)^T\right)
    \left(\left(\sum_{st} \Vec{Z}_{st}  \Vec{Z}_{st}^T \right)^{-1}\right)^T.
\end{align}

\subsection{Models with Logistic Link Function}\label{sec:appendix_sandwich_est_constant_tx}
The sandwich estimators under the GAMs with a logistic link are derived in a similar manner as in the models with a linear link. 
The main difference is that the unbiased estimating equation now gives:
\begin{align*}
    f(\theta) = \sum_{st} \Vec{Z}_{st} \left(Y_{st} - p_{st,\theta} \right),
\end{align*}
where $p_{st,\theta} = expit(\Vec{Z}_{st}^T\theta)$ and $\Vec{Z}_{st}$ is still given by Equations~\eqref{eq:z_for_constant_tx} or \eqref{eq:z_st_for_event_study}.
Under Assumption \ref{assumption:conditional_ind_btwn_states}, the sandwich estimator for the GAMs with a logistic link function is given by 
\begin{align}
    \hat{\Sigma} &= \frac{N}{N-d}\left(\sum_{st} \Vec{Z}_{st}  \Vec{Z}_{st}^T p_{st, \hat{\theta}}\left(1 - p_{st, \hat{\theta}}\right)\right)^{-1} \nonumber \\
    &\times
    \left(\sum_s \left( \sum_{t} \Vec{Z}_{st} \left(Y_{st} - expit\left(\Vec{Z}_{st}^T \hat{\theta}\right) \right)\right) \left( \sum_{t'} \Vec{Z}_{st'} \left(Y_{st'} - expit\left(\Vec{Z}_{st'}^T \hat{\theta}\right) \right)\right)^T \right)\nonumber \\
    &\times \left(\left(\sum_{st} \Vec{Z}_{st}  \Vec{Z}_{st}^T p_{st, \hat{\theta}}\left(1 - p_{st, \hat{\theta}}\right)\right)^{-1}\right)^T
\end{align}
where we estimate $\theta^*$ by $\hat{\theta}$.

\section{Comparing Event Study Model to Robust Treatment Effect Estimators}\label{appendix:robust_event_study_estimator}
As discussed in Section \ref{sec:gam_model_linear_link}, recent studies have shown that traditional event study models such as Equation~\eqref{eq:ols_model_dih} have limitations, especially in settings with treatment effect heterogeneity.\cite{borusyak2024revisiting, sun2021estimating, de2026difference}
Here, we empirically compare the treatment effect estimates of DIH prosecutions reported by the media from the traditional event study model to those from the robust estimator of de Chaisemartin and D'Haultf{\oe}uille.\cite{de2026difference} 
Divergence between the two estimators suggests that the traditional estimator is affected by these limitations in our setting, though we note that this comparison is illustrative rather than a formal test of those limitations.

We use the R package \texttt{did\_multiplegt\_dyn} to obtain the robust treatment effect estimates.\cite{de2025using}
To account for potential confounding, we include the other intervention variables from our analysis, such as Good Samaritan Laws and Naloxone Access Laws, to the estimation of the robust estimators. 
Note that the treatment effect estimates are similar if we do not include these confounding variables.
Furthermore, because the R package does not support a non-linear, non-parametric estimation of the time effects, we refit our event study model in Equation~\eqref{eq:ols_model_dih} using a fixed time effect, consistent with the traditional event study model. 
For both estimators, we estimate the effects of treatment for treatment duration $k$ from -20 to 37 (the maximum number of pre-treatment and post-treatment effects that the robust estimator can estimate), which corresponds to 10 years before treatment occurs up to 18.5 years after treatment occurs. 

\begin{figure}[!htb]
    \centering
\includegraphics{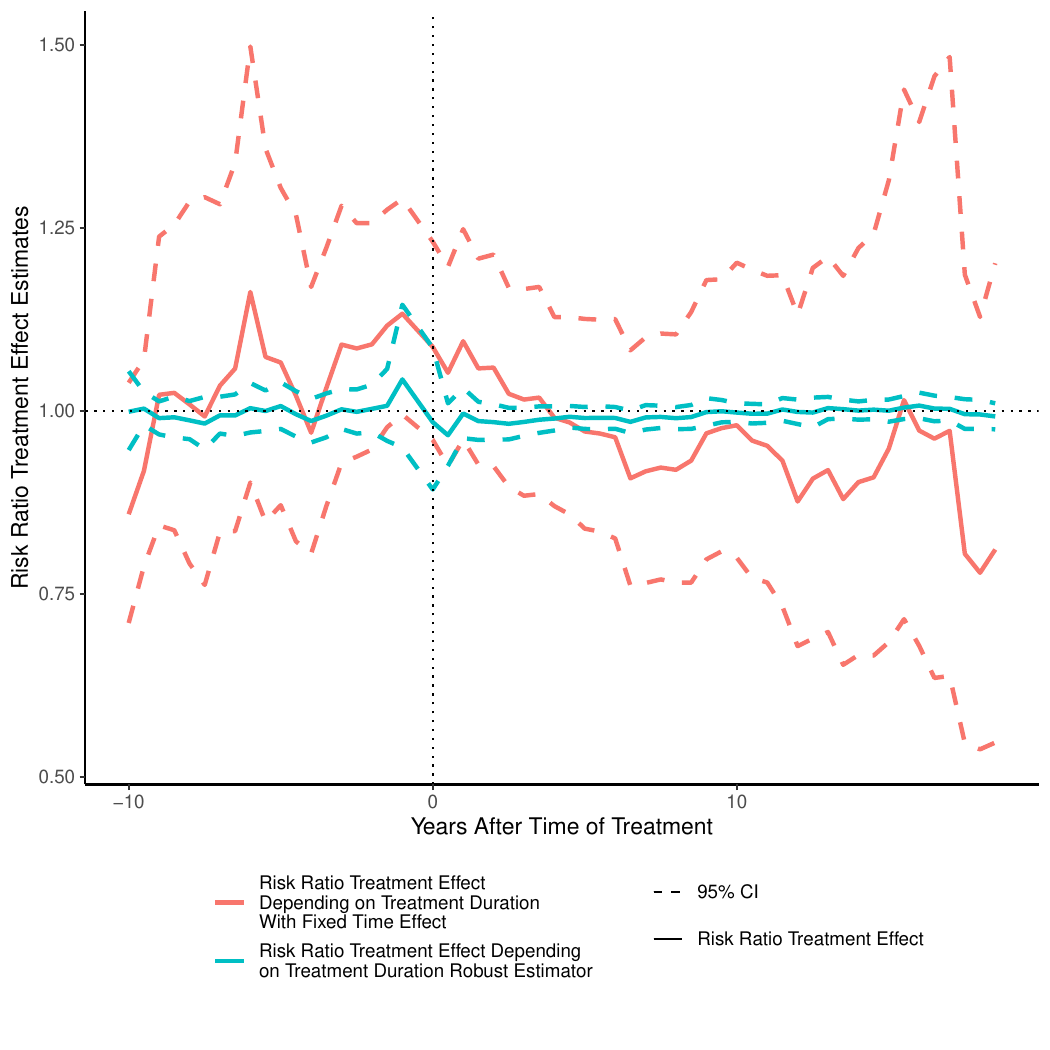}
    \caption{Estimated risk ratios with 95\% confidence intervals for each treatment duration before and after the first drug-induced homicide (DIH) prosecution was reported by the media in the state using the traditional event study model (red) and robust treatment effect estimator (blue).}
    \label{fig:robust_est_v_event_study_risk_ratios}
\end{figure}

Figure \ref{fig:robust_est_v_event_study_risk_ratios} shows the plot of the estimated risk ratios of the treatment effect with confidence intervals for the treatment durations from 10 years before treatment up to 18.5 years after treatment from the traditional event study model and the robust estimator, accounting for the confounding variables. 
Table \ref{tab:robust_est_v_event_study_risk_ratios} shows the corresponding treatment effect estimates and their 95\% confidence intervals.

Using the traditional event study model (left in Table \ref{tab:robust_est_v_event_study_risk_ratios}), the pre-treatment risk ratio estimates range from 0.859 to 1.162 with an average risk ratio of 1.040. 
The post-treatment risk ratio estimates range from 0.779 to 1.095 with an average risk ratio of 0.956. 
Although none of the pre-treatment point estimates are statistically significant, their deviation from 1, raises potential concerns about the validity of the parallel trends assumption under this estimator. 
The post-treatment point estimates suggest a harmful effect shortly after first treatment that becomes more protective at longer durations, though again none of the estimates are statistically significant.

The risk ratio estimates of the treatment effects from the robust estimator (right in Table \ref{tab:robust_est_v_event_study_risk_ratios}) of pre-treatment effects range from 0.983 to 1.043 with an average of 0.999, and the post-treatment risk ratio estimates range from 0.967 to 1.007 with an average risk ratio of 0.994.
The pre-treatment estimates being close to 1 and not statistically significant is consistent with the parallel trends and no anticipation assumptions.
The risk ratios from the robust estimator suggest that the parallel trends assumption and no anticipation effects assumption holds. 
Furthermore, the point estimates of post-treatment effects suggest a small initial protective effect of DIH prosecutions reported by the media, which diminishes as treatment duration increases.
However, none of the estimates are statistically significant and no causal conclusion can be drawn.

We do not include the results here, but results are similar when we do not include the confounding variables.

The divergence between the two estimators, especially in the pre-treatment period and at shorter post-treatment durations, is consistent with the traditional event study model being affected by the biases discussed in previous work in the presence of heterogeneous treatment effects.\cite{borusyak2024revisiting, sun2021estimating, de2026difference}
Despite this divergence between the traditional event study model and the robust estimator, both estimators produce treatment effect estimates that differ substantially from those of the constant treatment effect model. 
Notably, the constant treatment effect estimate falls outside the 5th and 95th percentiles of the estimated robust estimator treatment effects (0.984 and 1.004, respectively), suggesting that the bias introduced by imposing treatment effect homogeneity is substantial and not merely a consequence of the limitations of the traditional event study model.

\begin{table}[!htb]
    \caption{Estimated risk ratios with 95\% confidence intervals for each treatment duration before and after the first drug-induced homicide (DIH) prosecution was reported by the media in the state using the traditional event study model (left column) and robust treatment effect estimator (right column).} 
    \centering
    \begin{tabular}{|ccc|}
			\hline
			 \thead{Number of six-month time intervals after first \\DIH prosecution reported by the media} & \thead{Risk Ratio: Traditional Event Study Model \\ (95\% Confidence Interval)} & 
			 \thead{Risk Ratio: Heterogeneity-Robust Estimator \\ (95\% Confidence Interval)} \\
			\hline
			-20 & 0.859 (0.710, 1.038) & 0.999 (0.946, 1.054)\\
                -19  & 0.918 (0.788, 1.070) & 1.003 (0.980, 1.026)\\
                -18 & 1.022 (0.843, 1.238) & 0.990 (0.968, 1.013)\\
                -17 & 1.024 (0.837, 1.254) & 0.991 (0.964, 1.019)\\
                -16 & 1.009 (0.791, 1.286) & 0.987 (0.961, 1.013)\\
                -15 & 0.992 (0.762, 1.292) & 0.983 (0.947, 1.019)\\
                -14 & 1.034 (0.834, 1.282) & 0.994 (0.969, 1.019)\\
                -13 & 1.058 (0.836, 1.339) & 0.994 (0.966, 1.022)\\
                -12 & 1.162 (0.902, 1.498) & 1.004 (0.970, 1.038)\\
                -11 & 1.074 (0.849, 1.358) & 1.000 (0.972, 1.028)\\
                -10 & 1.066 (0.871, 1.304) & 1.006 (0.975, 1.039)\\
                -9 & 1.021 (0.822, 1.268) & 0.995 (0.965, 1.027)\\
                -8 & 0.970 (0.805, 1.169) & 0.986 (0.957, 1.016)\\
                -7 & 1.031 (0.870, 1.223) & 0.993 (0.964, 1.024)\\
                -6 & 1.090 (0.929, 1.280) & 1.002 (0.975, 1.030)\\
                -5 & 1.085 (0.937, 1.256) & 0.999 (0.969, 1.029)\\
                -4 & 1.091 (0.947, 1.256) & 1.003 (0.971, 1.036)\\
                -3 & 1.116 (0.978, 1.275) & 1.007 (0.959, 1.057)\\
                -2 & 1.133 (0.995, 1.289) & 1.043 (0.950, 1.145)\\
                0 & 1.087 (0.960, 1.231) & 0.984 (0.892, 1.086)\\
                1 & 1.052 (0.925, 1.197) & 0.966 (0.925, 1.010)\\
                2 & 1.095 (0.961, 1.248) & 0.996 (0.963, 1.031)\\
                3 & 1.058 (0.926, 1.208) & 0.986 (0.960, 1.012)\\
                4 & 1.059 (0.924, 1.213) & 0.984 (0.961, 1.009)\\
                5 & 1.023 (0.896, 1.168) & 0.982 (0.961, 1.004)\\
                6 & 1.015 (0.884, 1.166) & 0.985 (0.966, 1.004)\\
                7 & 1.018 (0.886, 1.169) & 0.988 (0.970, 1.006)\\
                8 & 0.991 (0.870, 1.128) & 0.989 (0.973, 1.006)\\
                9 & 0.984 (0.858, 1.128) & 0.992 (0.977, 1.007)\\
                10 & 0.972 (0.839, 1.125) & 0.990 (0.975, 1.005)\\
                11 & 0.969 (0.835, 1.124) & 0.990 (0.975, 1.006)\\
                12 & 0.964 (0.826, 1.125) & 0.990 (0.975, 1.005)\\
                13 & 0.908 (0.761, 1.083) & 0.985 (0.970, 1.000)\\
                14 & 0.917 (0.765, 1.100) & 0.991 (0.974, 1.008)\\
                15 & 0.922 (0.770, 1.106) & 0.992 (0.977, 1.007)\\
                16 & 0.919 (0.765, 1.104) & 0.990 (0.975, 1.005)\\
                17 & 0.932 (0.765, 1.135) & 0.991 (0.975, 1.008)\\
                18 & 0.969 (0.797, 1.179) & 0.998 (0.980, 1.017)\\
                19 & 0.976 (0.808, 1.180) & 0.999 (0.984, 1.015)\\
                20 & 0.980 (0.799, 1.202) & 0.998 (0.985, 1.010)\\
                21 & 0.959 (0.772, 1.193) & 0.996 (0.983, 1.009)\\
                22 & 0.952 (0.765, 1.184) & 0.996 (0.984, 1.009)\\
                23 & 0.932 (0.733, 1.185) & 1.002 (0.986, 1.017)\\
                24 & 0.876 (0.678, 1.131) & 0.998 (0.982, 1.015)\\
                25 & 0.907 (0.689, 1.195) & 0.998 (0.978, 1.018)\\
                26 & 0.919 (0.698, 1.210) & 1.004 (0.989, 1.019)\\
                27 & 0.879 (0.653, 1.184) & 1.002 (0.989, 1.015)\\
                28 & 0.903 (0.666, 1.222) & 1.000 (0.988, 1.013)\\
                29 & 0.909 (0.666, 1.242) & 1.002 (0.988, 1.015)\\
                30 & 0.948 (0.684, 1.315) & 1.000 (0.985, 1.015)\\
                31 & 1.015 (0.715, 1.439) & 1.004 (0.989, 1.020)\\
                32 & 0.973 (0.679, 1.395) & 1.007 (0.990, 1.025)\\
                33 & 0.962 (0.635, 1.457) & 1.003 (0.986, 1.021)\\
                34 & 0.973 (0.638, 1.483) & 1.003 (0.987, 1.019)\\
                35 & 0.805 (0.546, 1.186) & 0.995 (0.975, 1.016)\\
                36 & 0.779 (0.538, 1.128) & 0.995 (0.974, 1.015)\\
                37 & 0.811 (0.547, 1.202) & 0.992 (0.975, 1.010)\\
			\hline
    \end{tabular}	
    \label{tab:robust_est_v_event_study_risk_ratios}
\end{table}

\section{Sensitivity Analyses}\label{appendix:sensitivity_analysis}
\subsection{Excluding the Last Five Years}\label{sensitivity_anlys:exclude_last_five_years}
There may be biases in long-term treatment effects, especially since most states had at least one DIH prosecution reported by the media by the end of 2019. \cite{borusyak2024revisiting}
Because of the potential biased estimate of long-term treatment effects, we conduct a sensitivity analysis where we exclude the last five years from the analysis period. 
That is, we estimate the treatment effect of DIH prosecutions reported by the media from 2000-2014.
With the exclusion of the last five years, the control group consists of six never-treated states (Alaska, Delaware, Hawaii, Idaho, South Carolina, and South Dakota).

\begin{figure}[!htb]
    \centering
\includegraphics{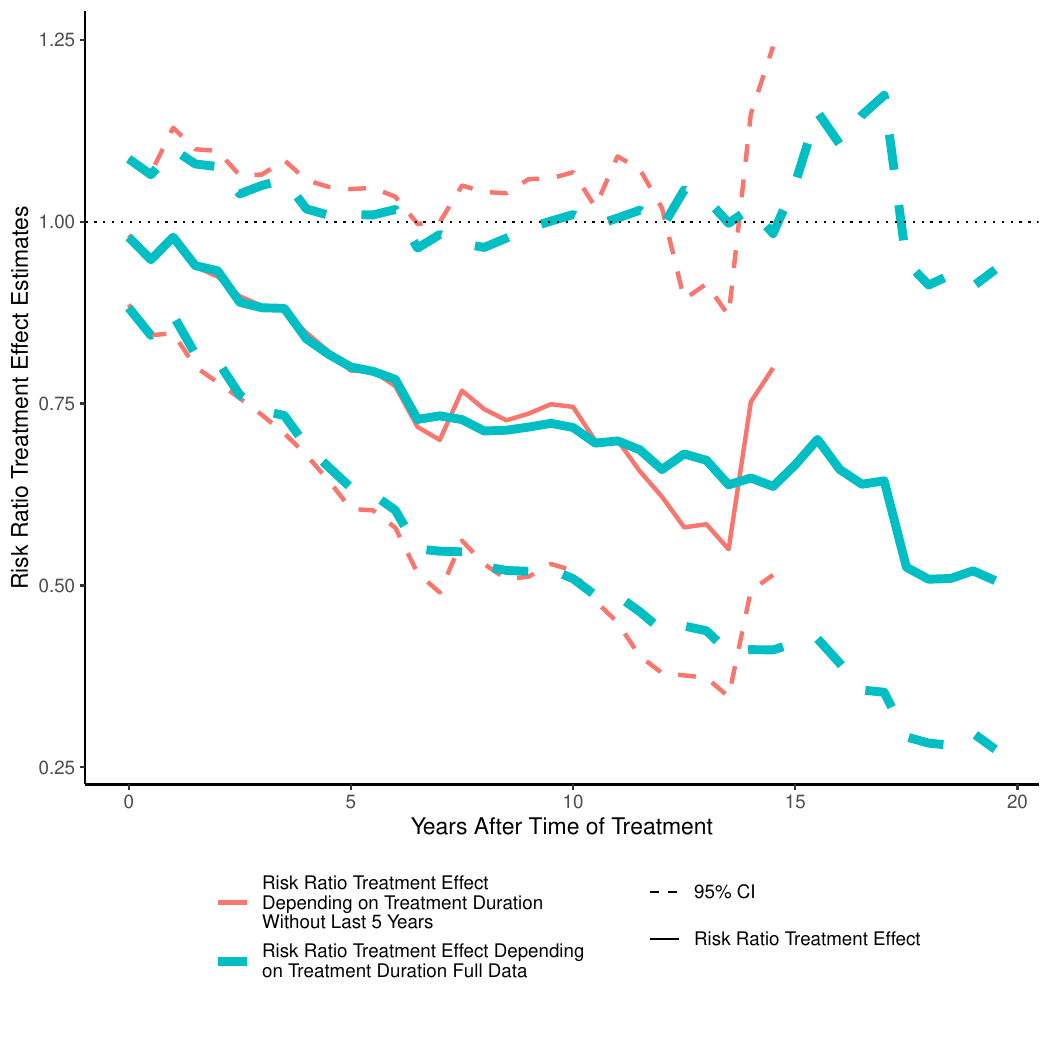}
    \caption{Sensitivity Analysis: Estimated risk ratios with 95\% confidence intervals for each time interval after the first drug-induced homicide (DIH) prosecution reported by the media in the state for the main analysis (red, thin lines) and sensitivity analysis (blue, thick lines), \textit{excluding the last five years}. Generalized additive model (GAM) with linear link function, treatment effect depending on the treatment duration.}
    \label{fig:pre_tx_trend_linear_sensitivity_v_main}
\end{figure}

Table \ref{appendix_tab:linear_link_models_RR_wo_last_five_years} shows the estimated risk ratios for the GAMs with a linear link function. 
Assuming a constant treatment effect, results when excluding the last five years are similar to the results from the main analysis. 
In general, the magnitudes of the risk ratios in this sensitivity analysis are smaller than the magnitudes of the risk ratios from the main analysis, and almost all of the risk ratios are not statistically significant. 
The naloxone access law where pharmacists are able to dispense without a prescription and the Prescription Drug Monitoring Program have statistically significant, protective effects.
The risk ratio for the DIH prosecution reported by the media is 0.956 (95\% CI: (0.834, 1.095)) (compared to the main analysis: 0.977 (95\% CI: (0.866, 1.101)). 
The risk ratios for the different relevant policy measures when excluding the last five years under a model where the treatment effect may depend on the treatment duration are similar to the risk ratios under the model where treatment effect is constant. 
Risk ratio estimates of DIH prosecutions reported by the media depending on the treatment duration when excluding the last five years are largely similar to risk ratio estimates when the last five years are not excluded (Table \ref{appendix_tab:event_study_models_coef_wo_last_five_years}, left). 
Figure \ref{fig:pre_tx_trend_linear_sensitivity_v_main} shows the estimated treatment effects for DIH prosecutions reported by the media depending on the treatment duration for the main analysis and when excluding the last five years. 
Similar to the main analysis, most of the risk ratios are not statistically significant.

\begin{table}[!htb]
	\caption{Estimated risk ratios and 95\% confidence intervals for relevant policy measures and drug-induced homicide (DIH) prosecutions reported by the media for generalized additive model (GAM) with linear link function when assuming a constant treatment effect and when assuming the treatment effect depends on the treatment duration, \emph{excluding the last five years}.} 
 \centering
\begin{tabular}{|lcc|}
			\hline
			 & \thead{Linear GAM with \\
			 Constant Treatment Effect \\ (95\% Confidence Interval)} & 
			 \thead{Linear GAM with Treatment Effect \\ 
    Depending
    on the Treatment Duration\\ (95\% Confidence Interval)} \\
			\hline
			Naloxone Access Law: pharmacists can dispense without prescription & 0.785 (0.638, 0.966) & 0.802 (0.643, 0.999)\\
            Naloxone Access Law: pharmacists cannot dispense without prescription & 0.951 (0.797, 1.135) & 0.959 (0.801, 1.150)\\
            Medical Marijuana Law & 1.244 (0.840, 1.843) & 1.236 (0.833, 1.835)\\
            Recreational Marijuana Law & 0.875 (0.624, 1.227) & 0.856 (0.612, 1.197)\\
            911 Good Samaritan Law & 1.023 (0.826, 1.267) & 1.042 (0.828, 1.312)\\
            Prescription Drug Monitoring Program & 0.841 (0.709, 0.997) & 0.822 (0.693, 0.976)\\
            Medicaid expansion & 0.999 (0.859, 1.161) & 1.003 (0.867, 1.159)\\
            DIH prosecutions reported by media & 0.956 (0.834, 1.095) & See Table \ref{appendix_tab:event_study_models_coef_wo_last_five_years}, Left\\
			\hline
		\end{tabular}	
  \label{appendix_tab:linear_link_models_RR_wo_last_five_years}
\end{table}

\begin{table}[!htb]
	\caption{Estimated risk ratios and 95\% confidence intervals for each six-month time interval after the first drug-induced homicide (DIH) prosecution was reported by the media in the state for generalized additive models (GAMs) with linear and logistic link function when assuming the treatment effect depends on the treatment duration, \textit{excluding the last five years}.} 
 \centering
\begin{tabular}{|ccc|}
			\hline
			 \thead{Number of six-month time intervals after first \\DIH prosecution reported by the media} & \thead{Risk Ratio: Linear GAM  \\ (95\% Confidence Interval)} & 
			 \thead{Risk Ratio: Logistic GAM \\ (95\% Confidence Interval)} \\
			\hline
			0 & 0.981 (0.886, 1.087) & 0.988 (0.923, 1.059)\\
                1  & 0.949 (0.844, 1.067) & 1.016 (0.929, 1.110)\\
                2 & 0.978 (0.847, 1.129) & 0.998 (0.905, 1.101)\\
                3 & 0.938 (0.800, 1.100) & 0.945 (0.854, 1.045)\\
                4 & 0.925 (0.779, 1.097) & 0.929 (0.834, 1.035)\\
                5 & 0.897 (0.757, 1.064) & 0.907 (0.812, 1.014)\\
                6 & 0.884 (0.734, 1.065) & 0.893 (0.786, 1.014)\\
                7 & 0.877 (0.709, 1.085) & 0.901 (0.778, 1.043)\\
                8 & 0.847 (0.679, 1.057) & 0.870 (0.751, 1.006)\\
                9 & 0.822 (0.644, 1.048) & 0.842 (0.716, 0.991)\\
                10 & 0.795 (0.605, 1.045) & 0.817 (0.681, 0.980)\\
                11 & 0.794 (0.603, 1.046) & 0.804 (0.671, 0.963)\\
                12 & 0.774 (0.579, 1.034) & 0.801 (0.657, 0.977)\\
                13 & 0.718 (0.517, 0.997) & 0.758 (0.610, 0.942)\\
                14 & 0.700 (0.490, 0.999) & 0.723 (0.568, 0.921)\\
                15 & 0.768 (0.562, 1.050) & 0.777 (0.622, 0.971)\\
                16 & 0.742 (0.529, 1.041) & 0.761 (0.601, 0.963)\\
                17 & 0.727 (0.509, 1.039) & 0.737 (0.573, 0.948)\\
                18 & 0.736 (0.512, 1.059) & 0.723 (0.557, 0.939)\\
                19 & 0.749 (0.530, 1.060) & 0.736 (0.579, 0.937)\\
                20 & 0.745 (0.520, 1.068) & 0.726 (0.569, 0.928)\\
                21 & 0.699 (0.479, 1.020) & 0.686 (0.534, 0.881)\\
                22 & 0.700 (0.449, 1.090) & 0.654 (0.494, 0.867)\\
                23 & 0.657 (0.403, 1.072) & 0.650 (0.480, 0.880)\\
                24 & 0.622 (0.380, 1.019) & 0.624 (0.460, 0.847)\\
                25 & 0.580 (0.376, 0.893) & 0.615 (0.460, 0.824)\\
                26 & 0.584 (0.373, 0.915) & 0.602 (0.441, 0.823)\\
                27 & 0.550 (0.347, 0.871) & 0.570 (0.413, 0.788)\\
                28 & 0.752 (0.493, 1.148) & 0.711 (0.524, 0.964)\\
                29 & 0.799 (0.514, 1.241) & 0.731 (0.533, 1.002)\\
			\hline
		\end{tabular}	
  \label{appendix_tab:event_study_models_coef_wo_last_five_years}
\end{table}

\begin{figure}[!htb]
    \centering
    \includegraphics{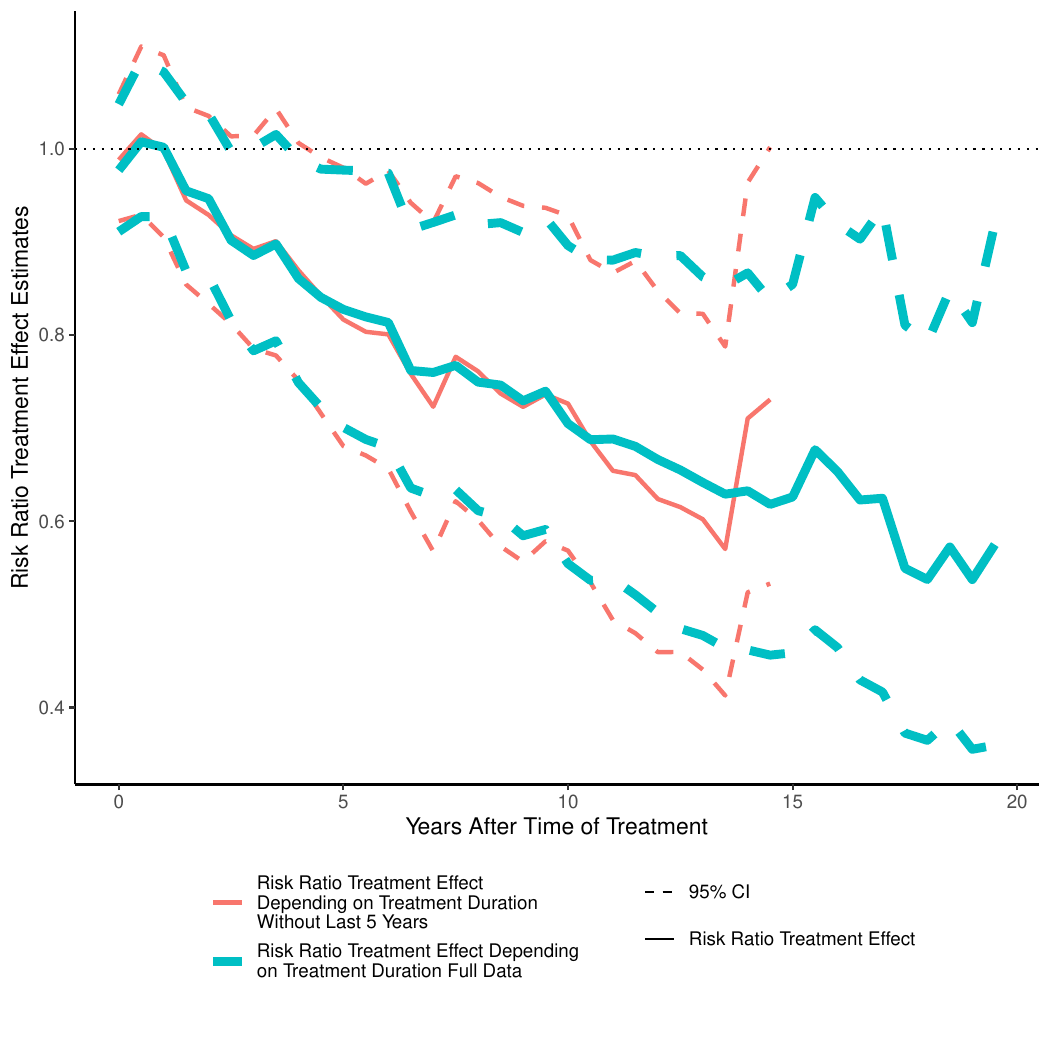}
    \caption{Sensitivity Analysis: Estimated risk ratios with 95\% confidence intervals for each time interval after the first drug-induced homicide (DIH) prosecution reported by the media in the state for the main analysis (red, thin lines) and sensitivity analysis (blue, thick lines), \textit{excluding the last five years}. Generalized additive model (GAM) with logistic link function, treatment effect depending on the treatment duration.}
    \label{fig:pre_tx_trend_logistic_sensitivity_v_main}
\end{figure}

Table \ref{appendix_tab:logistic_models_results_wo_last_five_years} shows the estimated risk ratios for the GAMs with a logistic link function. 
Under the constant treatment effect assumption, the estimated risk ratios excluding the last five years are smaller, but do not differ much from the estimated risk ratios from the main analysis. 
However, the 911 Good Samaritan Law is now associated with a protective, but not statistically significant effect. 
When we assume that the treatment effect of DIH prosecutions reported by the media depends on the treatment duration, almost all risk ratios for the DIH prosecutions reported by the media are below one (Table \ref{appendix_tab:event_study_models_coef_wo_last_five_years}, right), suggesting a protective effect. 
The overall conclusions for the treatment effect of DIH prosecutions reported by the media at each treatment duration remain largely similar to the conclusions from the main analysis.
Figure \ref{fig:pre_tx_trend_logistic_sensitivity_v_main} shows the estimated treatment effects for DIH prosecutions reported by the media depending on the treatment duration for the main analysis and when excluding the last five years under a logistic link function.
Under both the linear and logistic link functions, the constant treatment effect for DIH prosecutions reported by the media is still higher than almost all of the treatment effects for DIH prosecutions reported by the media when assuming treatment effects depend on the treatment duration.

\begin{table}[!htb]
	\caption{Estimated risk ratios and 95\% confidence intervals for the different relevant policy measures and drug-induced homicide (DIH) prosecutions reported by the media for generalized additive model (GAM) with logistic link function when assuming a constant treatment effect (left) and when assuming the treatment effect depends on the treatment duration (right), \textit{excluding the last five years}.} 
 \centering
\begin{tabular}{|lcc|}
			\hline
			 & \thead{Logistic GAM with \\
			 Constant Treatment Effect \\ (95\% Confidence Interval)} & 
			 \thead{Logistic GAM with Treatment Effect \\ 
    Depending
    on the Treatment Duration\\ (95\% Confidence Interval)} \\
			\hline
			Naloxone Access Law: pharmacists can dispense without prescription & 0.921 (0.803, 1.057) & 0.912 (0.800, 1.041)\\
            Naloxone Access Law: pharmacists cannot dispense without prescription & 1.028 (0.938, 1.126) & 1.025 (0.936, 1.123)\\
            Medical Marijuana Law & 1.069 (0.874, 1.306) & 1.029 (0.842, 1.256)\\
            Recreational Marijuana Law & 0.884 (0.700, 1.115) & 0.846 (0.677, 1.058)\\
            911 Good Samaritan Law & 0.921 (0.832, 1.019) & 0.954 (0.854, 1.065)\\
            Prescription Drug Monitoring Program & 0.952 (0.844, 1.074) & 0.925 (0.837, 1.021)\\
            Medicaid expansion & 1.020 (0.916, 1.136) & 1.035 (0.931, 1.151)\\
            DIH prosecutions reported by media & 1.069 (0.978, 1.169) & See Table \ref{appendix_tab:event_study_models_coef_wo_last_five_years}, Right\\
			\hline
		\end{tabular}	
  \label{appendix_tab:logistic_models_results_wo_last_five_years}
\end{table}

\subsection{Including the Number of States with at Least One Drug-Induced Homicide Prosecution as a Predictor}\label{sensitivity_anlys:num_tx_states}
\begin{figure}[!htb]
    \centering
    \includegraphics{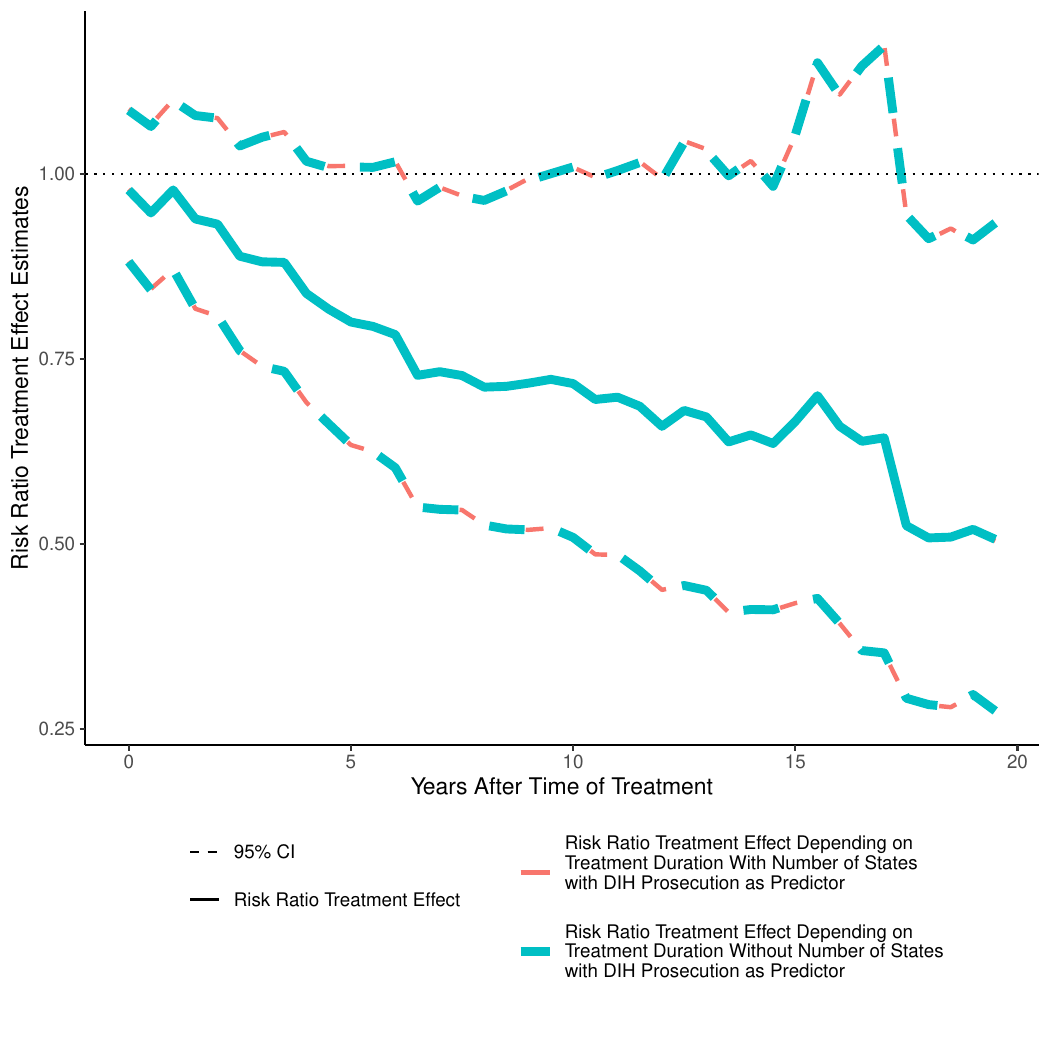}
    \caption{Sensitivity Analysis: Estimated risk ratios with 95\% confidence intervals for each time interval after the first drug-induced homicide (DIH) prosecution reported by the media in the state for the main analysis (red, thin lines) and sensitivity analysis (blue, thick lines), with the number of states at time interval $t$ with at least one DIH prosecution reported by the media included as a predictor. Generalized additive model (GAM) with linear link function, the treatment effect depending on the treatment duration.}
    \label{fig:linear_sensitivity_v_main_w_num_states_intervention}
\end{figure}

\begin{figure}[!htb]
    \centering
    \includegraphics{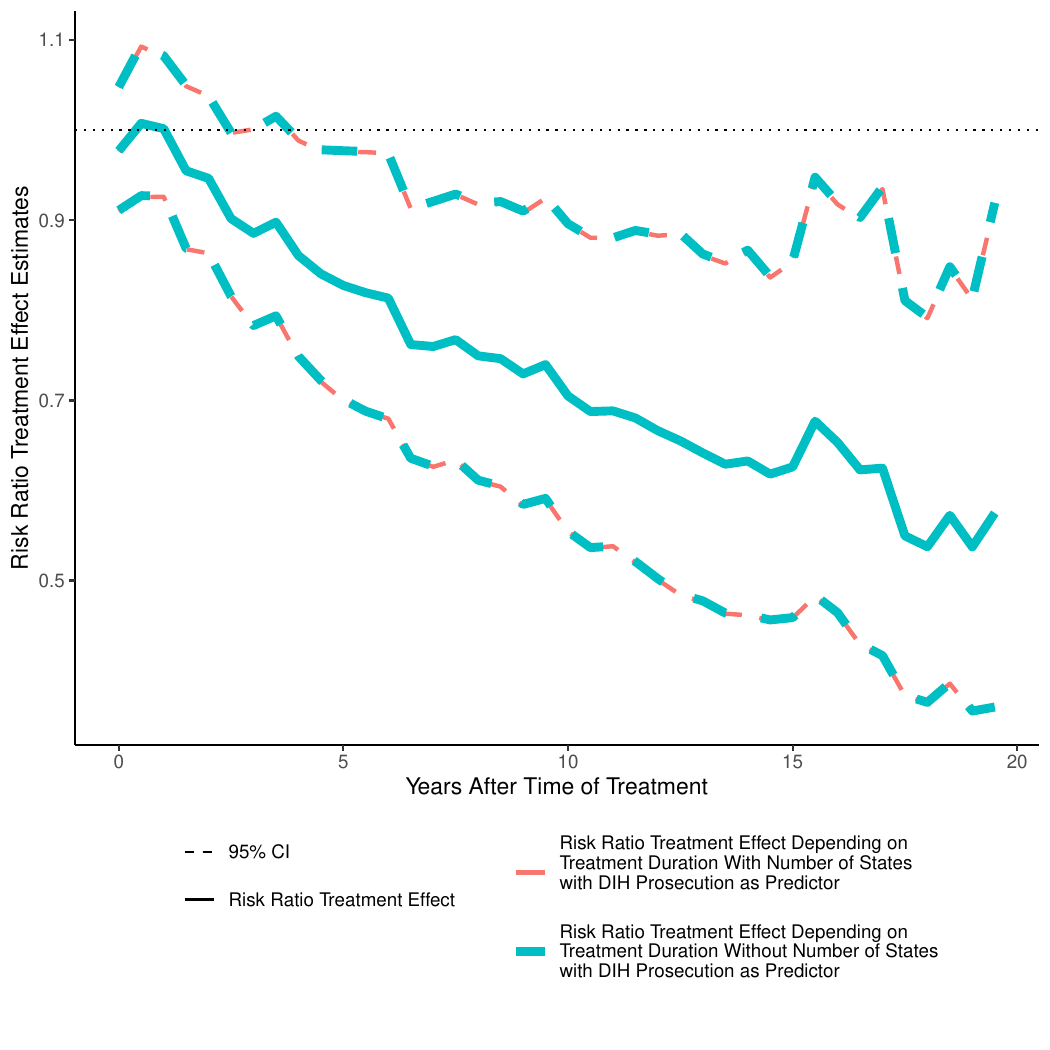}
    \caption{Sensitivity Analysis: Estimated risk ratios with 95\% confidence intervals for each time interval after the first drug-induced homicide (DIH) prosecution reported by the media in the state for the main analysis (red, thin lines) and sensitivity analysis (blue, thick lines), with the number of states at time interval $t$ with at least one DIH prosecution reported by the media included as a predictor. Generalized additive model (GAM) with logistic link function, treatment effect depending on the treatment duration.}
    \label{fig:logistic_sensitivity_v_main_w_num_states_intervention}
\end{figure}

We also conduct a sensitivity analysis where we include the number of states with at least one DIH prosecution reported by the media by the beginning of time interval $t$ as a variable to measure additional time effects to avoid treatment effects picking up time effects.
Results show that including the number of states with at least one DIH prosecution at time $t$ did not make a difference in estimating the treatment effects.
The estimated risk ratios using a GAM with a linear link function (Table \ref{appendix_tab:log_models_RR_w_num_states_w_intervention}) and using a GAM with a logistic link function (Table \ref{appendix_tab:logistic_models_results_w_num_states_intervention}) are very similar to the risk ratios of the corresponding main analyses. 
We can further see the similarity in results between this sensitivity analysis and the main analysis in Figures \ref{fig:linear_sensitivity_v_main_w_num_states_intervention} and \ref{fig:logistic_sensitivity_v_main_w_num_states_intervention} for the GAMs where treatment effect depends on the treatment duration (risk ratios in Table \ref{appendix_tab:event_study_models_coef_w_num_states_w_intervention}) using a linear link function and logistic link function, respectively.

\begin{table}[!htb]
	\caption{Estimated risk ratios and 95\% confidence intervals for the different relevant policy measures and drug-induced homicide (DIH) prosecutions reported by the media for generalized additive model (GAM) with linear link function when assuming a constant treatment effect (left) and when assuming the treatment effect depends on the treatment duration (right), with the number of states at time interval $t$ with at least one DIH prosecution reported by the media included as a predictor.} 
 \centering
\begin{tabular}{|lcc|}
			\hline
			 & \thead{Linear GAM with \\
			 Constant Treatment Effect \\ (95\% Confidence Interval)} & 
			 \thead{Linear GAM with Treatment Effect \\ 
    Depending
    on the Treatment Duration\\ (95\% Confidence Interval)} \\
			\hline
			Naloxone Access Law: pharmacists can dispense without prescription & 0.915 (0.805, 1.039) & 0.934 (0.831, 1.049)\\
            Naloxone Access Law: pharmacists cannot dispense without prescription & 0.996 (0.877, 1.131) & 1.002 (0.890, 1.128)\\
            Medical Marijuana Law & 1.204 (0.970, 1.493) & 1.225 (0.973, 1.542)\\
            Recreational Marijuana Law & 0.893 (0.754, 1.056) & 0.909 (0.762, 1.084)\\
            911 Good Samaritan Law & 1.059 (0.941, 1.193) & 1.067 (0.947, 1.203)\\
            Prescription Drug Monitoring Program & 0.859 (0.713, 1.035) & 0.841 (0.699, 1.012)\\
            Medicaid expansion & 1.095 (0.944, 1.271) & 1.086 (0.933, 1.266)\\
            DIH prosecutions reported by media & 0.977 (0.866, 1.103) & See Table \ref{appendix_tab:event_study_models_coef_w_num_states_w_intervention}, Left\\
            Number of states with at least one DIH prosecution & 0.996 (0.985, 1.007) & 0.996 (0.985, 1.007) \\
			\hline
		\end{tabular}	
  \label{appendix_tab:log_models_RR_w_num_states_w_intervention}
\end{table}

\begin{table}[!htb]
	\caption{Estimated risk ratios and 95\% confidence intervals for the different relevant policy measures and drug-induced homicide (DIH) prosecutions reported by the media for generalized additive model (GAM) with logistic link function when assuming a constant treatment effect (left) and when assuming the treatment effect depends on the treatment duration (right), with the number of states at time interval $t$ with at least one DIH prosecution reported by the media included as a predictor.} 
 \centering
\begin{tabular}{|lcc|}
			\hline
			 & \thead{Logistic GAM with \\
			 Constant Treatment Effect \\ (95\% Confidence Interval)} & 
			 \thead{Logistic GAM with Treatment Effect \\ 
    Depending
    on the Treatment Duration\\ (95\% Confidence Interval)} \\
			\hline
			Naloxone Access Law: pharmacists can dispense without prescription & 0.974 (0.887, 1.070) & 0.980 (0.904, 1.062)\\
            Naloxone Access Law: pharmacists cannot dispense without prescription & 1.007 (0.913, 1.111) & 1.017 (0.927, 1.116)\\
            Medical Marijuana Law & 1.057 (0.945, 1.183) & 1.046 (0.930, 1.178)\\
            Recreational Marijuana Law & 0.962 (0.847, 1.093) & 0.964 (0.841, 1.105)\\
            911 Good Samaritan Law & 1.035 (0.952, 1.125) & 1.054 (0.972, 1.144)\\
            Prescription Drug Monitoring Program & 0.981 (0.857, 1.123) & 0.958 (0.851, 1.077)\\
            Medicaid expansion & 1.104 (0.979, 1.245) & 1.103 (0.987, 1.231)\\
            DIH prosecutions reported by media & 1.064 (0.972, 1.165) & See Table \ref{appendix_tab:event_study_models_coef_w_num_states_w_intervention}, Right\\
            Number of states with at least one DIH prosecution & 1.004 (0.995, 1.013) & 1.004 (0.995, 1.013) \\
			\hline
		\end{tabular}	
  \label{appendix_tab:logistic_models_results_w_num_states_intervention}
\end{table}

\begin{table}[!htb]
	\caption{Estimated risk ratios and 95\% confidence intervals for each six-month time interval after the first drug-induced homicide (DIH) prosecution was reported by the media in the state for generalized additive model (GAM) with linear and logistic link function when assuming the treatment effect depends on the treatment duration, with the number of states at time interval $t$ with at least one DIH prosecution reported by the media included as a predictor.} 
 \centering
\begin{tabular}{|ccc|}
			\hline
			 \thead{Number of six-month time intervals after first \\DIH prosecution reported by the media} & \thead{Risk Ratio: Linear GAM \\ (95\% Confidence Interval)} & 
			 \thead{Risk Ratio: Logistic GAM \\ (95\% Confidence Interval)} \\
			\hline
			0 & 0.980 (0.882, 1.088) & 0.976 (0.910, 1.048)\\
                1  & 0.950 (0.845, 1.067) & 1.006 (0.926, 1.093)\\
                2 & 0.979 (0.872, 1.101) & 1.001 (0.926, 1.082)\\
                3 & 0.940 (0.818, 1.081) & 0.954 (0.868, 1.049)\\
                4 & 0.933 (0.808, 1.076) & 0.947 (0.863, 1.038)\\
                5 & 0.889 (0.761, 1.038) & 0.902 (0.816, 0.997)\\
                6 & 0.882 (0.740, 1.050) & 0.885 (0.783, 1.001)\\
                7 & 0.881 (0.733, 1.057) & 0.898 (0.794, 1.016)\\
                8 & 0.840 (0.691, 1.020) & 0.860 (0.748, 0.989)\\
                9 & 0.819 (0.663, 1.011) & 0.839 (0.721, 0.977)\\
                10 & 0.801 (0.634, 1.011) & 0.827 (0.700, 0.976)\\
                11 & 0.795 (0.625, 1.011) & 0.819 (0.687, 0.976)\\
                12 & 0.783 (0.603, 1.017) & 0.814 (0.680, 0.974)\\
                13 & 0.728 (0.550, 0.964) & 0.762 (0.635, 0.913)\\
                14 & 0.733 (0.547, 0.982) & 0.759 (0.626, 0.920)\\
                15 & 0.728 (0.546, 0.971) & 0.767 (0.634, 0.929)\\
                16 & 0.713 (0.526, 0.966) & 0.749 (0.611, 0.918)\\
                17 & 0.714 (0.521, 0.978) & 0.745 (0.604, 0.920)\\
                18 & 0.719 (0.519, 0.994) & 0.728 (0.583, 0.908)\\
                19 & 0.723 (0.522, 1.002) & 0.739 (0.590, 0.925)\\
                20 & 0.717 (0.509, 1.009) & 0.705 (0.554, 0.896)\\
                21 & 0.696 (0.486, 0.996) & 0.687 (0.536, 0.881)\\
                22 & 0.698 (0.485, 1.005) & 0.688 (0.538, 0.881)\\
                23 & 0.687 (0.464, 1.017) & 0.680 (0.520, 0.888)\\
                24 & 0.660 (0.438, 0.994) & 0.665 (0.501, 0.883)\\
                25 & 0.681 (0.444, 1.045) & 0.654 (0.484, 0.885)\\
                26 & 0.673 (0.438, 1.034) & 0.640 (0.476, 0.861)\\
                27 & 0.638 (0.408, 0.998) & 0.628 (0.463, 0.852)\\
                28 & 0.647 (0.411, 1.018) & 0.632 (0.461, 0.866)\\
                29 & 0.636 (0.411, 0.984) & 0.618 (0.456, 0.837)\\
                30 & 0.665 (0.420, 1.052) & 0.626 (0.458, 0.854)\\
                31 & 0.700 (0.426, 1.151) & 0.676 (0.483, 0.946)\\
                32 & 0.660 (0.393, 1.108) & 0.652 (0.463, 0.918)\\
                33 & 0.639 (0.356, 1.148) & 0.622 (0.429, 0.902)\\
                34 & 0.644 (0.352, 1.175) & 0.624 (0.416, 0.935)\\
                35 & 0.525 (0.291, 0.945) & 0.549 (0.372, 0.810)\\
                36 & 0.508 (0.283, 0.912) & 0.537 (0.364, 0.791)\\
                37 & 0.509 (0.280, 0.927) & 0.572 (0.386, 0.847)\\
                38 & 0.519 (0.296, 0.909) & 0.537 (0.355, 0.812)\\
                39 & 0.505 (0.274, 0.930) & 0.574 (0.359, 0.917)\\
			\hline
		\end{tabular}	
  \label{appendix_tab:event_study_models_coef_w_num_states_w_intervention}
\end{table}

\clearpage
\section{Estimation of Attributable Deaths} \label{sec:attributable_deaths_linear_link}
We also estimate the number of excess drug-overdose deaths attributable to DIH prosecutions reported by the media. 
To estimate the number of deaths in a state that are attributable to DIH prosecutions reported by the media, we derive an estimator that is similar to the etiologic fraction. \cite{miettinen1974proportion}
Let $n_{s, t, \text{attributable deaths}}$ be the number of deaths attributable to DIH prosecutions reported by the media in state $s$ for time intervals $t$ where there had been at least one DIH prosecution reported by the media. 
The number of observed deaths in state $s$ at time interval $t$ is denoted by $n_{s, t, \text{observed overdose deaths}}$, and $n_{s, t, \text{overdose deaths}}^{(a_{st} = 0)}$ denotes the number of unintentional drug-overdose deaths that would have occurred had the intervention never occurred in state $s$.
The number of drug-overdose deaths attributable to DIH prosecutions reported by the media is estimated by computing the difference between the observed number of drug-overdose deaths and the estimated number of deaths had there not been any DIH prosecutions reported by the media:
\begin{align}\label{num_attributable_deaths}
    \hat{n}_{\text{attributable deaths}, s, t} = n_{\text{observed overdose deaths}, s, t} - \hat{n}_{\text{overdose deaths}, s, t}^{(a_{st} = 0)},
\end{align}
where 
\begin{align}\label{eq:num_attr_deaths_no_tx}
    \hat{n}_{\text{overdose deaths}, s,t}^{(a_{st} = 0)} = n_{\text{population}, s, t} \times \hat{p}_{s,t}^{(a_{st} = 0)},
\end{align}
where $n_{s, t, \text{population}}$ is the population in state $s$ at time $t$ and $\hat{p}_{s, t}^{(a_{s, t} = 0)}$ is the estimated probability of unintentional drug-overdose deaths had the intervention not occurred.
That is, we estimate the number of drug-overdose deaths had there not been any DIH prosecutions reported by the media by multiplying the population in state $s$ at time interval $t$ by the estimated probability of drug-overdose deaths had there not been any DIH prosecutions reported by the media for state $s$ at time interval $t$, denoted as $\hat{p}_{s, t}^{(a_{st} = 0)}$.
We estimate this probability of drug-overdose deaths had there not been any DIH prosecutions reported by the media in two ways, depending on whether the treatment effect is constant (Section \ref{attributable_deaths_constant_tx}) or depends on the treatment duration (Section \ref{attributable_deaths_tx_dep_on_duration}).

\subsection{Constant Treatment Effect Model}\label{attributable_deaths_constant_tx}
First, assume that the treatment effect is constant.
We derive the estimates for the number of attributable deaths for the constant treatment effect model with a linear link function here, but these results can be easily extended to the model with a logistic link function.
The estimated number of attributable deaths in a state $s$ is given by Equations~\eqref{num_attributable_deaths} and \eqref{eq:num_attr_deaths_no_tx}:
\begin{align*}
    n_{\text{attributable deaths}, s, t} = n_{\text{observed overdose deaths}, s, t} - n_{\text{population}, s, t} \times \hat{p}_{s, t}^{(a_{s, t} = 0)}.
\end{align*} 
For states $s$ and time intervals $t$, our model (Equation~\eqref{eq:constant_tx_model_eq_lin_link_func}) implies that
\begin{align*}
    \E\left(\log Y_{st} \mid \Vec{Z}_{st}\right) &= \alpha_s + \gamma_{r(s)}(t) + A_{st} \beta + X_{st} \delta,
\end{align*}
where $Y_{st}$ is the risk of unintentional drug-overdose deaths in state $s$ at time interval $t$.
If no intervention occurred in state $s$ at time interval $t$, we have $A_{st} = 0$, and so 
\begin{align}
     \log p_{st}^{(a_{st} = 0)} = \alpha_s + \gamma_{r(s)}(t)  + X_{st} \delta.
\end{align}
With $\phi_{st} = \alpha_s + \gamma_{r(s)}(t) +  X_{st} \delta$, the log probability of an overdose death had the intervention not occurred is: 
\begin{align}\label{no_int_prob_constant_tx}
    \log p_{s, t}^{(a_{st} = 0)} = \phi_{st}.
\end{align} 
We can estimate $\log p_{s, t}^{(a_{st} = 0)}$ using $\hat{\beta}$ and $\hat{\phi}_{st}$, which are estimates for $\beta$ and $\phi_{st}$, respectively: note that we can also estimate $\log p_{st}$ by finding the log proportion of people who died from a drug overdose, i.e. $\log Y_{st}$.
Then, we have the following:
\begin{align}\label{attributable_death_calc_constant_tx}
    A_{st}\hat{\beta} + \hat{\phi}_{st} &= \log \hat{p}_{st} \approx \log Y_{st} \nonumber \\
    \Rightarrow \hat{\phi}_{st} &\approx  \log Y_{st} - A_{st} \hat{\beta} \nonumber\\
    \Rightarrow \hat{p}_{st}^{(a_{st} = 0)} = exp(\hat{\phi}_{st}) &\approx exp\left(\log Y_{st} - A_{st}\hat{\beta}\right) = Y_{st}exp\left(-A_{st} \hat{\beta}\right),
\end{align}
where in the last line we use Equation~\eqref{no_int_prob_constant_tx}.
Equation~\eqref{attributable_death_calc_constant_tx} provides an estimate for the probability of a drug-overdose death in state $s$, had the intervention not occurred. Combining Equations~\eqref{num_attributable_deaths}, \eqref{eq:num_attr_deaths_no_tx}, and \eqref{attributable_death_calc_constant_tx}:
\begin{align}\label{num_attributable_deaths_constant_tx}
    \hat{n}_{\text{attributable deaths}, s, t} &= n_{\text{observed overdose deaths}, s, t} - n_{\text{population}, s, t} Y_{st} exp\left(- A_{st}\hat{\beta}\right) \nonumber \\
    &= n_{\text{observed overdose deaths}, s, t} 
    \left( 1 - exp\left(- A_{st}\hat{\beta}\right)\right).
\end{align}
Similarly, in the case of a logistic link function, the number of attributable deaths is equal to:
\begin{align*}\label{num_attributable_deaths_constant_tx_logisitc}
    \hat{n}_{\text{attributable deaths}, s, t} &= n_{\text{observed overdose deaths}, s, t} - n_{\text{pop}, s, t} expit\left(logit(p_{s,t}) - A_{st}\hat{\beta}\right).
\end{align*}
We substitute the 95\% confidence interval limits of $\beta$ to find the 95\% confidence interval limits of the number of attributable deaths since the function for the number of attributable deaths is a strictly monotonic function.

\subsection{Treatment Effect That Depends on Exposure Duration}\label{attributable_deaths_tx_dep_on_duration}
We assume that the treatment effect depends on the exposure duration in this section. 
The estimated number of attributable deaths differ under the model with a linear link function (Section \ref{appendix:attributable_deaths_event_study}) and the model with a logistic link function (Section \ref{appendix:attributable_deaths_event_study_logistic}).

\subsubsection{Model with Linear Link Function}\label{appendix:attributable_deaths_event_study}
We first focus on the model with a linear link function.
For states $s$ and time intervals $t$, we assume from Equation~\eqref{eq:ols_model_dih} that:
\begin{align*}
    \E \left(\log Y_{st} \mid \Vec{Z}_{st}\right) &= \alpha_s + \gamma_{r(s)}(t) + \sum_{k=0}^{39} \mathbb{I}\{K_{st} = k\} \beta_k + X_{st} \delta.
\end{align*}
If no intervention occurred in state $s$ by time interval $t$, we have $A_{st} = 0$ and $\mathbb{I}\{K_{st} = k\} = 0$ for $k = 0, \dotsc, 39$, and so the log probability of unintentional drug-overdose deaths had no intervention occurred is
\begin{align}
      \log p_{st}^{(a_{st} = 0)} = \alpha_s + \gamma_{r(s)}(t) + X_{st}\delta = \phi_{st},
\end{align}
where $p_{st}^{(a_{st} = 0)}$ denotes the probability of unintentional drug-overdose deaths had no intervention occurred.
We estimate the probability of unintentional drug-overdose deaths had no intervention occurred in a similar manner as in Section \ref{attributable_deaths_constant_tx} using $\hat{\phi}_{st}$:
\begin{align}
    \sum_{k=0}^{39}\mathbb{I}\{K_{st} = k\}\hat{\beta}_k + \hat{\phi}_{st} &\approx \log Y_{st} \nonumber \\
    \Rightarrow \hat{\phi}_{st} &\approx  \log Y_{st} - \sum_{k=0}^{39}\mathbb{I}\{K_{st} = k\}\hat{\beta}_k  \nonumber\\
    \Rightarrow \hat{p}_{st}^{(a_{st} = 0)} = exp(\hat{\phi}_{st}) &\approx Y_{st} exp\left(- \sum_{k=0}^{39}\mathbb{I}\{K_{st} = k\}\hat{\beta}_k \right).
\end{align}
We then estimate the number of attributable deaths using $\hat{p}_{st}^{(a_{st} = 0)}$:
\begin{align}
    \hat{n}_{\text{attributable deaths}, s, t} &= n_{\text{observed overdose deaths}, s, t} - n_{\text{population}, s, t} Y_{st} exp\left(- \sum_{k=0}^{39}\mathbb{I}\{K_{st} = k\}\hat{\beta}_k\right) \nonumber \\
    &= n_{\text{observed overdose deaths}, s, t} \left(1 - exp\left(- \sum_{k=0}^{39}\mathbb{I}\{K_{st} = k\}\hat{\beta}_k\right)\right).
\end{align}

Since the number of attributable deaths is estimated using all $\hat{\beta}_k$, we need to account for the covariance between the coefficients when computing the lower and upper bounds of the 95\% confidence interval of the number of attributable deaths.
We first denote $\beta = \begin{pmatrix} \beta_0 & \beta_1 & \dotsc & \beta_{39}\end{pmatrix}^T$ and denote $\hat{\beta}$ as the estimate of $\beta$.
We also use $\Sigma_{{\beta}}^{as}$ to denote the asymptotic variance-covariance matrix of the coefficients.
From the theory of unbiased estimating equations, we have
\begin{align*}
    \sqrt{n_s}\left(\hat{\beta} - \beta\right) \overset{\mathcal{D}} \to \mathcal{N}(0, \Sigma_{{\beta}}^{as}),
\end{align*}
where $n_s$ is the number of states.
Since there are multiple treatment parameters, we use the Delta Method to derive the 95\% confidence interval of the number of attributable deaths:
\begin{align*}
    \sqrt{n_s}\left(m(\hat{\beta},t) - m(\beta,t)\right) \overset{\mathcal{D}} \to \mathcal{N}(0, J_{{\beta}, t} \Sigma_{{\beta}}^{as} J_{{\beta}, t}^T),
\end{align*}
where for a time interval $t$,
\begin{align*}
    m({\beta},t) &= \sum_{s} {n}_{\text{attributable deaths}, s, t} \\
    &= \sum_{s} {n}_{\text{population}, s, t} \left({p}_{st} - {p}_{st} exp\left(- \sum_{k=0}^{39}\mathbb{I}\{K_{st} = k\}{\beta}_k \right)\right),
\end{align*} 
where we estimate $p_{st}$ using $\hat{p}_{st}^{\text{obs}} = Y_{st}$, the observed risk of unintentional drug-overdose deaths, and $J_{{\beta}, t}$ is the Jacobian matrix. 
The Jacobian matrix is given by:
\begin{align*}
    J_{{\beta}, t} &= \begin{pmatrix} \frac{\partial m({\beta},t)}{\partial {\beta}_0} & \frac{\partial m({\beta},t)}{\partial {\beta}_1} 
    & \dotsc 
    & \frac{\partial m({\beta},t)}{\partial {\beta}_{39}} \end{pmatrix},
\end{align*}
where the $k$th element of the Jacobian matrix for $k \in \{0, \dotsc, 39\}$ is equal to
\begin{align*}
    \frac{\partial m({\beta},t)}{\partial {\beta}_k}
    &= \sum_{s} {n}_{\text{population}, s, t}  {p}_{st} exp\left(- \sum_{k=0}^{39}\mathbb{I}\{K_{st} = k\}{\beta}_k \right) \mathbb{I}\{K_{st} = k\}.
\end{align*}
Hence, the lower and upper limits of the 95\% confidence interval of the attributable deaths in at time interval $t$ are given by:
\begin{align*}
    \sum_{s} \hat{n}_{\text{attributable deaths}, s, t} \pm 1.96 \sqrt{J_{{\beta}, t} \hat{\Sigma}_{{\beta}} J_{{\beta}, t}^T}, 
\end{align*}
where $\hat{\Sigma}_{{\beta}}$ is obtained from the sandwich estimator of the variance of the parameters as described in Section \ref{appendix:sandwich_est_event_study}. 

\subsubsection{Model with Logistic Link Function}\label{appendix:attributable_deaths_event_study_logistic}
In the case of using the logistic link function, for states $s$ and time intervals $t$, we assume from Equation~\eqref{eq:logistic_event_study} that:
\begin{align*}
    logit \left(\E \left(Y_{st} \mid \Vec{Z}_{st} \right)\right) &= \alpha_s + \gamma_{r(s)}(t) + \sum_{k=0}^{39} \mathbb{I}\{K_{st} = k\} \beta_k + X_{st} \delta,
\end{align*}
where $Y_{st}$ is the risk of unintentional drug-overdose deaths in state $s$ at time interval $t$.
Similar to the GAM with a linear link, if no intervention occurred in state $s$ at time interval $t$, the logit of probability of unintentional drug-overdose deaths had no intervention occurred is
\begin{align}
      logit \left(p_{st}^{(a_{st} = 0)}\right) = \alpha_s + \gamma_{r(s)}(t) + X_{st}\delta = \phi_{st},
\end{align}
where $p_{st}^{(a_{st} = 0)}$ denotes the probability of unintentional drug-overdose deaths had no intervention occurred.
We estimate the probability of unintentional drug-overdose deaths had no intervention occurred using $\hat{\phi}_{st}$:
\begin{align}
    \sum_{k=0}^{39}\mathbb{I}\{K_{st} = k\}\hat{\beta}_k + \hat{\phi}_{st} &\approx logit \left(p_{st}\right) \nonumber \\
    \Rightarrow \hat{\phi}_{st} &\approx  logit \left(p_{st}\right) - \sum_{k=0}^{39}\mathbb{I}\{K_{st} = k\}\hat{\beta}_k  \nonumber\\
    \Rightarrow \hat{p}_{st}^{(a_{st} = 0)} = expit(\hat{\phi}_{st}) &\approx expit \left(logit \left(p_{st}\right) - \sum_{k=0}^{39}\mathbb{I}\{K_{st} = k\}\hat{\beta}_k \right).
\end{align}
The estimated number of attributable deaths using $\hat{p}_{st}^{(a_{st} = 0)}$ is then:
\begin{align}
    \hat{n}_{s, t, \text{attributable deaths}} &= n_{s, t, \text{observed overdose deaths}} \nonumber \\
    &- n_{s, t, \text{population}} expit \left(logit \left(p_{st}\right) - \sum_{k=0}^{39}\mathbb{I}\{K_{st} = k\}\hat{\beta}_k \right).
\end{align}

As in Section \ref{appendix:attributable_deaths_event_study}, we need to use the Delta Method to derive the 95\% confidence interval of the number of attributable deaths.
As before, we denote the variance-covariance of the parameters by $\Sigma_{{\beta}}$, where $\beta = \begin{pmatrix} \beta_0 & \beta_1 & \dotsc & \beta_{39}\end{pmatrix}^T$ .
However, we now have
\begin{align*}
    m({\beta}, t) &= \sum_{s} {n}_{\text{attributable deaths}, s, t} \\
    &= \sum_{s} {n}_{\text{population}, s, t} \left({p}_{st} - expit \left(logit (p_{st}) - \sum_{k=0}^{39}\mathbb{I}\{K_{st} = k\}{\beta}_k \right)\right),
\end{align*} 
where we estimate $p_{st}$ using $\hat{p}_{st}^{\text{obs}} = Y_{st}$, the observed risk of unintentional drug-overdose deaths. 
The $k$th element of the Jacobian matrix for $k \in \{0, \dotsc, 39\}$, is now equal to:
\begin{align*}
    \frac{\partial m({\beta}, t)}{\partial {\beta}_k}
    &=  \sum_{s} \frac{\partial }{\partial {\beta}_k}{n}_{\text{population}, s, t} \left({p}_{st} - expit \left(logit  (p_{st}) - \sum_{k=0}^{39}\mathbb{I}\{K_{st} = k\}{\beta}_k \right)\right) \\
    &=  -\sum_{s} \frac{\partial }{\partial {\beta}_k}{n}_{\text{population}, s, t}\left(1 + exp\left(-logit  (p_{st}) +\sum_{k=0}^{39}\mathbb{I}\{K_{st} = k\}{\beta}_k \right)\right)^{-1} \\
    &=  \sum_{s}{n}_{\text{population}, s, t}\left(1 + exp\left(-logit  (p_{st}) +\sum_{k=0}^{39}\mathbb{I}\{K_{st} = k\}{\beta}_k \right)\right)^{-2} \\
    &\times exp\left(-logit  (p_{st}) +\sum_{k=0}^{39}\mathbb{I}\{K_{st} = k\}{\beta}_k \right) \mathbb{I}\{K_{st} = k\} \\
    &=  \sum_{s}{n}_{\text{population}, s, t}\frac{expit\left(-logit  (p_{st}) +\sum_{k=0}^{39}\mathbb{I}\{K_{st} = k\}{\beta}_k \right)\mathbb{I}\{K_{st} = k\}}{1+exp\left(-logit \ (p_{st}) +\sum_{k=0}^{39}\mathbb{I}\{K_{st} = k\}{\beta}_k \right)}.
\end{align*}
The lower and upper limits of the 95\% confidence interval of the attributable deaths in at time interval $t$ are given by:
\begin{align*}
    \sum_{s} \hat{n}_{\text{attributable deaths}, s, t} \pm 1.96\sqrt{J_{{\beta}, t} \hat{\Sigma}_{{\beta}} J_{{\beta}, t}^T},
\end{align*}
where $\hat{\Sigma}_{{\beta}}$ is obtained from the sandwich estimator of the parameters as described in Section \ref{sec:appendix_sandwich_est_constant_tx}. 

\subsubsection{Sum of Attributable Deaths in All Years}
We also estimate the number of attributable deaths in the study for all years from 2000 to 2019. 
To find the total number of attributable deaths in the study, we take the sum of attributable deaths across all $t$:
$\sum_t \sum_s \hat{n}_{\text{attributable deaths}, s, t}$.
To find the 95\% confidence interval of the total number of attributable deaths from 2000 to 2019 under the assumption of constant treatment effect, we substitute $\beta$ with its 95\% confidence interval limits, as we have done previously in Section \ref{attributable_deaths_constant_tx}.
To find the 95\% confidence interval of the total number of attributable deaths from 2000 to 2019 under the assumption that the treatment effect depends on the treatment duration, we use the Delta Method. 

Following similar derivations as in the above sections, the 95\% confidence interval for the total number of attributable deaths when the treatment effect depends on the treatment duration is
$$\sum_t \sum_{s} \hat{n}_{\text{attributable deaths}, s, t} \pm 1.96 \sqrt{J_{{\beta}} \hat{\Sigma}_{{\beta}} J_{{\beta}}^T},$$
where the Jacobian entries for $\beta_k$ are:
\begin{align}
    \frac{\partial m({\beta})}{\partial {\beta}_k}
    &= \sum_t \sum_{s} {n}_{\text{population}, s, t}  {p}_{st} exp\left(- \sum_{k=0}^{39}\mathbb{I}\{K_{st} = k\}{\beta}_k \right) \mathbb{I}\{K_{st} = k\}
\end{align}
under a linear link function and 
\begin{align}
    \frac{\partial m({\beta})}{\partial {\beta}_k}
    &=  \sum_t \sum_{s}{n}_{\text{population}, s, t}\frac{expit\left(-logit  (p_{st}) +\sum_{k=0}^{39}\mathbb{I}\{K_{st} = k\}{\beta}_k \right)\mathbb{I}\{K_{st} = k\}}{1+exp\left(-logit \ (p_{st}) +\sum_{k=0}^{39}\mathbb{I}\{K_{st} = k\}{\beta}_k \right)}
\end{align}
under a logistic link function.

\subsection{Results for Estimating Attributable Deaths}
\begin{figure}[!htb]
    \centering
    \includegraphics{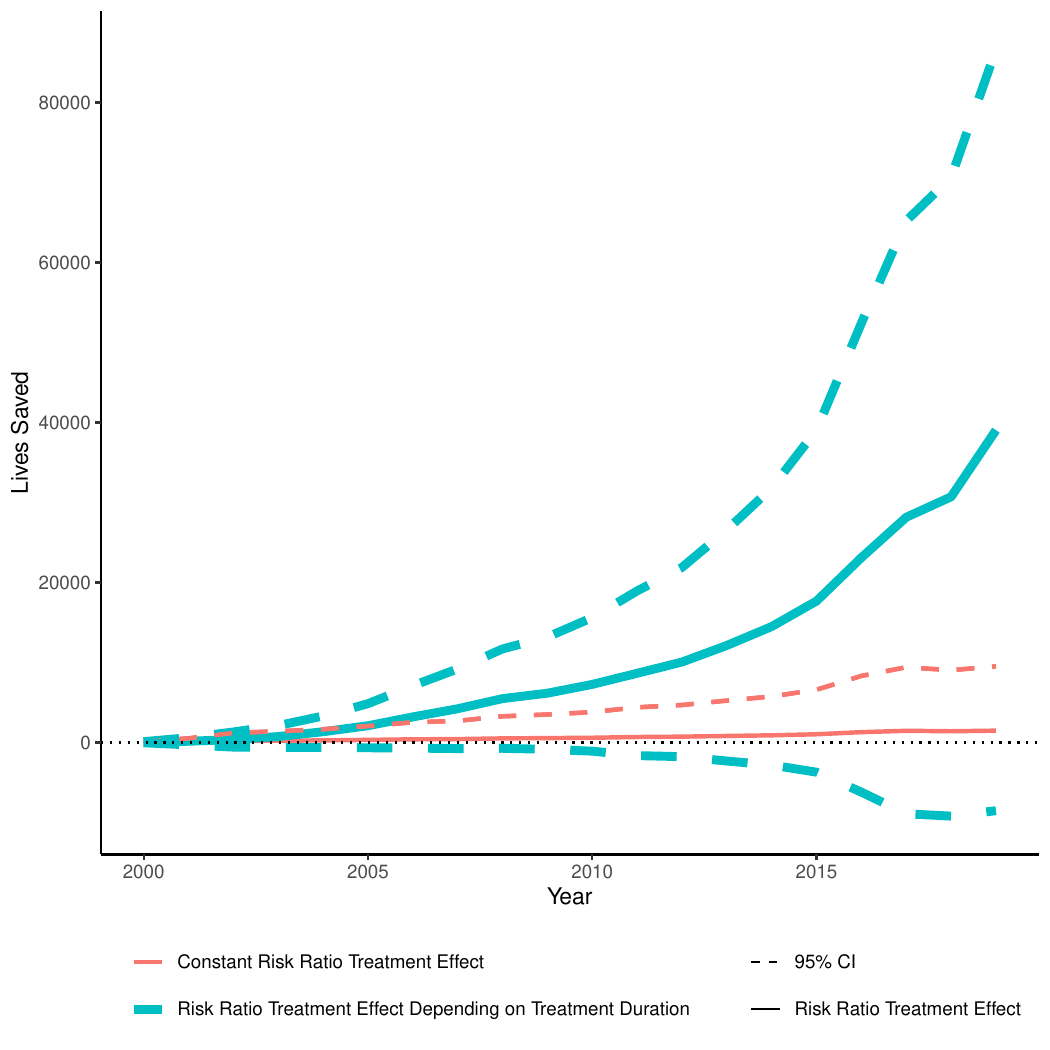}
    \caption{Estimated yearly number of lives saved due to drug-induced homicide (DIH) prosecutions reported by the media using generalized additive models (GAMs) with linear link function under constant treatment effect (red, thin lines) or the treatment effect depends on the duration (blue, thick lines). A negative number means estimated number of deaths attributable to drug-induced homicide prosecutions reported by the media.}
    \label{fig:attr_deaths_lin_link}
\end{figure}

Figure \ref{fig:attr_deaths_lin_link} shows the estimated yearly number of drug-overdose deaths attributable to DIH prosecutions reported by the media using a linear link function when assuming a constant treatment effect and when assuming the treatment effect depends on the treatment duration.
Since we found protective effects of the DIH prosecutions reported by the media, the ``deaths attributable'' become ``lives saved''.
Note that a negative number for the "lives saved" is interpreted as "attributable deaths".
From 2000 to 2019, we estimate a total of 13,234 (95\% CI: (-50,885, 85,514)) lives saved due to DIH prosecutions reported by the media when we assume that the treatment effect is constant.
When we assume that the treatment effect depends on the treatment duration, we estimate a total of 214,943 (95\% CI: (-48,337, 478,223)) lives saved attributable to DIH prosecutions reported by the media.
This is approximately 2.0\% (95\% CI: (-7.7\%, 13.0\%)) and 32.7\% (95\% CI: (-7.3\%, 72.7\%)) of the total number of  unintentional drug-overdose deaths in the U.S. from 2000 to 2019, under the constant treatment assumption and the assumption that the treatment effect depends on the treatment duration, respectively.
Since the confidence intervals include negative numbers (i.e. number of deaths attributable to DIH prosecutions), we cannot definitively say that DIH prosecutions reported by the media reduced the number of unintentional drug-overdose deaths.

\begin{figure}[!htb]
    \centering
    \includegraphics{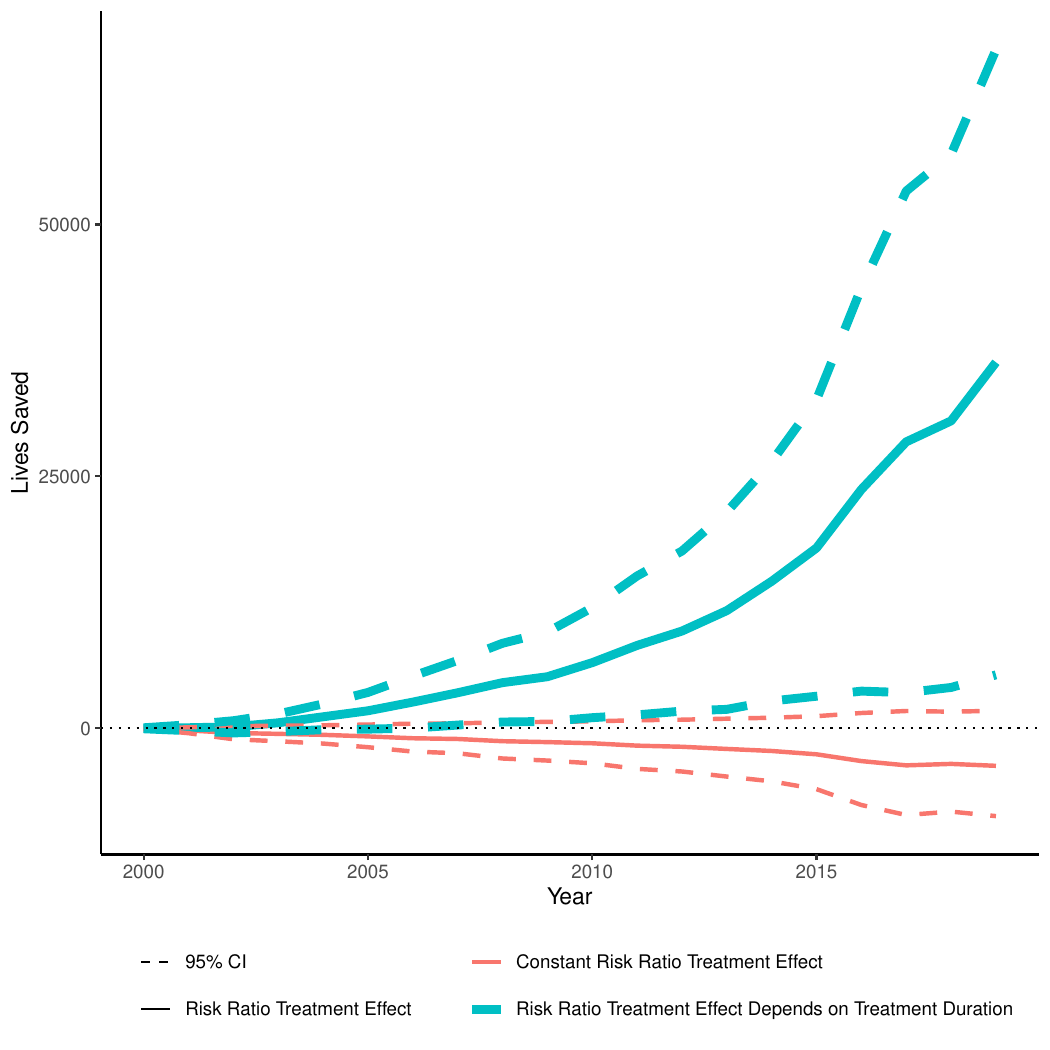}
    \caption{Estimated yearly number of lives saved due to drug-induced homicide (DIH) prosecutions reported by the media using generalized additive models (GAMs) with logistic link function under constant treatment effect (red, thin lines) or the treatment effect depends on the duration (blue, thick lines). A negative number means estimated number of deaths attributable to drug-induced homicide prosecutions reported by the media.}
    \label{fig:attr_deaths_logistic_link}
\end{figure}

Figure \ref{fig:attr_deaths_logistic_link} shows the estimated yearly number of deaths attributable to the DIH prosecutions reported by the media when assuming a model with a logistic link function. 
Note that under the assumption of a constant treatment effect, we estimate a harmful effect of the DIH prosecutions reported by the media.
Under the assumption of constant treatment effect, we estimate a total of 33,436 (95\% CI: (-15,651, 78,297)) \textit{unintentional drug-overdose deaths} are attributable to DIH prosecutions reported by the media in all states from 2000 to 2019, which is approximately 5.1\% (95\% CI: (-2.4\%, 11.9\%)) of the total number of unintentional drug-overdose deaths from 2000 to 2019.
When we assume that the effect of DIH prosecutions reported by the media depends on the treatment duration, we estimate that a total of 206,937 (95\% CI: (31,846, 382,028)) \textit{lives are saved} due to DIH prosecutions reported by the media, approximately 31.4\% (95\% CI: (4.8\%, 58.0\%)) of the total number of unintentional drug-overdose deaths from 2000 to 2019.

\clearpage
\begin{sloppypar}
{\doublespacing
\bibliography{Documents/appendix_bib}
}
\end{sloppypar}